\documentclass[11pt,a4paper]{article}
\pdfoutput=1
\usepackage{amsmath,amssymb,graphicx,subfig,bm,euscript,cancel,mathtools,setspace,color,mathrsfs,hyperref}
\usepackage[utf8]{inputenc}
\usepackage[sort&compress,square,numbers]{natbib}
\usepackage[left=35mm,right=35mm,top=35mm,bottom=40mm]{geometry}
\numberwithin{equation}{section}
 \def\eq#1{Eq.~(\ref{#1})}

 \def\Fig#1{Fig.~{\ref{#1}}}
 \def\beq{\begin{equation}}
 \def\eeq{\end{equation}}
 \def\beqa{\begin{eqnarray}}
 \def\eeqa{\end{eqnarray}}
 
 \newcommand{\bs}{\boldsymbol}
 \def\one{\!\!{\hbox{ 1\kern-.8mm l}}}
 
 \newcommand{\ket}[1]{|{#1}\rangle}
 
 \newcommand{\ex}[1]{{\rm e}^{#1}}
 \renewcommand{\d}{{\rm d}}
 \def\ii{{\rm i}}

\makeatletter
\newcommand{\dotminus}{\mathbin{\text{\@dotminus}}}
\newcommand{\@dotminus}{%
  \ooalign{\hidewidth\raise1ex\hbox{.}\hidewidth\cr$\m@th-$\cr}%
}
\makeatother

\newcommand{\tran}{^{\text{t}}}

\DeclareMathAlphabet\EuRoman{U}{eur}{m}{n}
\SetMathAlphabet\EuRoman{bold}{U}{eur}{b}{n}

\newcommand{\ie}{\emph{i.e.}}
\newcommand{\eg}{\emph{e.g.}}

\newcommand{\psl}{\ensuremath{\text{PSL}(2)}}

\newcommand{\wt}{\ensuremath{\widetilde}}
\newcommand{\wh}{\ensuremath{\widehat}}

\newcommand{\osp}{\ensuremath{\text{OSp}(1|2)}}
\newcommand{\ve}{\varepsilon}
\newcommand{\cf}{\emph{cf.}}

\allowdisplaybreaks
\begin{document}
\begin{titlepage}
\begin{flushright}
\vspace*{-25pt}
\end{flushright}
\begin{center}
\vspace{1cm}
{\Large \bf Pinching parameters for open (super) strings}

\vspace{8mm}
Sam Playle
\footnote{email: {\tt playle@to.infn.it}}\, \,
Stefano Sciuto
\footnote{email: {\tt sciuto@to.infn.it}}
\vskip .5cm
{\sl Dipartimento di Fisica, Universit\`a di Torino  \\
and INFN, Sezione di Torino}\\
{\sl Via P. Giuria 1, I-10125 Torino, Italy}\\
\vskip 1.2cm
\begin{abstract}
\noindent
We present an approach to the parametrization of (super) Schottky space obtained by sewing together three-punctured discs with strips. Different cubic ribbon graphs classify distinct sets of pinching parameters; we show how they are mapped onto each other. The parametrization is particularly well-suited to describing the region within (super) moduli space where open bosonic or Neveu-Schwarz string propagators become very long and thin, which dominates the IR behaviour of string theories. We show how worldsheet objects such as the Green's function converge to graph theoretic objects such as the Symanzik polynomials in the $\alpha ' \to 0$ limit, allowing us to see how string theory reproduces the sum over Feynman graphs. The (super) string measure takes on a simple and elegant form when expressed in terms of these parameters.
\end{abstract}
\end{center}

\vfill

\end{titlepage}
\tableofcontents
\section{Introduction}
\label{intro}

In this article, we lay out all the details of the approach introduced in \cite{Playle:2016hie}. There we suggested a new set of `pinching parameter' co-ordinates for (super) Schottky space of multi-loop open string worldsheets. These pinching parameters are useful in a number of ways.

One benefit of the pinching parameters is that they clarify how various Feynman diagram topologies with arbitrary numbers of loops emerge from the string theory measure on (super) moduli space, in the zero slope limit.
It has been known since the early days of string theory \cite{Scherk:1971xy,Scherk:1974ca} that as the string tension goes to infinity (\ie~$\alpha' \to 0$), string scattering amplitudes reproduce those of a quantum field theory as the stringy states become infinitely massive and decouple. In the case of bosonic strings or type I/II superstrings, the open string sector yields gauge theories in the $\alpha ' \to 0$ limit while the closed string sector gives gravitation theories.

This fact has been used to gain insight about the computation of particle scattering amplitudes (without assuming any position on the validity of superstring theory itself as an accurate description of the world). The added complication due to the replacement of particles with extended objects is offset by a number of simplifications, for example, traces over colour factors and integrals over loop momenta are done for free in the string-based approach, with Schwinger-parametrized Feynman integrands arising naturally as the limits of the string measure on moduli space. An analysis of how this works in terms of tropical geometry was proposed in \cite{Tourkine:2013rda}.

 Starting in the late 1980's, string-inspired approaches to the computation of one-loop amplitudes were developed for gauge theories  \cite{Bern:1987tw,Bern:1990cu,Bern:1990ux,Bern:1991an,Bern:1993mq}
and gravitation \cite{Bern:1993wt}. It was shown in \cite{Bern:1991an} how to relate the one-loop string-inspired rules to the Feynman graph based approach --- but only if the Feynman diagrams were computed using a particular non-Hermitian gauge condition for the Yang-Mills gauge fields, identified by Gervais and Neveu in \cite{Gervais:1972tr}. The reasons for this gauge preference have recently been explained \cite{Pesando:2017mct}. In \cite{DiVecchia:1996uq} it was established which regions of the moduli space contributed to which Feynman diagram topologies.

Going beyond one loop, it becomes important to choose a useful set of worldsheet moduli to integrate over. The benefits of using Schottky groups for string-inspired QFT calculations were laid out in \cite{Roland:1992cc}. Schottky groups had arisen as a parametrization of the moduli space of multi-loop string worldsheets in the earliest days of string theory, emerging automatically from an approach to amplitudes based on sewing together string interaction vertices \cite{Carbone:1970sy,Kosterlitz:1970up,Lovelace:1970sj,Kaku:1970ym,Alessandrini:1971cz,Alessandrini:1971dd}. The use of Schottky groups in string theory was put on a surer footing in the 1980's with the use of methods such as BRST quantization \cite{DiVecchia:1986uu,DiVecchia:1987uf,DiVecchia:1988cy,DiVecchia:1988jy,DiVecchia:1989id},
conformal field theory \cite{Martinec:1986bq} and the `group theoretic approach' \cite{Neveu:1987xb,Neveu:1988am}, and a super-geometric extension of Schottky groups \cite{Manin1986} allowed the analysis to be generalized to the Neveu-Schwarz (NS) sector of the superstring.

Detailed two-loop scalar QFT computations using Schottky groups have been presented in \cite{DiVecchia:1996kf} and \cite{Frizzo:1999zx}.  Similar techniques have been used to compute multi-loop open string amplitudes with twisted periodicity conditions, allowing the analysis of states between D-branes at an angle or with a constant background gauge field switched on, allowing the computation of the two-loop Euler-Heisenberg effective action \cite{Russo:2003tt,Russo:2003yk,Magnea:2004ai}. In fact, with such approaches, it is possible to examine the various terms in the string moduli space integrand and identify not only separate Feynman graph \emph{topologies}, but to see which fields are propagating through which edges in the graphs, allowing string-based computations to be used on a diagram-by-diagram basis 
\cite{Magnea:2013lna,Magnea:2015fsa}.

Canonically, Schottky space is co-ordinatized by the fixed points and multipliers of a set $\{ \gamma_i \}$ of hyperbolic M\"obius maps which generate the Schottky group. The presence of two types of parameter complicates the measure and makes it difficult to find a symmetric mapping onto the Schwinger parameters $t_i$ of a Feynman graph in the zero-slope limit. This problem was solved for the `sunset' two-loop vacuum Feynman graph in
\cite{Magnea:2013lna,Magnea:2015fsa}. At two loops there is a symmetry between three of the Schottky group multipliers \cite{DiVecchia:1987uf} which allows us to write down the three pinching parameters straightforwardly. But for Feynman graphs of more complicated topologies, there are no longer enough constraints to obtain a set of `pinching parameter' moduli from graph symmetries alone.

Our approach is based on the idea of building open string worldsheets by sewing together three-punctured discs with plumbing fixtures. For the bosonic string, the `lengths' of the plumbing fixtures are the only parameters, and they become the moduli of the resulting surfaces. The construction turns out to be equivalent to a particular parametrization of Schottky groups. At two loops, our approach precisely reproduces the pinching parameters obtained by symmetry in \cite{Magnea:2013lna,Magnea:2015fsa}, but our approach has the advantage of being defined for general Feynman graph topologies, with arbitrary numbers of loops and external edges.

 The basic idea of our analysis resembles a number of other approaches which have appeared in the Physics and Mathematics literature. Fock gives an approach very similar to ours in spirit, in which the geometry of a ribbon graph is translated into M\"obius maps between its vertices, although the mapping is not the same \cite{Fock:1993pr}. The idea of building string Feynman diagrams by joining cubic vertices with propagators was used in Witten's open string field theory \cite{Witten:1985cc} which is known to give a single cover of moduli space \cite{Zwiebach:1990az}. But the Witten vertex yields a complicated mapping between strip lengths and Koba-Nielsen variables \cite{Giddings:1986iy}, whereas in our approach the Koba-Nielsen variables are rational functions of the pinching parameters.
 Recently, Sen has constructed a gauge-invariant 1PI effective action for superstring theory by separating the contribution from string diagrams which can be obtained by gluing lower-order diagrams with plumbing fixtures \cite{Sen:2014pia,Sen:2014dqa,Sen:2015hha}.
  Penner described how to use trivalent ribbon graphs to give a cell decomposition of moduli space \cite{Penner:1986hy}, and Marden gave a coordinatization of Teichm\"uller space derived from the starting point of sewing three-punctured spheres, although the transition functions are not all M\"obius maps (unlike ours) \cite{Marden:1986hz}.

The use of the pinching parameters provides a link between worldsheet objects defined in terms of Riemann surfaces such as the period matrix and the Green's function, and graph theoretic objects such as the Symanzik polynomials  (\cf~sections 3.4 and 4.1 of \cite{Tourkine:2013rda}). To be precise, we can express pinching parameters $p_i$ in terms of dimensionful Schwinger parameters $t_i$ by setting $p_i = \ex{ - t_i / \alpha'}$, and then in taking the $\alpha ' \to 0$ limit we find that geometric objects asymptote to graph theoretic objects. We focus on the worldline Green's function, which is used in string-inspired calculations of scalar QFT amplitudes. Dai and Siegel described in an elegant paper how the worldline Green's function could be found for arbitrary multi-loop diagrams using methods from electric circuit theory  \cite{Dai:2006vj}. We show how the building blocks used in their analysis can be arrived at from the limiting behaviour of Riemann surface objects. This illustrates, in particular, how Feynman graphs for scalar QFTs can arise from the asymptotics of string theory diagrams.

Our approach also defines coordinates on the super-moduli space of super Riemann surfaces (SRS), near corners corresponding to NS open string degenerations.
In recent years there has been a resurgence of interest in SRS and super moduli space \cite{Witten:2012ga,Witten:2012bh,Witten:2013cia,Witten:2013tpa,Donagi:2013dua,Donagi:2014hza}.
Superstring perturbation theory contains integrals which are only conditionally convergent in the IR region, and there can be ambiguities arising from the choice of which even supermoduli to hold fixed when integrating over the odd supermoduli. For example, consider the integral $\int_{y = \epsilon}^{1-\epsilon} \d \theta\, \d \phi \, \d \log y$, where $y$ is an even supermodulus and $\theta$, $\phi$ are odd supermoduli. This seems to vanish on the face of it, but after
making the change of variables $ y = 1 - u - \theta\,\phi$ this is no longer the case. The choice of which even supermoduli to hold fixed while integrating over the odd supermoduli is thus non-arbitrary, and a complete formulation of perturbative superstring theory requires a prescription of which even supermoduli to use. Witten shows that the correct even supermoduli to fix near the corners of supermoduli space are the so-called \emph{canonical parameters}
(see, for example, section 6.3 of \cite{Witten:2012ga} and section 2.4 of \cite{Witten:2013cia}).
 The pinching parameters that we have written down for the NS sector are defined by gluing pairs of punctures \emph{\`a la} Eq.~(6.31) of \cite{Witten:2012ga}, which means they constitute a set of canonical parameters (canonical parameters, in general, are only defined modulo coordinate choices around the punctures). Thus our pinching parameters allow the correct computation of conditionally convergent integrals. The NS pinching parameters are, roughly, like square roots of the bosonic pinching parameters.
 The three-punctured discs we use for the construction only have NS punctures. Such discs have a Grassmann-odd super-projective invariant; these invariants become the odd supermoduli.

One benefit of the pinching parameterization is that the string measure on (super) Schottky space is given a very elegant and symmetric form. In particular, there is a direct correspondence between poles in the measure and edges in the corresponding Feynman graph. This differs from the canonical coordinatization of Schottky space for which the geometric meaning of the poles can be unintuitive. Furthermore, it is interesting that the measure includes a product over the borders of the worldsheet, meaning that, for example, the measure for a planar two-loop vacuum diagram (which has three borders) is qualitatively different from the measure for the non-planar two-loop vacuum diagram (which has just one border) --- a distinction that is completely obscured with the canonical coordinates.

The construction given in section \ref{constr} requires the use of a cyclic ordering of the three punctures on the discs that are used to construct the worldsheets. This is appropriate for open strings because they have a fixed ordering at interactions. Conversely, closed strings do not have a fixed ordering at interactions, and so they cannot be described by these pinching parameters in a natural way. It would be interesting to find an extension of the pinching parametrization that exhibits the symmetries of closed string amplitudes.

The organization of the paper is as follows. In section \ref{constr} we present the construction of the pinching parametrization, first in the bosonic case in section \ref{param} and then for the Neveu-Schwarz sector in section \ref{supparam}. In section \ref{graph} we describe how the graph period matrix and worldline Green's function arise as the $\alpha ' \to 0$ limit of objects on moduli space, given a straightforward mapping between pinching parameters and Schwinger parameters. Section \ref{DualitySect} describes the relations between sets of pinching parameters that correspond to pairs of graphs differing by a duality transformation on one of the internal edges. In section \ref{measu} we give the leading part of the (super) string measure in terms of the pinching parameters, and show how it can be proven. In appendix \ref{schotapp} we provide proofs of several claims in the paper: we derive the leading behaviour of pinching parameter expressions for the (semi) multiplier in section \ref{multiplierproof}, for the period matrix in section \ref{pmproof} and for the Green's function in section \ref{greenfapp}.

\section{Construction of the pinching parametrization}
\label{constr}
Our goal is to construct a set of (super) moduli for string worldsheets that have a simple geometric interpretation in terms of degenerating surfaces, and thus can be used gainfully to analyse the $\alpha ' \to 0$ limit of string amplitudes.
\subsection{Bosonic strings}
\label{param}
First we consider the simpler bosonic case. In section \ref{koban} we describe how sewing discs can lead to an expression for Koba-Nielsen variables in terms of pinching parameters, and then in section \ref{schotk} we show how the construction can be extended to incorporate loops with the use of Schottky groups.
\subsubsection{Local coordinates at the punctures}
\label{koban}
\begin{figure}
\centering
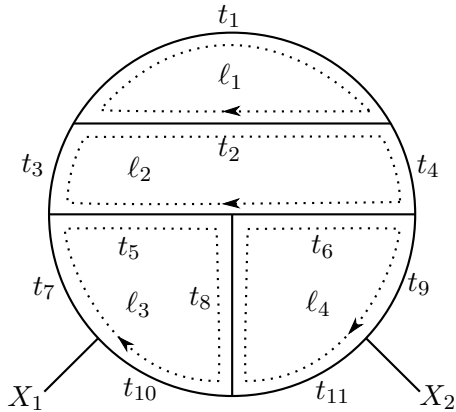 \label{fig:ssb}
\caption{
A graph with $E = 11$ internal edges labelled with Schwinger parameters $t_i$, $n=2$ external edges labelled $X_j$ and $g=4$ loops with a basis $\{ \ell_i\}$. }\label{examplegraph1}
\end{figure}
Let the \emph{target graph} $\Gamma$ be a connected ribbon graph (\ie~the edges connect to vertices with a fixed cyclic ordering) whose vertices are all either three-valent or one-valent (we call edges joining a 1-valent vertex \emph{external edges}). Let all edges whose endpoints are both three-valent vertices (possibly the same one) be called \emph{internal edges} and be labelled with a \emph{Schwinger parameter} $t_k$. If the graph has $g$ loops and $n$ external edges, then we can label the external edges $X_i$ ($i=1, \ldots, n$), label the cubic vertices $v_j$ ($j=1,\ldots, 2g+n-2$), and label the internal edges $E_k$ ($k=1,\ldots,3g+n-3$). Furthermore, we choose a homology basis of $g$ loops in $\Gamma$, \ie~a set of $g$ closed paths $\ell_i$ in $\Gamma$ which are independent in the sense that no non-trivial composition of the $\ell_i$ is homologous to a trivial path. An example target graph with 4 loops, 11 internal edges and 2 external edges is shown in \Fig{examplegraph1}.

Given such a graph, we can build a corresponding bordered Riemann surface $\Sigma$ parametrized by $3g-3+n$ real moduli $\{ p_i\}$ (\ie~coordinates on Schottky space). When we take all of the $p_i \to 0$, the surface degenerates into $\Sigma^0$: a union of 3-punctured discs joined at nodes, so we call the $p_i$'s \emph{pinching parameters}. The topology of $\Sigma^0$ is classified by the target graph $\Gamma$.

We build $\Sigma$ from two ingredients: 3-punctured discs and open-string plumbing fixtures. On each 3-punctured disc, we put three standard coordinate charts: one centred at each puncture. If we label the punctures $a_1$, $a_2$ and $a_3$ with a clockwise ordering, then we want three coordinate charts $z_i$, $i=1,2,3$ satisfying \begin{align}
z_i(a_i) & = 0 \, . \label{disccoorddef1}
\end{align}
The image of the disc under $z_i$ is the upper-half plane with the projective real line as its boundary, so let us fix the $z_i$ coordinates of the other two coordinates to $1$ and $\infty$:
\begin{align}
z_i(a_{i+1}) & = \infty &
z_i(a_{i-1}) & = 1 \, ,  \label{disccoorddef2}
\end{align}
where the indices are mod 3. Then there is a unique M\"obius map $\rho$ which serves as a transition function cycling the three charts.
In order to satisfy
\begin{align}
z_i & = \rho(z_{i+1}) \, , \label{rhomeaning}
\end{align}
we need to have
\begin{align}
\rho(0) & = \infty \, ; &
\rho(\infty) & = 1 \, ; &
\rho(1) & = 0 \, .
\end{align}
This is satisfied by
\begin{align}
\rho(z) & \equiv 1 - \frac{1}{z} \, , \label{rhoanalytic}
\end{align}
which can be written as the \psl~matrix
\begin{align}
\rho & \equiv \left(\begin{array}{cc} 1 & - 1 \\ 1 & 0 \end{array}\right) \, .  \label{rhodef}
\end{align}
The inverse transformation is given by
\begin{align}
\rho^{-1} (z) & = \frac{1}{1-z} \, ; &
\rho^{-1} & \equiv \left( \begin{array}{cc} 0 & 1 \\ -1 & 1 \end{array}\right) \, . \label{rhoi}
\end{align}
Note, of course, that $\rho^3 = 1$.

\begin{figure}
\centering
\subfloat[]{ 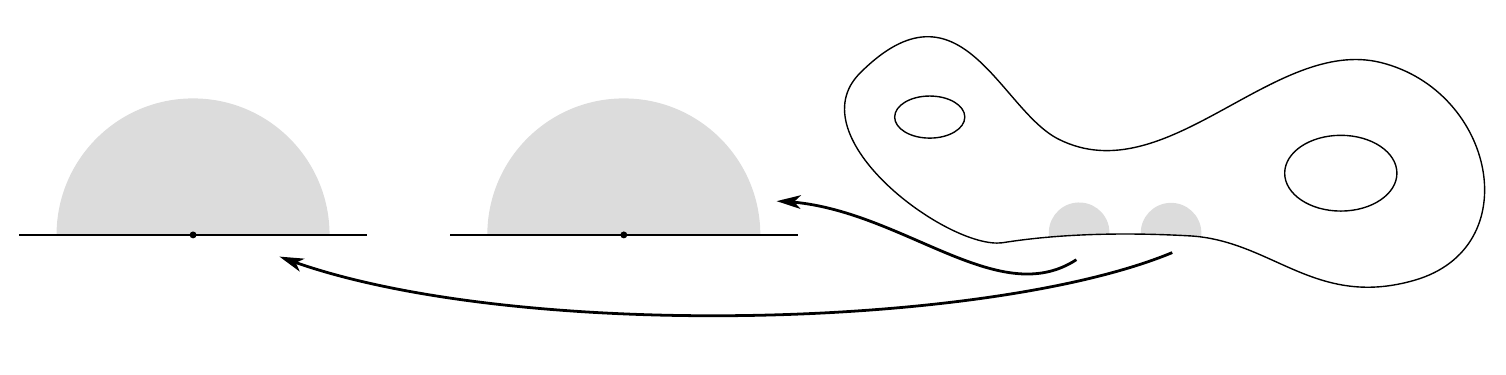 \label{fig:plumb1} }\\
\subfloat[]{ 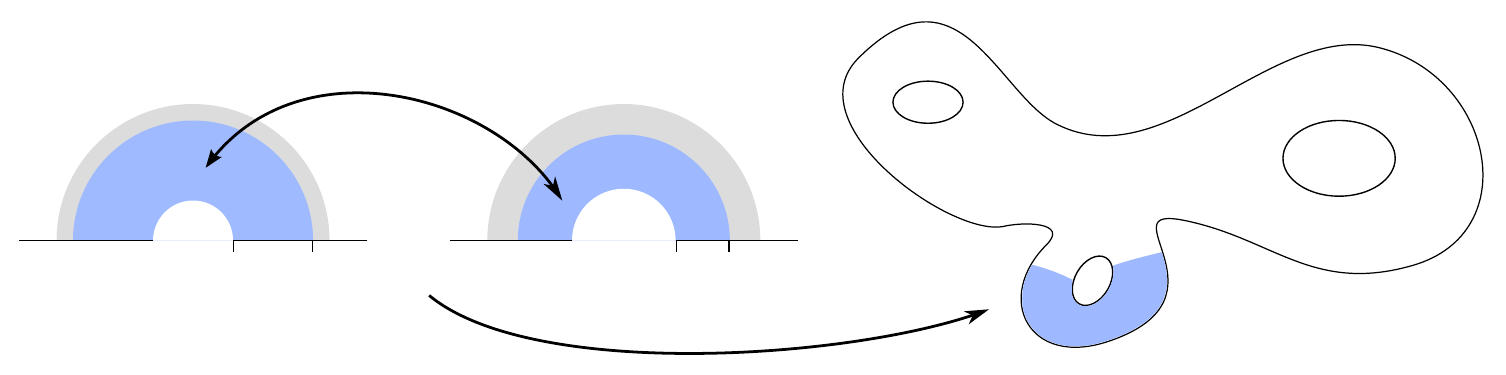 \label{fig:plumb2} }
\caption{
Given a bordered Riemann surface $\Sigma$ with two charts $z$ and $w$ which map the border of $\Sigma$ onto the real line and whose images include a neighbourhood of 0 (\Fig{fig:plumb1}), we can add an \emph{open string plumbing fixture} to $\Sigma$ (\Fig{fig:plumb2}) by imposing \eq{plumbeq} on a suitable domain.
}
\label{fig:plumb}
\end{figure}

The other ingredient is the `open string plumbing fixture'. Suppose a Riemann surface $\Sigma$ has two charts $w$ and $z$, both of which map its border to the real line, and whose images each include a neighbourhood around 0 (\Fig{fig:plumb1}). Then for a given pinching parameter $p \in (0,1)$, we can \emph{insert a plumbing fixture} by cutting out two semi-discs $|w| < \epsilon_w $ and $|z| < \epsilon_z$ and imposing the relation
\begin{align}
z w & = - p \label{plumbeq}
\end{align}
for $ \epsilon_w < |w | \leq p / \epsilon_z$ and $ \epsilon_z< |z| \leq p / \epsilon_w$ (\Fig{fig:plumb2}).
The effect of the plumbing fixture is to add a `strip' between the two charts on $\Sigma$, either by adding a loop or by joining together two connected components.

Once a plumbing fixture is in place, we can use it to get a transition function between the two charts we used to define it: we can say
\begin{align}
w & = \sigma_p(z) \, \label{sigmameaning}
\end{align}
where
\begin{align}
\sigma_p(z) & \equiv - \frac{p}{z} \, \, ;  &
\sigma_p & \equiv \frac{1}{\sqrt{p}} \left(\begin{array}{cc} 0 & - p  \\ 1 & 0 \end{array}\right) \, .  \label{sigmadef}
\end{align}

The two types of transition function we've defined, \eq{rhodef} and \eq{sigmadef}, are the two `building blocks' we'll use to build the Riemann surface $\Sigma$. In order to define $\Sigma$, we need to specify the transition functions between any pair of coordinate charts, as well as the positions of all the punctures. To achieve this, we lay out the three-punctured discs and plumbing fixtures in the arrangement defined by the target graph $\Gamma$. Each disc comes with three local coordinate systems as defined by \eq{disccoorddef1} and \eq{disccoorddef2}, related to each other by \eq{rhomeaning}. Furthermore, two coordinate systems $w$ and $z$ which lie at opposite ends of the same open string plumbing fixture can be related to each other by \eq{sigmameaning}. Thus, by finding a path between two charts, we can write down a transition function as an alternating product of $\sigma_{p_i}$'s and $\rho^{\pm 1}$'s.

\begin{figure}
\centering
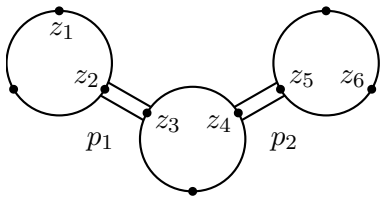
\caption{Three 3-punctured discs sewn to give a surface with $n=5$ punctures. The $z_i$'s label coordinate charts centred at the indicated points. We can use the pinching parametrization to find transition functions between any pair of them in terms of the $p_i$'s. Note that the centres of $z_2, \ldots, z_5$ get excised from the surface by the sewing procedure.
}
\label{fig:sewingex}
\end{figure}

For example, consider the surface indicated by the diagram in \Fig{fig:sewingex}; suppose we want to find the transition function between the charts $z_1$ and $z_6$. We can build it piece-by-piece: first we note that since $z_1$ and $z_2$ are on the same disc with the centre of $z_1$ situated \emph{anti-clockwise} from the centre of $z_2$, we can use \eq{rhomeaning} to see $z_1  = \rho(z_2)$. Next, since $z_2$ and $z_3$ are at opposite ends of the open string plumbing fixture labelled by the pinching parameter $p_1$, we have $z_2 = \sigma_{p_1}(z_3)$. We can combine this with the previous equation to get a transition function straight from $z_3$ to $z_1$: $z_1 = (\rho \cdot \sigma_{p_1})( z_3)$. The next two pairs of consecutive coordinate systems are related to each other by $z_3 = \rho(z_4)$ and $z_4 = \sigma_{p_2}(z_5)$. Lastly, because $z_5$ is on the same disc as $z_6$ but in this case centred \emph{clockwise} from it, we have $z_5 = \rho^{-1} (z_6)$. Putting this all together we find $z_1 = \big(\rho \cdot \sigma_{p_1} \cdot \rho \cdot \sigma_{p_2} \cdot \rho^{-1}\big) (z_6)$.

Clearly, it was not necessary to write down the intermediate coordinate charts, and in fact, the transition function depended only on the path taken to get from the puncture at $z_6 = 0$ to the puncture at $z_1 = 0$. This is the case in general: the procedure for getting a surface $\Sigma$ from a target graph $\Gamma$ can be defined in terms of a mapping from \emph{paths} on $\Gamma$ into the space \psl~of M\"obius maps.

\begin{figure}
\centering
\subfloat[]{ 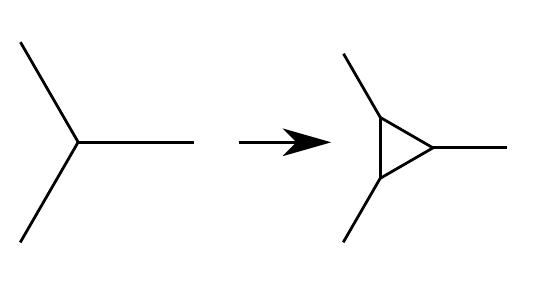 \label{fig:blowup} }
\subfloat[]{ 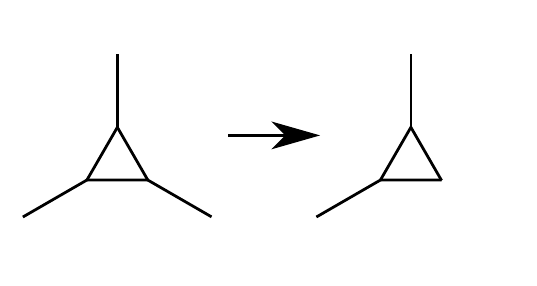 \label{fig:cutedge} }
\caption{$\widehat \Gamma$ is obtained from $\Gamma$ by first blowing up all the cubic vertices into triangles (\Fig{fig:blowup}) and then deleting any external edges but reusing their labels (\Fig{fig:cutedge}).
}
\label{fig:gammahatdef}
\end{figure}

In order to define the parametrization precisely, instead of using paths on $\Gamma$ we must construct a new graph $\widehat{\Gamma}$ in the following way: Firstly, we `blow up' every 3-valent vertex into a triangle whose corners are 3-valent vertices as shown in \Fig{fig:blowup}. Secondly, we delete every external edge, leaving a now 2-valent vertex which we label with the label $X_j$ that belonged to the deleted external edge, as shown in \Fig{fig:cutedge}.
Let us distinguish the edges in triangles and the edges inherited from $\Gamma$ as \emph{short edges} and \emph{long edges}, respectively.
The triangles in $\widehat \Gamma$ correspond to the three-punctured discs used to construct $\Sigma$; the edges correspond to the open string plumbing fixtures.

Every vertex in $\widehat \Gamma$ corresponds to a local coordinate chart on $\Sigma$, and every path in $\widehat \Gamma$ corresponds to a transition function between the charts at its two ends. There is a homomorphism $\phi$ from the groupoid of paths on $\widehat \Gamma$ to \psl; we can define it as follows.

\newcommand{\cw}{\ensuremath{\text{\textsc{cw}}}}
\newcommand{\acw}{\ensuremath{\text{\textsc{acw}}}}
\newcommand{\tra}[1]{\ensuremath{\text{\textsc{t}}_{#1}}}
Every path $P$ can be decomposed into three elementary types of paths.
The first type of elementary path is the traversal of one of the long edges $E_k$ that was inherited from $\Gamma$ with Schwinger parameter $t_k$. Let us denote this type of path by (\tra{k}). The second type of elementary path is simply a `rotation' from one vertex to another vertex on the same triangle. We can distinguish further between rotations depending on whether the end vertex is immediately clockwise or anti-clockwise of the start vertex: let us call them (\cw) and (\acw) respectively. Thus any path in $\widehat \Gamma$ can be written as a sequence composed of (\cw), (\acw) and (\tra{k}).
\begin{figure}
\centering
\subfloat[]{ 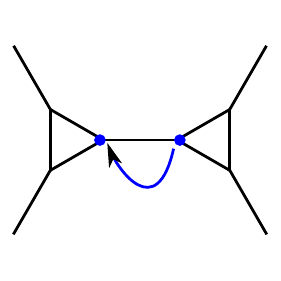 \label{fig:Ek} }
\subfloat[]{ 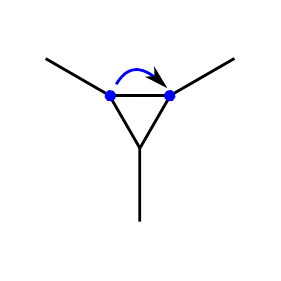 \label{fig:cw} }
\subfloat[]{ 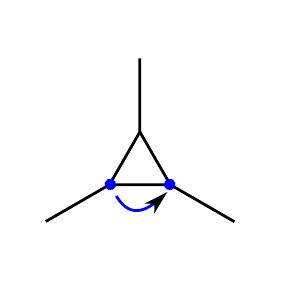 \label{fig:acw} }
\caption{
A path in $\widehat \Gamma$ can be decomposed into three elementary types of path: traversal of a long edge $E_k$ (\tra{k}) (\Fig{fig:Ek}), clockwise moves around a triangle (\cw) (\Fig{fig:cw}) and anti-clockwise moves around a triangle (\acw) (\Fig{fig:acw}).
}
\label{fig:basicmoves}
\end{figure}

Once decomposed in this way, a path $P$ can be written as a M\"obius map $\phi(P) \in \psl$ with the homomorphism $\phi$ where:
\begin{align}
\phi(\acw) & =  \rho \, , &
 \phi(\cw) & =  \rho^{-1} \, ,   &
\phi(\tra{k}) & =  \sigma_{p_k} \, , \label{phidef}
\end{align}
in terms of the \psl~maps defined in \eq{rhodef}, \eq{rhoi} and \eq{sigmadef}.

Note that to do this for a graph $\Gamma$ we have implicitly introduced a set $\{p_k\}$ of pinching parameters with one associated to each long edge $E_k$. These will become the coordinates on Schottky space.

Also note that because $\rho^3 = \sigma_{p_k}^2 = \text{id}$, the mapping $\phi$ is insensitive to sub-paths which traverse an edge then reverse, or which visit all three nodes at a cubic vertex in succession.

Given a graph $\Gamma$, let us choose a vertex in $\widehat \Gamma$ and fix it as the `base point' $B$. We'll use the coordinate chart defined by $B$ as a canonical chart and write down all the other charts with respect to it. In particular, we want to find the transition functions to the charts which vanish at the punctures on the surface $\Sigma$ (corresponding to the external edges on $\Gamma$). To achieve this, for each external edge $X_j$, we find a path $P_{X_j}$ from $X_j$ to $B$ on the blown-up graph $\widehat \Gamma$. Then the $n$ sets of local coordinates at the external punctures are given by $V_{X_j}^{-1}$, where
 \begin{align}
 V_{X_j} & \equiv \phi(P_{X_j}) \, , \label{PtoV}
 \end{align}
 in terms of the mapping $\phi$ defined by \eq{phidef}.
  The Koba-Nielsen variables corresponding to the positions of the external punctures as seen in the chart defined by the base point $B$ are given as
 \begin{align}
 z_{j} & = V_{X_j}(0) \, . \label{KNfromV}
 \end{align}

\begin{figure}
\centering
\subfloat[]{ 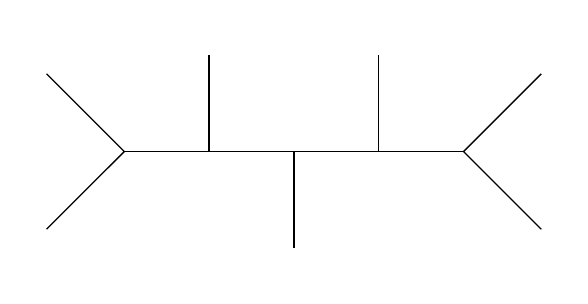 \label{fig:treeex} }
\subfloat[]{ 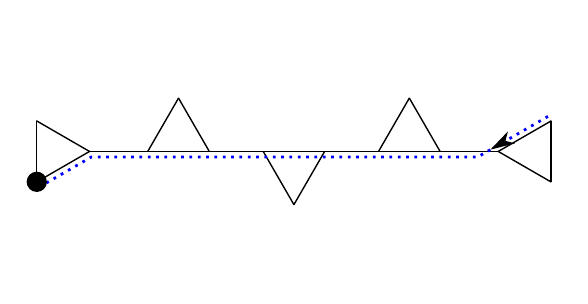 \label{fig:treeexh} }
\caption{
\Fig{fig:treeex} shows an $n=7$ tree graph $\Gamma$; the external edges are labelled $X_i$, $i=1,\ldots, 7$ and the internal edges are labelled $E_j$, $j=1, \ldots , 4$. The blown-up graph $\widehat \Gamma$ (\Fig{fig:treeexh}) has a base point at $B= X_1$ (indicated with a dot); a path $P_{X_5}$ from $X_5$ to $B$ is indicated with the blue dotted line.
}
\label{fig:treeexfig}
\end{figure}

  For example, consider the $n=7$ tree graph $\Gamma$ shown in \Fig{fig:treeex} and its blow-up $\widehat \Gamma$ in \Fig{fig:treeexh}.
A path $P_{X_5}$ from $X_5$ to the base point $B = X_1$ is indicated on $\widehat \Gamma$, and it can be decomposed into the basic moves of \Fig{fig:basicmoves} as
  \begin{align}
  P_{X_5} & = \cw \, \cdot \,\tra{1} \, \cdot \,\cw \, \cdot \,\tra{2} \, \cdot \,  \acw \, \cdot \, \tra{3}  \, \cdot \, \cw  \, \cdot \, \tra{4} \, \cdot \,\acw .
  \end{align}
  Using \eq{phidef}, this gives us
  \begin{align}
  V_{5} = \phi(P_{X_5}) & = \rho^{-1} \, \cdot \, \sigma_{p_1} \, \cdot \, \rho^{-1} \, \cdot \, \sigma_{p_2} \, \cdot \, \rho \, \cdot \, \sigma_{p_3} \, \cdot \, \rho^{-1} \, \cdot \, \sigma_{p_4} \, \cdot \, \rho \, ;
  \end{align}
  plugging this in \eq{KNfromV} we find that the Koba-Nielsen variable for the corresponding point on $\Sigma$ is given in terms of the pinching parameters as
  \begin{align}
  z_5 & = \frac{1 + p_3}{1 + p_1 + p_3 + p_1 p_3 + p_1 p_2 p_3} \, .
  \end{align}
  Since the base point is at $B=X_1$, we have $V_1 = \text{id}$ and so the Koba-Nielsen variable corresponding to the first puncture is at $z_1 = 0$. The Koba-Nielsen variables for the other five punctures $z_i(p_j)$ can be computed similarly.

\subsubsection{The Schottky group}
\label{schotk}
We've introduced a mapping $\phi$ from the set of paths in $\widehat \Gamma$ into \psl, and said that it can be used to find transition functions between charts on $\Sigma$. But for multiply-connected graphs $\Gamma$, there are many independent paths joining any pair of vertices in $\widehat \Gamma$. This would seem to violate the cocycle condition, but in fact is not a problem once we understand all the charts as being defined modulo a \emph{Schottky group} $G$.

\begin{figure}
\centering
\subfloat[]{ 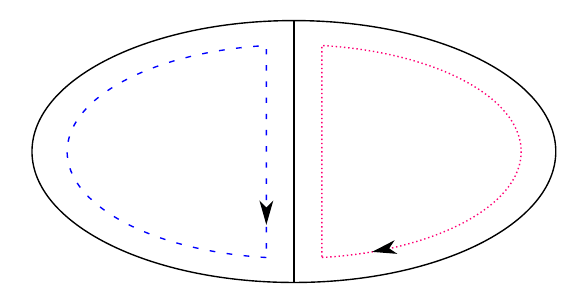 \label{fig:halfappg} }
\subfloat[]{ 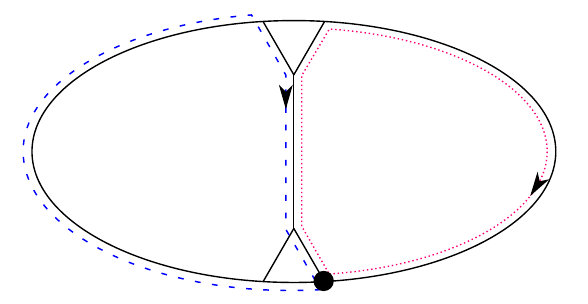 \label{fig:halfapphg} }
\caption{The graph $\Gamma$ with $n=0$, $g = 2$ comes with a homology basis of two loops $\ell_1$ and $\ell_2$ (\Fig{fig:halfappg}); these can be used to define paths $P_{\ell_1}$ and $P_{\ell_2}$ on the blown-up graph $\widehat \Gamma$ which start and end at the base point $B$ (marked as a black dot on \Fig{fig:halfapphg}).
}
\label{fig:halfappex}
\end{figure}

Recall that $\Gamma$ came with a homology basis $\{ \ell_i\}_{i=1}^g$; each loop $\ell_i$ can be used to define a closed path $P_{\ell_i}$ on $\widehat \Gamma$ which starts and ends at a chosen base point $B$. See \Fig{fig:halfappex} for a two-loop example.

Then we define $g$ M\"obius maps $\gamma_i$ with
\begin{align}
\gamma_i & \equiv \phi(P_{\ell_i}) \, . \label{PtoSchot}
\end{align}
The $\gamma_i$'s are hyperbolic M\"obius maps---that is to say, they are conjugate to dilatations $z \mapsto k_i\, z$, and thus they can be parametrized by their \emph{multiplier} $k_i$ with $|k_i|<1$, and their \emph{attractive} and \emph{repulsive} fixed points $u_i$ and $v_i \in \mathbf{R} \cup \{\infty\} $. The $\gamma_i$'s are related to their parameters via
\begin{align}
\frac{\gamma_i(z) - u_i}{\gamma_i(z) - v_i} & = k_i \, \frac{z - u_i}{z - v_i }\, .
\end{align}
From another point of view, the vectors $\ket{u_i} \equiv (u_i, 1)\tran$ and $\ket{v_i} \equiv (v_i, 1)\tran$ are eigenvectors of the matrix $\gamma_i$, satisfying
\begin{align}
\gamma_i \ket{u_i} & = \frac{1}{\sqrt{k_i}} \ket{u_i} \,
&
\gamma_i \ket{v_i} & = \sqrt{k_i} \ket{v_i} \, ;
\end{align}
these equations can be solved to express the parameters $k_i$, $u_i$ and $v_i$ as functions of the pinching parameters $\{p_j\}$.

 The rank-$g$ group $G$ freely generated by the $\gamma_i$'s is a \emph{Schottky group} of genus $g$. Schottky groups give a classical construction of Riemann surfaces as quotient spaces; in the early days of string perturbation theory they arose naturally by sewing $n$-string vertices \cite{Kaku:1970ym,Lovelace:1970sj,Alessandrini:1971cz,Alessandrini:1971dd}. Let us briefly recall the Schottky group construction for open strings; a more detailed discussion is given, for example, in section 2.1 of \cite{Playle:2015sxa}.

\begin{figure}
\centering
\subfloat[]{ 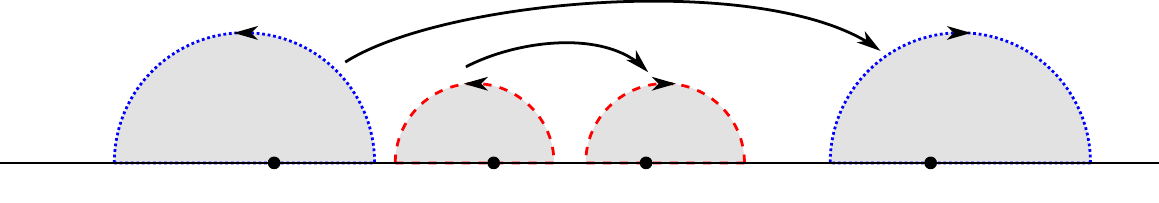 \label{fig:osschot} } \\
\subfloat[]{ \small 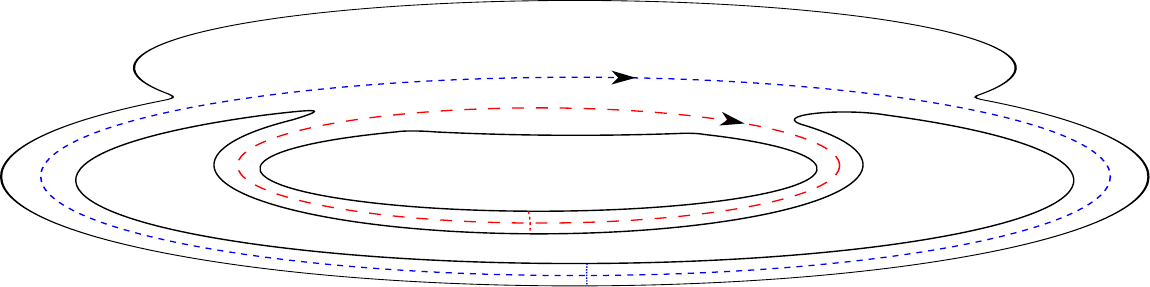 \label{fig:osschotb} }
\caption{The Schottky group construction (\Fig{fig:osschot}) of a planar $g=2$ surface with $b=3$ borders (\Fig{fig:osschotb}). The construction induces a loop basis $\{ \ell_1, \ell_2\}$ on the surface.}
\label{fig:schottky}
\end{figure}

 The construction uses the fact that M\"obius maps take circles to circles. In particular, any semicircle $\cal C$ that is centred on the real axis and that separates the two fixed points $u$, $v$ of a hyperbolic M\"obius map $\gamma$ is mapped by $\gamma$ to another such semicircle $\cal C' $ (see \Fig{fig:osschot}). This cuts the upper-half plane into three regions: one containing $u$, one containing $v$, and one (say $\cal F$) that contains neither fixed point.
 Quotienting the closure $\overline{\cal F}$ by $\gamma$ has the effect of identifying the two semicircles on its boundary $\cal C \sim \cal C '$; the resulting surface is an annulus.

The construction makes sense with any number $g$ of M\"obius maps $\gamma_i$ as long as we can find an appropriate set of $2g$ semicircles $\{ {\cal C}_i, {\cal C}_i ' \}$ such that $\gamma_i({\cal C}_i) = {\cal C}_i'$ and that all lie on the boundary of a single region $ \cal F$ (see \Fig{fig:osschot} for an example with $g=2$). Then quotienting the closure $\overline{\cal F}$ by $G$ has the effect of identifying pairs of semicircles ${\cal C}_i \sim {\cal C}_i'$, where $G$ is the group generated by the $g$ $\gamma_i$'s. The resulting surface is a bordered Riemann surface of genus $h = (g + 1 - b)/{2}$ where $b$ is the number of border components (note that $h=0$ for planar topologies, which have $b = g+1$ borders). $\cal F$ can be treated as a single chart covering the surface. The example in \Fig{fig:osschotb} has $g=2$, $h=0$.

The surface doesn't depend on the choice of semicircles; any set of semicircles with empty intersection that satisfies $\gamma_i({\cal C}_i) = {\cal C}_i'$ gives the same surface $\Sigma$. It's not strictly necessary for ${\cal C}_i$ and $\gamma_i({\cal C}_i)$ to be geometric semicircles, only that they have the right topology; there exist \emph{non-classical} Schottky groups such that the ${\cal C}_i$ can't all be chosen as geometric semicircles, but the Schottky construction remains valid.

 The $g$ M\"obius maps $\{ \gamma_i \}$ thus define a surface, but there is some redundancy because two Schottky groups that are related by a \psl~change of coordinates define the same surface. We can choose three specific parameters, say $u_1$, $v_1$ and $v_2$, to `gauge fix' to some chosen values; this saturates the reparametrization symmetry. The remaining $3g-3$ parameters $\{ u_i \}_{i = 2}^g \cup \{  v_i  \} _{i = 3}^g \cup\{k_i \}_{ i = 1}^g$ can be used as moduli. The space described by these parameters is called \emph{Schottky space}.

 The $3g-3$ Schottky parameters can themselves be expressed in terms of pinching parameters via \eq{PtoSchot}. Note that for graphs $\Gamma$ with no external edges $n=0$, the number of internal edges of $\Gamma$ is $|E| = 3g-3$, so the pinching parameters can be used as a local coordinate system on Schottky space near the corresponding degeneration.

 Recall that for graphs $\Gamma$ with $n \geq 1$ external edges, we need to find paths $P_{X_i}$ from the external edges to the base point $B$ on $\widehat \Gamma$. But with $g \geq 1$ loops, there isn't a unique way to choose such a path. Two different paths $P_{X_i} \neq \wt P_{X_i} $ will give two different inverse charts $V_{X_i} = \phi(P_{X_i})$ and $\wt V_{X_i}= \phi(\wt P_{X_i})$, and therefore two different values for the Koba-Nielsen variable of $X_i$: one will give $z_{X_i} = V_{X_i}(0)$ and the other will give $\wt z_{X_i} = \wt V_{X_i}(0)$. But $z_{X_i}$ is defined only modulo the Schottky group $G$, so this isn't a problem. We can find a path $P_{\ell_\alpha } \equiv P_{X_i} \cdot (\wt P_{X_i} )^{-1}$; because this path starts and ends at $B$, it can be expressed in terms of the homology basis $\{ P_{\ell_i} \}$. Thus the M\"obius map $\gamma_\alpha \equiv \phi(P_{\ell_\alpha})$ can be written as a product of the $\gamma_i = \phi(P_{\ell_i})$, so it is a member of the Schottky group, $\gamma_\alpha \in G$. So $V_{X_i} = \gamma_\alpha \cdot \wt V_{X_i} $, and thus $z_{X_i} = \gamma_\alpha( \wt z_{X_i})$; in other words, the procedure defines the Koba-Nielsen variable unambiguously modulo the Schottky group $G$.

  So the $g$ Schottky group generators and the $n$ Koba-Nielsen variables, which can be used as a set of moduli for the surface, can be written as functions of the $3g+n-3$ gluing parameters $p_k$, and thus the $p_k$'s can be used as coordinates on a suitable region of Schottky space.

\subsubsection{Properties of the parametrization}
\label{bosonprops}
\begin{enumerate}
\item \label{nochoice} Since $\phi$ is a homomorphism, the Schottky group $G$ doesn't depend on the choice of homology basis $\{ \ell_i \}$, \ie~on which of its elements are treated as generators, but only on the graph $\Gamma$.
\item \label{basechange} Changing the base node $B \mapsto \wh B$ is equivalent to a global change of coordinates: let $P_{B\wh B}$ be a path from $B$ to $\wh B$ then the parametrization with the base node $\wh B$ is given by taking
    \begin{align}
    \wh \gamma_i  & = \phi(P_{B\wh B}) \,\cdot\, \gamma_i \,\cdot\, \phi(P_{B\wh B})^{-1} \, ;
    &
    \wh V_{X_j} & = \phi(P_{B\wh B}) \,\cdot\, V_{X_j} \, .
    \end{align}
\item  \label{multone} Let us use the term \emph{loop} to refer to a closed path in $\Gamma$ defined modulo cyclic permutations, \ie~the path $E_{j_n} \cdot \ldots \cdot E_{j_2} \cdot E_{j_1}$ defines the same loop $\ell$ as $E_{j_1} \cdot E_{j_n } \cdot \ldots \cdot E_{j_2}$ (see \Fig{fig:path} for an illustration).
     Every closed path $P_\alpha $ in $\widehat \Gamma$ corresponds to a loop $\ell_\alpha$ in $\Gamma$.
      To obtain $\ell_\alpha$, we first list the edges traversed by $P_\alpha$. If the first and last edges listed are the same then we can use invariance under cyclic permutations to delete them from the list, since they must be traversed in opposite directions. This can be repeated until the first and last edges that remain are distinct.

If $\ell_\alpha$ is a loop corresponding to a closed path $P_\alpha$ then the multiplier of the Schottky group element $\gamma_\alpha \equiv \phi ( P _{\ell _\alpha})$ is given to leading order as
 \begin{align}
 k_\alpha & = \Big( \prod_i \, p_i^{\, n_\alpha{}^{\!\! i}} \, \Big)\, (1 + {\cal O}(p_j) ) \, , \label{eq:multexpr}
 \end{align}
 where $n_\alpha{}^{\!\! i}$ counts the number of times that the loop $\ell_\alpha$ crosses the edge $E_i$.
 \begin{figure}
\centering
\subfloat[]{ 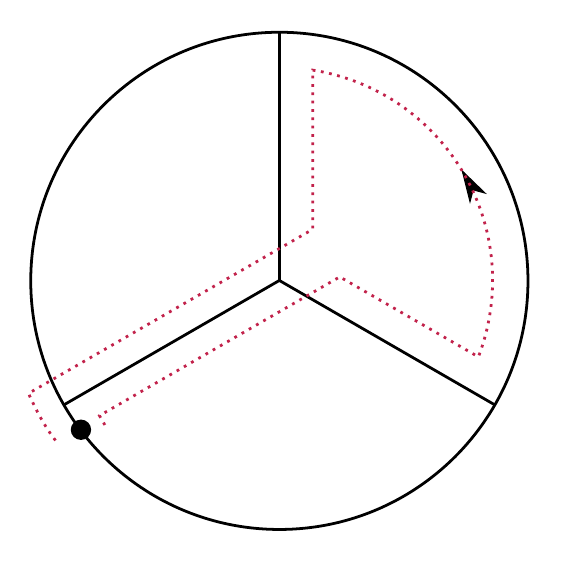 \label{fig:path1} }
\subfloat[]{ 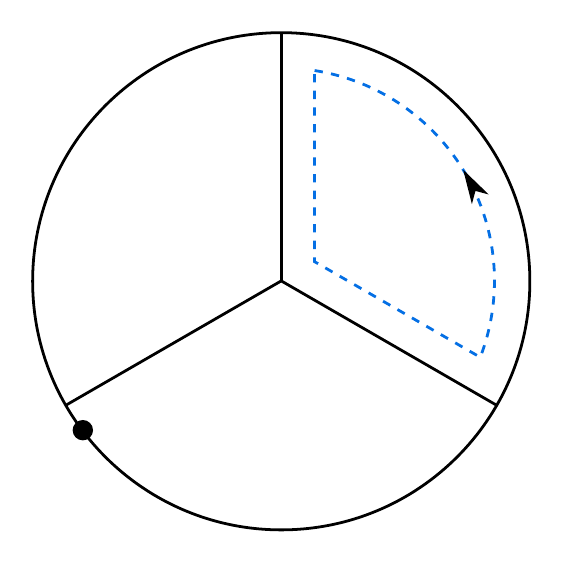 \label{fig:path2} }
\caption{
We distinguish between closed paths $P_\alpha$ in the graph (\Fig{fig:path1}) and the corresponding loops $\ell_\alpha$ (\Fig{fig:path2}), which are equivalence classes of closed paths defined modulo conjugation, and which are independent of the base point.
}
\label{fig:path}
\end{figure}
\item  \label{multtwo} If the path $P_\alpha$ is conjugate to a loop $\ell_\alpha$ that consists solely of either clockwise or anticlockwise turns at the vertices, then the multiplier is precisely the product of the gluing parameters of the edges traversed by $\ell_\alpha$:
 \begin{align}
 k_\alpha & =  \prod_i \, p_i^{\, n_\alpha{}^{\!\! i}} \, .
 \end{align}
\item   \label{dmcomp} The moduli $p_i$ characterize some of the boundaries of the Deligne-Mumford compactification of moduli space: the boundary corresponding to the pinching of an edge with gluing parameter $p_i$ is given by taking $p_i \to 0$.
\item
Our construction is only suitable for open strings, not closed strings, because the transition functions on the three-punctured discs can only be written down by assuming a cyclic ordering of the punctures, meaning that strings must be assumed to interact with a fixed cyclic ordering.
\end{enumerate}

\subsubsection{Examples}
\begin{figure}
\centering
\subfloat[]{ 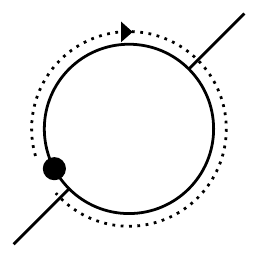 \label{fig:eg1c} }
\subfloat[]{ 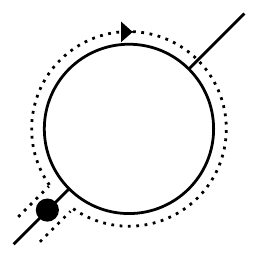 \label{fig:eg1a} }
\subfloat[]{ 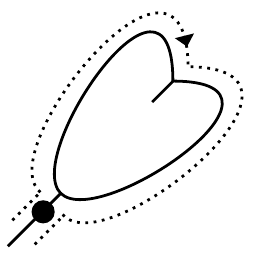 \label{fig:eg1b} }
\subfloat[]{ 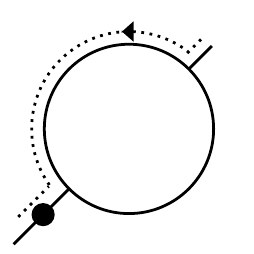 \label{fig:eg1d} }
\subfloat[]{ 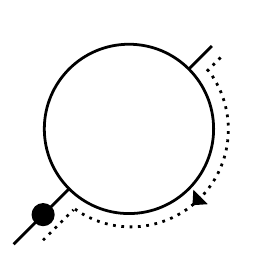 \label{fig:eg1e} }
\caption{Some examples of graphs $\Gamma$ with paths corresponding to loops (\Fig{fig:eg1c}--\Fig{fig:eg1b}) and local coordinates at punctures (\Fig{fig:eg1d}--\Fig{fig:eg1e}). The base points $B$ are marked as large dots.}\label{fig:eg1}
\end{figure}
Some illustrative examples are given in \Fig{fig:eg1}. Note that we have drawn the paths $P_i$ on the target graphs $\Gamma$ instead of on the blown-up graphs $\widehat\Gamma$, because the target graphs $\Gamma$ are simpler but we can unambiguously represent paths on $\widehat \Gamma$ in this way, as long as we indicate the base point. To indicate the base point (which is a vertex of $\widehat \Gamma$) on $\Gamma$, we draw a black dot near the appropriate vertex on the appropriate incident edge.
\begin{enumerate}
\item
\Fig{fig:eg1c} corresponds to a path which can be expressed as follows:
\begin{itemize}
\item Traverse the edge labelled $E_1$.
\item Move anticlockwise around the next vertex onto the edge labelled $E_2$.
\item Traverse the edge labelled $E_2$.
\item Having arrived at the starting vertex, move \emph{anti}clockwise around the vertex to arrive back at the base point.
\end{itemize}
More concisely, the path is $P_{1} = \acw\cdot \tra{2} \cdot \acw \cdot \tra{1} $, so the M\"obius map $\gamma_1 = \phi(P_1)$ is given by
\begin{align}
\gamma_{1} & = \rho \, \sigma_{2} \, \rho \, \sigma_{1}   = \frac{1}{\sqrt{p_1}\sqrt{p_2}}\left(\begin{array}{cc} 1 & p_1(1+p_2) \\  0 &p_1 p_2 \end{array} \right) \, .  \label{gam1c}
\end{align}
\item
The loop shown in \Fig{fig:eg1a} is the same as the one in \Fig{fig:eg1c} except that the base point $B$ has been moved to a different leg of the same vertex. In this case, the base point is on a different edge from the one the path starts on, so the first step must be a clockwise move about the vertex:
\begin{itemize}
\item Starting from the base point $B$, move clockwise about the vertex onto the edge labelled $E_1$.
\item Traverse the edge labelled $E_1$.
\item 
Move anticlockwise around the next vertex onto the edge labelled $E_2$.
\item Traverse the edge labelled $E_2$.
\item Having arrived at the starting vertex, move clockwise around the vertex to arrive back at the base point.
\end{itemize}
More concisely, the path can be written right-to-left as $P_2 = \cw\cdot \tra{2} \cdot \acw \cdot \tra{1} \cdot \cw$, so the M\"obius map $\gamma_2 = \phi(P_2)$ corresponding to this path is given as
\begin{align}
\gamma_2 & = \rho^{-1}\, \sigma_{2} \, \rho \, \sigma_{1}  \, \rho^{-1}  = \frac{1}{\sqrt{p_1}\sqrt{p_2}}\left(\begin{array}{cc} p_1 & -(1+p_1) \\ p_1(1+p_2) & -(1+p_1(1+p_2)) \end{array} \right) \, .  \label{gam1a}
\end{align}
Note that this can be obtained from the map in \eq{gam1c} after a conjugation by $\rho$. This means, in particular, that their multipliers are the same.
\item
The loop depicted in \Fig{fig:eg1b} is almost the same as the one in \Fig{fig:eg1a} except that the path must cross the second vertex by making a clockwise turn, not an anticlockwise turn. So in this case, we must change the third step
to
\begin{itemize}
\item Move \emph{clockwise} around the next vertex onto the edge labelled $E_2$.
\end{itemize}
while the other steps are unchanged.
The path is now $P_{3} = \cw\cdot \tra{2} \cdot \cw \cdot \tra{1} \cdot \cw$ and thus the M\"obius map $\gamma_3 = \phi(P_3)$ corresponding to this path is given as
\begin{align}
\gamma_{3} & = \rho^{-1}\, \sigma_{2} \, \rho^{-1} \, \sigma_{1}  \, \rho^{-1}  =\frac{1}{\sqrt{p_1}\sqrt{p_2}}\left(\begin{array}{cc} 0 & 1 \\  - p_1 p_2 & 1+p_2(1+p_1)\end{array} \right)  \, .
\end{align}
\item
The path shown in \Fig{fig:eg1d} corresponds to a local coordinate at the external puncture. The path, as always, starts at the external puncture $X_1$ and finishes at the base point $B$. It is composed of the following steps:
\begin{itemize}
\item Move anticlockwise about the vertex from the external puncture $X_1$ onto the edge labelled $E_1$.
\item Traverse the edge labelled $E_1$.
\item Move anticlockwise about the vertex onto the base point $B$.
\end{itemize}
The path can be written as $P_4 = \acw \cdot \tra{1} \cdot \acw$, and thus the corresponding local coordinate system $V_4 = \phi(P_4)$ is given by
\begin{align}
V_4 & = \rho \, \sigma_1 \, \rho \, = \frac{1}{\sqrt{p_1}}\left(\begin{array}{cc} -1 -p_1 & 1 \\  -p_1 & 0 \end{array} \right) \, .
\end{align}
The Koba-Nielsen variable $z_4$ corresponding to this puncture is given by
\begin{align}
z_4 & = V_4 (0) = \infty \, .
\end{align}
\item
Similarly, the path shown in \Fig{fig:eg1e} is an alternative choice of local coordinates for the same puncture, equivalent modulo the action of the Schottky group. In this case the path is given by $P_5 = \cw \cdot \tra{2} \cdot \cw$, so the associated coordinate chart $V_5 = \phi(P_5)$ is given by the M\"obius map
\begin{align}
 V_5 & = \rho^{-1} \, \sigma_2 \, \rho^{-1} \, = \frac{1}{\sqrt{p_2}}\left(\begin{array}{cc}0 & 1 \\  -p_2 & 1 + p_2 \end{array} \right) \, ,
\end{align}
so the Koba-Nielsen variable is given by
\begin{align}
 z_5 & = V_5 (0) = \frac{1}{1+p_2} \, .
\end{align}
The two coordinate choices in the last two examples are related to each other by the action of the Schottky generator $\gamma_2$ given in \eq{gam1a} (since these examples share a base point with the graph in \Fig{fig:eg1a}). We have
\begin{align}
V_5 & = \gamma_2 \cdot V_4 \,
\end{align}
and in particular
\begin{align}
z_5 & = \gamma_2(z_4) \, .
\end{align}
 In general, alternative choices of path $P_a$ which give different Koba-Nielsen variables for the same puncture actually just correspond to different representatives of the same Schottky group equivalence class.
 \end{enumerate}

\subsection{Neveu-Schwarz superstrings}
\label{supparam}
The description of Riemann surfaces by Schottky groups can be extended to describe non-split super Riemann surfaces, with the restriction that only $2^g$ of the $2^{2g}$ possible spin-structures are admissible---roughly, those in which the odd coordinate is single-valued on the Schottky covering space, \ie~the $a_i$ components of the $\vartheta$-characteristics are fixed. In physics terms, these \emph{super Schottky groups} \cite{Martinec:1986bq,Manin1986} can be used to describe RNS superstring worldsheets near degenerations in which all the pinched string channels belong to the NS sector.

See section 2.2 of \cite{Playle:2015sxa} for a discussion of super Schottky groups; we will use the conventions and notation of that reference. Then to extend the approach we used in section \ref{param}, we need to describe how to give $g$ super Schottky generators $\bs \gamma_i$ and $n$ sets of local coordinates $\bs V_j$, now from $\osp$ instead of \psl.

As in section \ref{param}, we give a homomorphism $\bs \phi$ from the semigroup of paths between nodes in $\Gamma$ into $\osp$. To do this, we endow each of the $(2g+n-2)$ 3-valent vertices $v_i$ in the target graph $\Gamma$ with a Grassmann-odd parameter $\theta_i$. Also, we must pick a direction (arbitrarily) for each edge in $\Gamma$. Each path must be decomposed into \emph{four} types of building block: now the traversal of an edge with Schwinger parameter $t_k$ must take into account whether its sense is the same as, or opposite from, the edge's direction marking (denoted $E_k$ and $E_k{}^{-1}$, respectively), and the two types of turn, $\acw_i$ and $\cw_j$, must also be labelled by the corresponding 3-valent vertex.

Then $\bs \phi$ is given by
\begin{eqnarray}
\bs \phi(\acw_i) & = & \bs\rho_{\theta_i} \equiv \left(\begin{array}{cc|c} - 1 & 1 & -\theta_i \\ - 1 & 0 & 0 \\ \hline -\theta_i & 0 & 1 \end{array}\right) \label{brhodef} \\
\bs \phi(\cw_i) & = &\bs \rho_{\theta_i}^{-1} = \left(\begin{array}{cc|c} 0 & - 1 & 0  \\ 1 & - 1 & \theta_i \\ \hline 0 & -\theta_i & 1 \end{array}\right)
\label{brhoidef} \\
\bs \phi(E_j{}^{-1}) & = & \bs \sigma_{\ve_j}^{-1} \equiv \left(\begin{array}{cc|c} 0 &  \ve_j & 0 \\ -\ve_j^{-1} & 0 & 0  \\ \hline 0 & 0 & 1 \end{array}\right) \, \\
\bs \phi(E_j) & = & \bs \sigma_{\ve_j} \equiv \left(\begin{array}{cc|c} 0 & - \ve_j & 0 \\ \ve_j^{-1} & 0 & 0  \\ \hline 0 & 0 & 1 \end{array}\right) \, , \label{bsigdef}
\end{eqnarray}
(note that $\bs \sigma_{\ve_k} \in \osp$ is not its own inverse, unlike $\sigma_{p_k} \in \psl$). Just as in section \ref{param}, we choose a base node $B$ and then write down an appropriate path for each loop and each external edge. Now using $\bs \phi$, we can write each path as an element of $\osp$, giving us the $g$ super Schottky group generators $\bs \gamma_i$ and the $n$ local coordinates $\bs V_j$.

The derivation of the parametrization is similar to the bosonic case, but some changes need to be made. Firstly, instead of 3-punctured discs, the surfaces are to be built from SRS discs with three NS punctures (NNN discs). Whereas 3-punctured discs have no moduli, NNN discs have one odd supermodulus: if the three punctures have superconformal coordinates $\bs x_i = x_i| \xi_i$, then
\begin{align}
\Theta_{\bs x_1 \bs x_2 \bs x_3} & \equiv \pm \frac{\xi_1 (\bs x_2 \dotminus \bs x_3) + \xi_2(\bs x_3 \dotminus \bs x_1) + \xi_3(\bs x_1 \dotminus \bs x_2) + \xi_1 \xi_2 \xi_3}{\sqrt{(\bs x_1 \dotminus \bs x_2)(\bs x_2 \dotminus \bs x_3)(\bs x_3 \dotminus \bs x_1)} }\, \label{oddinvariant}
\end{align}
is a superprojective (pseudo)-invariant. Here $\dotminus$ denotes the NS difference of two superpoints: if $\bs a = a|\alpha$ and $\bs b = b|\beta$ then
\begin{align}
\bs a \dotminus \bs b \equiv a - b - \alpha \beta \, . \label{superdiff}
\end{align}
 $\Theta_{\bs x_1 \bs x_2 \bs x_3}$ can be used as the odd supermodulus of the NNN disc. We can fix a global superconformal chart so the three punctures have homogeneous coordinates
\begin{align}
&(0,1 |0)\tran \, ,  & &(1,0|0)\tran \, , & &(1,1|\Theta_{\bs x_1 \bs x_2 \bs x_3})\tran \, .
\end{align}
Then \eq{brhodef} and \eq{brhoidef} can be used to cycle between the three charts on an NNN disc whose pseudo-invariant is $\theta_i$, in the same way that \eq{rhodef} and \eq{rhoi} cycle between the three charts on a 3-punctured disc.

The second change that needs to be made to the bosonic derivation is that the plumbing fixture \eq{plumbeq} needs to be replaced with a NS plumbing fixture. If $\bs z = z |\zeta$ and $\bs w = w |\psi$ are two superconformal charts with small discs cut out around $0|0$, then we can sew them together by identifying
\begin{align}
z\, w & = - \ve^2 \, ,  &
z \, \psi & = \ve \, \zeta \, ,  &
w \, \zeta & = - \ve \, \psi \, , &
\psi \, \zeta & = 0 \, \label{NSplumb}
\end{align}
(Eq.~(6.31) of \cite{Witten:2012ga}). Here $\ve$ is a Grassmann-even parameter which behaves roughly as the square root of a bosonic pinching parameter; we call it the ``NS pinching parameter'' of the plumbing fixture. \eq{NSplumb} is satisfied by taking $\bs w = \bs \sigma _{\ve}(\bs z)$, where $\bs \sigma _{\ve}$ is the superconformal function
\begin{align}
\bs \sigma _{\ve}(\bs z) & \equiv  - \ve^2 / z \, \big| \, \ve \, \zeta / z  \, ;
\end{align}
this function is equivalent to the \osp~matrix in \eq{bsigdef}.

The $(3g+n-3)$ even parameters $\ve_i$ and the $(2g+n-2)$ odd parameters $\theta_j$ can be used as local coordinates for super Schottky space.

\subsubsection{Properties of the parametrization of supermoduli}
The obvious analogues of properties \ref{nochoice} and \ref{basechange} from section \ref{bosonprops} also hold in the NS case. For hyperbolic \osp~maps (such as super Schottky group elements), it is more useful to talk about \emph{semimultipliers} than multipliers: each such map is conjugate to one of the form $z | \zeta \mapsto q^2 z | q \zeta$, then $q$ is the semimultiplier (see section 2.2.3 of \cite{Playle:2015sxa}). Thus properties \ref{multone} and \ref{multtwo} need to be replaced with expressions for the \emph{semi}-multipliers of loops.
\begin{itemize}
    \item  Just as in the bosonic case, every super Schottky group element $\bs \gamma_\alpha$ can be associated to a closed path $P_\alpha \in \pi_1( \Gamma, B)$ in the fundamental group of $ \Gamma$.
    Furthermore, we can reduce the closed path $P_\alpha$ to a \emph{loop} $\ell_\alpha$  by cancelling out any consecutive pairs of edges which are inverses of each other.
    The semimultiplier $q_\alpha$ of $\bs \gamma_\alpha$ is given to leading order as
    \begin{align}
    q_\alpha & = (-1)^{n_\alpha} \Big(  \prod_{E_i \in \Gamma} \ve_i^{m_\alpha^i} \Big) (1 + {\cal O}(\ve_j)) \, , \label{loopsemi}
    \end{align}
    where
       \begin{align} n_\alpha & = \# \text{ of \emph{clockwise} turns in }\ell_\alpha\\
    &\, \, \, \, \, \, \, \,  + \, \, \, \, \#\text{ of edges that }\ell_\alpha\text{ crosses \emph{against} the marked direction} \nonumber
    \shortintertext{and}
    m_\alpha^i & =\#\text{ of times that }\ell_\alpha\text{ crosses the edge }E_i \, .
    \end{align}
    This formula is proved in Appendix \ref{multiplierproof}.
    \item For paths $P_\alpha$ which are composed solely of \cw~or \acw~turns, $q_\alpha$ is given exactly by the leading term of \eq{loopsemi}:
    \begin{align}
    q_\alpha & = (-1)^{n_\alpha}\,  \prod_{E_i \in \Gamma} \ve_i^{m_\alpha^i} \, . \label{loopsemiex}
    \end{align}
\item   \label{sdmcomp} The supermoduli $\ve_i$ characterize some of the boundaries of the Deligne-Mumford compactification of supermoduli space: the boundary corresponding to the pinching of an edge with gluing parameter $\ve_i$ is given by taking $\ve_i \to 0$.
\end{itemize}

\section{Graph polynomials from Riemann Surfaces}
\label{graph}
Dai and Siegel \cite{Dai:2006vj} showed how Feynman graph integrands for scalar field theories can be computed in terms of geometric properties of the skeleton graph. In this section we show how the objects they used can be obtained as the $\alpha ' \to 0$ limit of objects defined on (super) Riemann surfaces in terms of the parametrization given above, so long as we make the following identification between Schwinger times and gluing parameters:
\begin{align}
p_i & = \ex{ - t_i / \alpha ' } \, . \label{pteq}
\end{align}
 Let $\Gamma$ be a $g$-loop Schwinger-parametrized cubic Feynman graph with a chosen homology basis $\ell_i$, then we can define the $g \times g$ \emph{graph period matrix} $(\theta_{ij})$ of $\Gamma$. First define an inner product measuring the degree of intersection of two paths:
 \begin{align}
\langle P, \wh{P}  \rangle^k  & \equiv \begin{cases}
1 & \text{ if }P \text{ and } \wh P \text{ cross }E_k \text{ in the same direction}\\
- 1 & \text{ if }P \text{ and } \wh P \text{ cross }E_k \text{ in opposite directions}\\
0 & \text{ if }P \text{ and } \wh P\text{ do not both cross }E_k.
\end{cases}
\end{align}
 Then the graph period matrix is given by
\begin{align}
\theta_{ij} & = \sum_k \langle \ell_i , \ell_j \rangle ^k \, t_k\, , \label{thetadef}
\end{align}
where $i,j = 1 , \ldots , g$.
Let $X_a$ and $X_b$ be two external edges on $\Gamma$, then choose a path $P_{ab}$ from $X_b$ to $X_a$. Then define a vector $\vec v$ by
\begin{align}
v_i & = \sum_{k} \langle \ell_i , P_{ab}\rangle^ k\,  t_k \, , \label{vdef}
\end{align}
where the sum runs over all the edges.
Finally, define a scalar $s$ by
\begin{align}
s & = \sum_{k} \langle P_{ab} , P_{ab}\rangle^ k\,  t_k \, . \label{sdef}
\end{align}
Dai and Siegel compute the Green's function as
\begin{align}
G(X_a,X_b) & = - \frac{1}{2} s + \frac{1}{2} \vec v\,\tran \cdot \theta^{-1} \cdot \vec v \, . \label{DSgreen}
\end{align}
 This formula is analogous to the worldsheet Green's function commonly used in string theory \cite{DiVecchia:1996uq}
 \begin{align}
{\cal G}(x_1,x_2) & = \log | E(x_1,x_2) | - \frac{1}{2} \Big( \int_{x_1}^{x_2} \vec{\omega} \Big) \cdot ( 2 \pi \text{Im} \tau)^{-1} \cdot  \Big( \int_{x_1}^{x_2} \vec{\omega} \Big) \, .
\label{wsgf}
 \end{align}
 Here  $E(x_1,x_2)$ is the Schottky-Klein prime form, given by
\begin{align}
E(x_1,x_2) & \equiv \frac{x_1-x_2}{\sqrt{\d x_1}{\sqrt{\d x_2}}}  {\prod_\alpha}' \frac{x_1 - \gamma_\alpha(x_2)}{x_1 - \gamma_\alpha(x_1)} \frac{x_2 - \gamma_\alpha(x_1)}{x_2 - \gamma_\alpha(x_2)}  \, , \label{primeform}
\end{align}
with the Schottky group product including one from each pair of inverse elements $\{\gamma_\alpha, \gamma_\alpha^{-1}\}$.

$\omega_i$, $i = 1, \ldots , g$ is a basis of Abelian differentials and $\tau$ is the period matrix, defined by
 \begin{align}
 \frac{1}{2 \pi \ii} \oint_{a_j} \omega_i & = \delta_{ij} \, ; \,
 &
 \tau_{ij} & =  \frac{1}{2 \pi \ii} \oint_{b_j} \omega_i \, .
 \end{align}
 where $a_i$ and $b_j$ are a basis of homology cycles normalized in the standard way \cite{DiVecchia:1988cy}.
Using the Schottky parametrization, $\omega_i(z)$ are given by the formula:
\begin{align}
\omega_i (z) & = {\sum_\alpha}{}^{(i)} \Big( \frac{1}{z - \gamma_\alpha(u_i)} - \frac{1}{z - \gamma_\alpha(v_i)} \Big) \, \d z \, \label{abdifeq}
\end{align}
where the sum is over all Schottky group elements whose right-most factor is not $\gamma_i^{\pm n}$.
There is a Schottky group formula for the period matrix:
\begin{align}
\tau_{ij} & = \frac{1}{2 \pi \ii } \Big( \delta_{ij} \, \log k_i \, - \, {}^{(i)} {\sum_{\gamma_\alpha}}'{}^{(j)} \log \frac{u_i - \gamma_\alpha(v_j)}{u_i - \gamma_\alpha(u_j)} \frac{v_i - \gamma_\alpha(u_j)}{v_i - \gamma_\alpha(v_j)} \Big) \label{pmseries}
\end{align}
where the summation is over all Schottky group elements $\gamma_\alpha$ whose left-most factor is not $\gamma_i^{\pm n}$ and whose right-most factor is not $\gamma_j^{\pm n}$; the identity is excluded for $i=j$.

 When using the pinching parametrization described in section \ref{constr}, and choosing the paths from the external edges to the base node such that
 \begin{align}
 P_{X_b} & = P_{X_a} P_{ab} \, ,
 \end{align} we find that
 graph-theoretic objects defined on $\Gamma$ arise as limits of objects defined on $\Sigma$:
 \begin{align}
 s & = \lim_{\alpha' \to 0} 2 \alpha' \log \frac{ E(z_a, z_b) }{\sqrt{V_a'(0) V_b'(0)}} \, \label{sClaim} \\
  v_i & = \lim_{\alpha ' \to 0} \alpha' \,  \int_{z_a}^{z_b} \vec \omega_i \label{vClaim} \\ \theta_{ij}& = \lim_{\alpha ' \to 0}  2 \pi \alpha'\, \text{Im}\, \tau_{ij}\, , \label{thetaClaim}
 \end{align}
  if we make the identification \eq{pteq}.

 This makes clear how Feynman graph integrands arise as limits of the string measure on moduli space (at least for scalar fields).

 \subsection{Example}
 Let us work out an example. We will pick a Feynman graph and show first of all in section \ref{exampleinspection} how its period matrix and worldline Green's function can be written down by inspecting the graph \emph{\`a la} Dai and Siegel, and then in section \ref{examplelimit} how the same quantities arise asymptotically from objects on moduli space.
 \subsubsection{By inspection}
 \label{exampleinspection}
\begin{figure}
\centering
 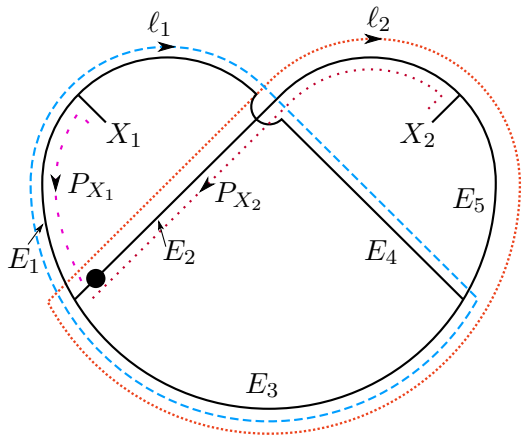
\caption{
A non-planar graph $\Gamma$ with $g=2$ loops and $n=2$ external edges (labelled $X_1$ and  $X_2$). The internal edges are labelled $E_i$, $i=1,\ldots, 5$. A loop basis $\{ \ell_1, \ell_2\}$ is indicated. The base point is marked as a black dot; the chosen paths from the external edges to the base point are marked as $P_{X_1}$ and $P_{X_1}$.
}\label{fig:pretzel}
\end{figure}

  Consider the $n=2$, $g=2$ non-planar graph $\Gamma$ shown in \Fig{fig:pretzel}.
The paths $P_{X_1}$, $P_{X_2}$ from the two external edges $X_1$ and $X_2$ to the base point are indicated; the paths $P_{\ell_1}$ and $P_{\ell_2}$ corresponding to the loop basis can be taken as the natural choice (since both loops pass through the vertex at which the base point lies).

First of all, let us compute the graph period matrix $(\theta_{ij})$. The diagonal entries $\theta_{ii}$ are just the sums of the Schwinger parameters of the edges in the loops $\ell_i$, so we have
\begin{align}
\theta_{11} & = t_1 + t_3 + t_4 \, ,
&
\theta_{22} & = t_2 + t_3 + t_5 \, . \label{pretzq1}
\end{align}
The two loops $\ell_1$ and $\ell_2$ intersect on one edge, $E_3$, and they both cross it in the same direction so we have
\begin{align}
\theta_{12} = \theta_{21} & = t_3 \, . \label{pretzq2}
\end{align}
The first Symanzik polynomial can be computed as
\begin{align}
\det \theta & = (t_1+t_4)(t_2 + t_5) +  t_3\,(t_1 + t_2 + t_4 + t_5 ) \, .
\end{align}
Next let us compute the Green's function \emph{\`a la} Dai and Siegel~\cite{Dai:2006vj}. We need to pick a path from $X_2$ to $X_1$; let us use $P_{X_1X_2} \equiv P_{X_1}^{-1}\cdot P_{X_2}$. Then since $P_{X_1X_2}$ crosses the edges $E_2$ and $E_1$, we have
\begin{align}
s & = t_1 + t_2 \, . \label{pretzs}
\end{align}
$P_{X_1X_2}$ intersects $\ell_1$ along the edge $E_1$, which they both traverse in the same direction, while $P_{X_1X_2}$  intersects $\ell_2$ along the edge $E_2$ which they both traverse in opposite directions. Therefore the vector $\vec v$ is given by
\begin{align}
v_1 & = t_1 \, , &
v_2 & = - t_2 \, . \label{pretzv}
\end{align}
Plugging Eqs.~(\ref{pretzq1})--(\ref{pretzv}) into \eq{DSgreen}, we get the following expression for the worldline Green's function:
\begin{align}
G & = - \frac{ t_1 t_2 t_4 + t_1 t_3 t_4 + t_2 t_3 t_4 + t_1 t_2 t_5 + t_1 t_3 t_5 + t_2 t_3 t_5 + t_1 t_4 t_5 + t_2 t_4 t_5 }{2 \det \theta} \, .
\end{align}
\subsubsection{From the zero-slope limit}
\label{examplelimit}
\begin{figure}
\centering
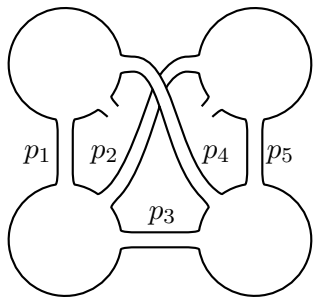
\caption{
The Feynman graph in \Fig{fig:pretzel} can be used to construct a Riemann surface with pinching parameters arranged in such a way that the graph is recovered in the $\alpha ' \to 0$ limit, after matching pinching parameters to Schwinger parameters according to \eq{pteq}.
}  \label{fig:pretzelrs}
\end{figure}
Let us now verify that our approach gives the same answer. We can use the graph $\Gamma$ in \Fig{fig:pretzel} to construct a bordered Riemann surface \Fig{fig:pretzelrs} parametrized by pinching parameters $\{ p_i\}$ corresponding to the graph edges $E_i$. Then we can write down expressions for the period matrix and the terms in the Green's function, use \eq{pteq} to take the zero-slope limit, and verify that Eqs.~(\ref{sClaim}) -- (\ref{thetaClaim}) hold.

The paths in \Fig{fig:pretzel} can be expressed as
\begin{align}
P_{X_1} & = \cw \cdot \tra{1} \cdot \cw \\
P_{X_2} & = \tra{2} \cdot \cw \\
P_{\ell_1} & = \acw \cdot \tra{3} \cdot \acw \cdot \tra{4} \cdot \cw \cdot \tra{1} \cdot \acw \\
P_{\ell_2} & = \acw \cdot \tra{3} \cdot \cw \cdot \tra{5} \cdot \cw \cdot \tra{2} \, .
\end{align}
Then according to our prescription, the corresponding bordered Riemann surface (\Fig{fig:pretzelrs}) can be parameterized by the M\"obius maps $V_i = \phi(P_{X_i})$, $\gamma_j = \phi(P_{\ell_j})$, giving
\begin{align}
V_1 & = \rho^{-1} \cdot \sigma_1 \cdot \rho^{-1} \, , \\
V_2 & = \sigma_2 \cdot \rho^{-1} \, ,  \\
\gamma_1 & = \rho \cdot \sigma_3 \cdot \rho \cdot \sigma_4 \cdot \rho^{-1} \cdot \sigma_1 \cdot \rho \, , \\
\gamma_2 & = \rho \cdot \sigma_3 \cdot \rho^{-1} \cdot \sigma_5 \cdot \rho^{-1} \cdot \sigma_2 \, .
\end{align}
The six Schottky group moduli can be computed from $\gamma_1$ and $\gamma_2$; they are given in terms of the $p_i$'s by
\begin{align}
k_1 & = \frac{F_{413} - \sqrt{F_{413}^2 - 4\, p_1\, p_3\, p_4}}{F_{413} + \sqrt{F_{413}^2 - 4\, p_1\, p_3\, p_4}} = p_1\, p_3 \, p_4 + {\cal O}(p_i^2) \, , \label{pretzk1} \\
k_2 & = \frac{F_{352} - \sqrt{F_{352}^2 - 4\,p_2 \,p_3\,p_5}}{F_{352} + \sqrt{F_{352}^2 - 4\,p_2 \,p_3\,p_5}} = p_2\, p_3 \, p_5  + {\cal O}(p_i^2) \, ,
\label{pretzk2}  \\
u_1 & =\frac{ 1}{1+p_1} + \frac{F_{413} + \sqrt{F_{413}^2 - 4\, p_1\, p_3\, p_4}}{2\, (1+p_1)\,p_3\,p_4 } \, = \frac{1}{p_3 p_4} \big(1 + {\cal O}(p_i) \, \big) \, , \label{pretzu1} \\
v_1 & =\frac{ 1}{1+p_1} + \frac{F_{413} - \sqrt{F_{413}^2 - 4\, p_1\, p_3\, p_4}}{2\, (1+p_1)\,p_3\,p_4 } = 1 + {\cal O}(p_i) \, ,  \label{pretzv1}  \\
u_2 & = - \frac{p_2 \, p_5}{1+p_5} + \frac{F_{352} + \sqrt{F_{352}^2 - 4\,p_2 \,p_3\,p_5}}{2 p_3 \, (1 + p_5)} \, = \frac{1}{p_3} \big( 1 + {\cal O}(p_i) \, \big) \, , \label{pretzu2} \\
v_2 & = - \frac{p_2 \, p_5}{1+p_5} + \frac{F_{352} - \sqrt{F_{352}^2 - 4\,p_2 \,p_3\,p_5}}{2 p_3 \, (1 + p_5)}  = - p_2 p_3 p_5 + {\cal O}(p_i^2) \, ,  \label{pretzv2}
\end{align}
where we've used the notation
\begin{align}
F_{abc\cdots} & \equiv 1 + p_a \, F_{bc \cdots} \, ;
&
F_{a} & \equiv 1 + p_a \, .
\end{align}
The Koba-Nielsen variables $z_i = V_i(0)$ are given by
\begin{align}
z_1 & = \frac{1}{1+p_1} \, ,
&
z_2 & = - p_2 \, . \label{pretKN}
\end{align}
We also have
\begin{align}
V_1'(0) & = \frac{p_1}{(1 + p_1)^2} \, ,
&
V_2 '(0) & = p_2 \, . \label{pretweights}
\end{align}

First of all we can compute the period matrix using \eq{pmseries}. At leading order, the diagonal elements come only from the Schottky group multipliers: from \eq{pretzk1} and \eq{pretzk2} we get
\begin{align}
\tau_{11} & =\frac{1}{2 \pi \ii }  \log \big( p_1 \,p_3\, p_4 \big)+ {\cal O}(p_i) \, ,
&
\tau_{22} & = \frac{1}{2 \pi \ii }  \log \big( p_2\, p_3\, p_5 \big) + {\cal O}(p_i) \, . \label{prettdiag}
\end{align}
To compute the diagonal entries $\tau_{12} = \tau_{21}$, we use the second term in \eq{pmseries}. For this graph, the only non-vanishing contribution to the sum at leading order in the $p_i$ comes from the identity element of the Schottky group. After substituting the expressions for the fixed points we computed in \eq{pretzu1} -- \eq{pretzv2} we get
\begin{align}
\tau_{12} = - \frac{1}{2 \pi \ii } \log \Big( \frac{u_1 - v_2}{u_1 - u_2} \frac{v_1 - u_2}{v_1 - v_2} \Big) \,  + \,  {\cal O}(p_i) \, & =  - \frac{1}{2 \pi \ii } \log \Big( - \frac{1}{p_3} \Big)  \,  + \,  {\cal O}(p_i) \, \nonumber \\
& = \frac{1}{2 \pi \ii} \log p_3 \, + \, n + \frac{1}{2} \, + \, {\cal O}(p_i ) \, ,\label{prettoffd}
\end{align}
where we've written $\log(-1) = (2 n + 1) \pi$.
Then we can check that \eq{thetaClaim} holds. We have
\begin{align}
2 \pi \,\text{Im} (\tau_{ij}) & = \left(\begin{array}{cc} -  \, \log \big( p_1 p_3 p_4 \big) & - \, \log p_3 \\ - \, \log p_3 & - \, \log\big( p_2 p_3 p_5\big)
\end{array}\right) \, + \, {\cal O}(p_i) \, .
\end{align}
After making the substitution in \eq{pteq} and multiplying by $\alpha'$, we get
\begin{align}
2 \pi \,\alpha ' \,\text{Im} (\tau_{ij}) & = \left(\begin{array}{cc}t_1 + t_3 + t_4 & t_3  \\t_3 & t_2 + t_3 + t_5
\end{array}\right) \, + \, {\cal O}(\ex{- t_i/ \alpha'}) \, .
\end{align}
In the $\alpha ' \to 0$ limit, the correction vanishes and we arrive at the graph period matrix $(\theta_{ij})$ given in \eq{pretzq1} and \eq{pretzq2}, showing that \eq{thetaClaim} holds in this case.

Next let's check \eq{sClaim}, for the `scalar' term $s$ in Dai and Siegel's approach to the worldline Green's function. First we can compute the numerator. To leading order in the $p_i$'s the only contribution to the Schottky-Klein prime form \eq{primeform} comes from the first factor, so we have
\begin{align}
\log E(z_1, z_2) & = \log(z_1 - z_2) + {\cal O}(p_i) \, .
\end{align}
The Koba-Nielsen variables $z_i$ can be substituted from \eq{pretKN} giving
\begin{align}
\log E(z_1, z_2) & =  \log\Big(\frac{1}{1 + p_1} + p_2 \Big) + {\cal O}(p_i) = {\cal O}(p_i) \, , \label{pretsnum}
\end{align}
\ie~the numerator vanishes at leading order and the only contribution comes from the weights in the denominators. Substituting from \eq{pretweights}, we get
\begin{align}
\log \frac{1}{\sqrt{V_1'(0) V_2'(0)}} & = \log\frac{1+ p_1}{(p_1 p_2)^{1/2}} \, + \, {\cal O}(p_i) \,
\\
& = - \frac{1}{2} \log p_1 p_2 \, + \, {\cal O}(p_i) \, . \label{pretsdenom}
\end{align}
Combining \eq{pretsnum} and \eq{pretsdenom}, making the substitution in \eq{pteq} and multiplying by $2 \alpha' $ we get
\begin{align}
2\alpha' \, \log \frac{ E(z_1, z_2)}{\sqrt{V_1'(0) V_2'(0)}}  & = t_1 + t_2+ {\cal O}(\ex{ - t_i / \alpha ' }) \, .
\end{align}
Taking $\alpha ' \to 0$, the correction vanishes and we recover $s$ from \eq{pretzs} as expected, verifying \eq{sClaim} for this example.

Last of all, we need to check that \eq{vClaim} holds. We can integrate \eq{abdifeq} to get
\begin{align}
\int_{z_2}^{z_1} \omega_i(z) & = {\sum_{\alpha}}^{(i)} \Big[ \log \frac{z - \gamma_\alpha(u_i)}{z - \gamma_\alpha(v_i)} \Big]_{z_2}^{z_1} \, .
\end{align}
At leading order in the $p_i$'s, the only Schottky group element that gives a non-vanishing contribution to the sum is the identity, so we obtain
\begin{align}
\int_{z_2}^{z_1} \omega_i(z) & = \log  \frac{z_1 -u_i}{z_1 - v_i} \frac{z_2 -v_i}{z_2 - u_i} \, + \, {\cal O}(p_i) \, .
\end{align}
Substituting the Koba-Nielsen variables from \eq{pretKN} and the fixed points from \eq{pretzu1} -- \eq{pretzv2}, we find
\begin{align}
\int_{z_2}^{z_1} \vec \omega(z) & = \big(-\,\log p_1, \, \log p_2 + (2n+1) \ii \, \pi\big)^{\text t} \, + \, {\cal O}(p_i) .
\end{align}
Multiplying by $\alpha' $ and replacing the pinching parameters with Schwinger parameters according to \eq{pteq}, we get
\begin{align}
\alpha ' \, \int_{z_2}^{z_1} \vec \omega(z) & =  \big(t_1, \, -\, t_2 +  \ii \, \pi\, \alpha' (2n+1)  \big)^{\text t} \, + \, {\cal O}(\ex{- t_i / \alpha}) .
\end{align}
Taking the limit $\alpha ' \to 0$, we recover $\vec v$ as given in \eq{pretzv}, so we've also seen that \eq{vClaim} holds too as expected.
This completes the computation demonstrating that Eqs.~(\ref{sClaim}) -- (\ref{vClaim}) all hold for our example.

There is nothing special about our choice of example. A proof that this type of computation would give the expected results in the general case is given in sections \ref{pmproof} and \ref{greenfapp}.

 \section{Duality transformations}
 \label{DualitySect}
\begin{figure}
\centering
\subfloat[]{ 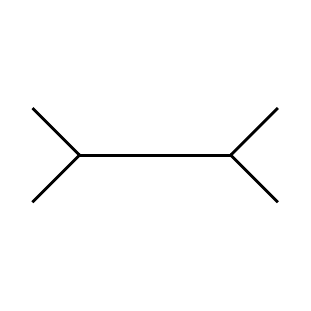 \label{fig:dualitya} }
\subfloat[]{ 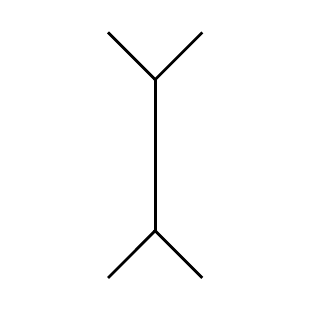 \label{fig:dualityb} }
\caption{Two four-point tree graphs which are dual to each other (in the sense of dual-resonance models).}\label{fig:duality1}
\end{figure}

We say two cubic skeleton graphs are `dual' to each other (in the sense of dual resonance models) if one can be obtained from the other by (repeatedly) cutting out a subgraph of the form of \Fig{fig:dualitya} and replacing it with one of the form of \Fig{fig:dualityb}, respecting the labels of the cut edges. This is sometimes known as a \emph{Whitehead move}.

If two graphs are dual to each other, then their corresponding pinching parametrizations can be mapped onto each other with a change of variables, which we compute in this section. We work out the change of variables needed in the super-geometric case, since the bosonic case follows as a corollary by ignoring the odd parameters and taking $p_i = \ve_i^2$.

\begin{figure}
\centering
\subfloat[]{ 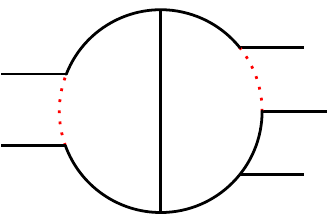 \label{fig:duality2a} }
\subfloat[]{ 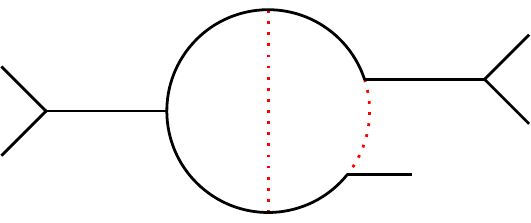 \label{fig:duality2b} }\\
\subfloat[]{ 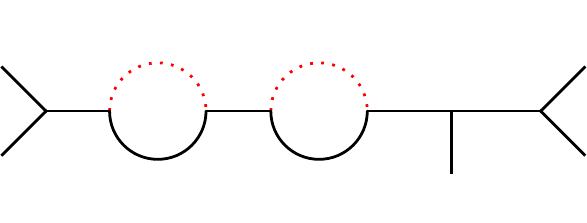 \label{fig:duality2d} }
\subfloat[]{ 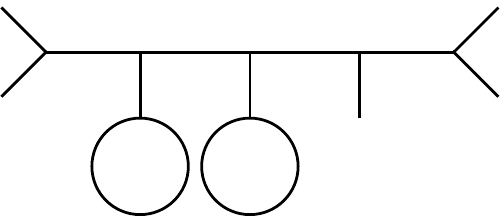 \label{fig:duality2e} }
\caption{An example cubic graph with $n=5$, $g=2$ (\Fig{fig:duality2a}) and the intermediate graphs (\Fig{fig:duality2b}, \Fig{fig:duality2d}) obtained via duality transformations in order to put it in the canonical form (\Fig{fig:duality2e}). In Figs.~\ref{fig:duality2a}--\ref{fig:duality2d}, edges drawn with red dots are those which are duality-transformed to arrive at the successive graph.}\label{fig:duality2}
\end{figure}

Any two planar cubic graphs with the same number of loops and the same number of external edges labelled in the same order are dual to each other. This means that any planar cubic graph is dual to one of the form in \Fig{fig:planarex}. \Fig{fig:duality2} shows an example with $n=5$, $g=2$, illustrating the duality transformations needed to get from the graph in \Fig{fig:duality2a} to the canonical one in \Fig{fig:duality2e}.

As described in section \ref{param}, each $g$-loop, $n$-point skeleton graph induces a set of (super) moduli for a worldsheet with the same number of loops and $n$ punctures. Thus, the notion of duality between graphs gives a natural notion of duality between sets of pinching moduli $p_i$. We can derive the algebraic relationship between the two sets of moduli.

\subsection{Duality on internal edges}
\begin{figure}
\centering
\subfloat[]{ 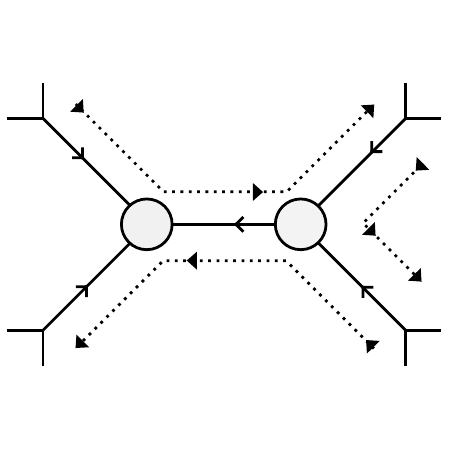 \label{fig:duality3a} }
\subfloat[]{ 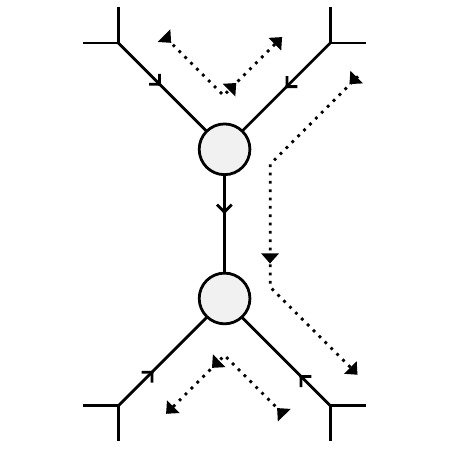 \label{fig:duality3b} }
\caption{Two dual sub-graphs of a cubic graph. We can find the relationship between the two sets of supermoduli ($t_1, \ldots, t_5|\varphi_a, \varphi_b$ for \Fig{fig:duality3a} and $t_a, \ldots, t_e|\varphi_1,\varphi_2$ for \Fig{fig:duality3b}) by identifying the corresponding paths as described in the text.}\label{fig:duality3}
\end{figure}

First of all, let us consider the case where all of the edges involved in the duality transformation are internal edges with associated Schwinger times $t_i$. Consider a subgraph of a target graph, as shown in \Fig{fig:duality3a}, and the subgraph it is dual to, shown in \Fig{fig:duality3b}. We are considering the supergeometric case, so we have labelled the vertices involved in the duality transformation with odd supermoduli $\varphi_a$, $\varphi_b$, $\varphi_1$ and $\varphi_2$, and we have given the involved edges an orientation.

We have indicated three paths $P_\alpha$, $P_\beta$ and $P_\gamma$ on \Fig{fig:duality3a}. Any path which traverses this subgraph may be written in terms of these three paths. On \Fig{fig:duality3b}, we have indicated three paths $P_{\widetilde{\alpha}}$, $P_{\widetilde{\beta}}$ and $P_{\widetilde{\gamma}}$, which correspond to the matching ones on \Fig{fig:duality3a} in the sense that they join the same pairs of vertices. If we can find a relationship between the dual supermoduli which solves
\begin{align}
\bs \phi(P_\alpha) & = \bs \phi(P_{\widetilde{\alpha}}) \, ,
&
\bs\phi(P_\beta) & =\bs \phi(P_{\widetilde{\beta}}) \, ,
&
\bs\phi(P_\gamma) & = \bs\phi(P_{\widetilde{\gamma}}) \, , \label{eq:dualityreq}
\end{align}
then the super Schottky group generators $\bs \gamma_\ell$ and local co-ordinate charts $\bs V_i$ induced by the two dual graphs will be identical, and thus describe the same surface. Thus, \eq{eq:dualityreq}, gives us a mapping between dual sets of pinching supermoduli.

According to the prescription in section \ref{supparam}, \eq{eq:dualityreq} we have
\begin{align}
\phi(P_\alpha)  & = \bs \sigma_2^{-1} \, \bs \rho_a^{-1} \, \bs \sigma_3 \, \bs \rho_b^{-1} \, \bs \sigma_5 \label{phiPalpha} \\
& =
\frac{1}{\ve_2 \,  \ve_3 \,  \ve_5} \left(
\begin{array}{cc|c}
p_2  \, (1 + p_3 - \ve_3  \, \varphi_a  \, \varphi_b) & p_2 \,  p_3 \,  p_5 & p_2 \,  \ve_3  \, \ve_5 ( \varphi_a - \ve_3  \, \varphi_b) \\
-1 & 0 & 0 \\ \hline
\ve_2\, (\varphi_a - \ve_3 \,  \varphi_b) & 0 & \ve_2  \, \ve_3 \,  \ve_5 \end{array}
\right) \label{phiPalphamat}
\\
  \bs \phi(P_{\widetilde{\alpha}}) & = \bs \sigma_d ^{-1} \, \bs \rho_2^{-1} \, \bs \sigma_e  \, \\
  & = \frac{1}{\ve_d \, \ve_e} \left(
  \begin{array}{cc|c}
  - p_d & - p_d \, p_e &  p_d \, \ve_e\, \varphi_2 \\
  1 & 0 & 0 \\ \hline
  - \ve_d \, \varphi_2 & 0 & \ve_d \, \ve_e
  \end{array}
  \right) \,
\\
  \bs \phi(P_\beta) & = \bs \sigma_5 ^{-1} \, \bs \rho_b^{-1} \, \bs \sigma_4  \, \\
  & = \frac{1}{ \ve_4\,\ve_5} \left(
  \begin{array}{cc|c}
  - p_5 & - p_4 \, p_5 &  p_5 \, \ve_4\, \varphi_b \\
  1 & 0 & 0 \\ \hline
  - \ve_5 \, \varphi_b & 0 &  \ve_4\,\ve_5
  \end{array}
  \right) \,
  \label{Pbeta}
  \\
\phi(P_{\widetilde{\beta}})  & = \bs \sigma_e^{-1} \, \bs \rho_2^{-1} \, \bs \sigma_c \, \bs \rho_1^{-1} \, \bs \sigma_b \\
& =
\frac{1}{\ve_b \,  \ve_c \,  \ve_e} \left(
\begin{array}{cc|c}
p_e  \, (1 + p_c + \ve_c  \, \varphi_1  \, \varphi_2) & p_b \,  p_c \,  p_e & p_e \,  \ve_b  \, \ve_c ( \varphi_2 - \ve_c  \, \varphi_1) \\
-1 & 0 & 0 \\ \hline
\ve_e\, (\varphi_2 - \ve_c \,  \varphi_1) & 0 & \ve_b  \, \ve_c \,  \ve_e \end{array}
\right)
\label{Ptbeta}
\\
\phi(P_\gamma)  & = \bs \sigma_4^{-1} \, \bs \rho_b^{-1} \, \bs \sigma_3 ^{-1} \, \bs \rho_a^{-1} \, \bs \sigma_1 \\
& =
\frac{1}{\ve_1 \,  \ve_3 \,  \ve_4} \left(
\begin{array}{cc|c}
- p_4  \, (1 + p_3 - \ve_3\, \varphi_a  \, \varphi_b  ) & - p_1 \,  p_3 \,  p_4 & p_4 \, \ve_1 \,  \ve_3  ( \varphi_b + \ve_3  \, \varphi_a) \\
1 & 0 & 0 \\ \hline
- \ve_4\, (\varphi_b + \ve_3 \,  \varphi_a) & 0 &  \ve_1  \, \ve_3 \,  \ve_4 \end{array}
\right)
\\
  \bs \phi(P_{\widetilde{\gamma}}) & = \bs \sigma_b ^{-1} \, \bs \rho_1^{-1} \, \bs \sigma_a  \, \\
  & = \frac{1}{\ve_a \, \ve_b} \left(
  \begin{array}{cc|c}
  - p_b & - p_a \, p_b &  p_b \, \ve_a\, \varphi_1 \\
  1 & 0 & 0 \\ \hline
  - \ve_b \, \varphi_1 & 0 & \ve_a \, \ve_b
  \end{array}
  \right) \, , \label{phiPgammahatmat}
\end{align}
where we have written $p_i \equiv \ve_i^2$. Using \eq{phiPalphamat}--\eq{phiPgammahatmat}, we see that \eq{eq:dualityreq} requires the simultaneous solution of 15 equations (9 Grassmann even and 6 Grassmann odd); since we have only $5|2$ supermoduli to solve for, this is a non-trivial check of the consistency of our approach.
The solution exists and is given by
\begin{align}
\ve_a & =\pm\, \frac{\ve_1 \, \ve_3}{M} \, ,
&
\ve_b & = \pm\,\ve_4\, M \, ,
&
\ve_c & = \pm \, \frac{1}{\ve_3} \, ,
&
\ve_d & = \ve_2 \, M \, ,
&
\ve_e & = - \, \frac{ \ve_3 \, \ve_5}{M} \, ,
\nonumber \end{align}
\begin{align}
\varphi_1 & = \pm \,  \frac{\varphi_b + \ve_3\, \varphi_a}{M} \, ,
&
\varphi_2 & = \frac{\varphi_a - \ve_3\, \varphi_b}{M}\, , \label{dualsupermod}
\end{align}
where $M$ is one of the two roots of
\begin{align}
M^2 & = 1 + p_3 - \ve_3\, \varphi_a \, \varphi_b \, , \label{Mdef}
\end{align}
giving four branches to the solution (the $\pm$ signs must be chosen consistently).
Thus, once we have found a set of pinching moduli $(\ve_i | \varphi_j)$ corresponding to some skeleton graph $\Gamma$, we can compute the pinching supermoduli $(\hat\ve_{{i}} | \hat\varphi_{{j}})$ for a dual graph $\hat{\Gamma}$ with the (repeated) use of \eq{dualsupermod}.

We have assumed in this calculation that all of the edges involved in the duality transformation are distinct from each other. In section \ref{loopduality} we describe the case in which two edges are identified such that there is a loop (as in the transformation from \Fig{fig:duality2d} to \Fig{fig:duality2e}), which has a slightly different solution.
\subsection{Duality with external edges}
If some of the edges involved in the duality transformation are external edges, then a modified approach is needed. Recall that each external edge on the skeleton graph $\Gamma$ is associated to a puncture on the surface $\bs \Sigma$ located at $\bs z_i = \bs V_i(0|0)$, where $\bs V_i$ is computed as described in section \ref{supparam}. Thus, we want to find a relationship between the dual sets of pinching supermoduli $(\ve_i|\varphi_j)$, $(\hat \ve_i |\hat \varphi_j)$ such that we leave unchanged all paths traversing the subgraph and also such that the positions of the punctures on $\bs \Sigma$ are unchanged,
$\bs V_i(0|0) = \hat {\bs V}_i(0|0)$.

\begin{figure}
\centering
\subfloat[]{ 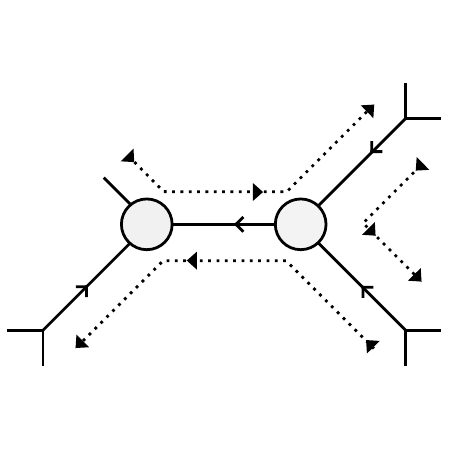 \label{fig:duality4a} }
\subfloat[]{ 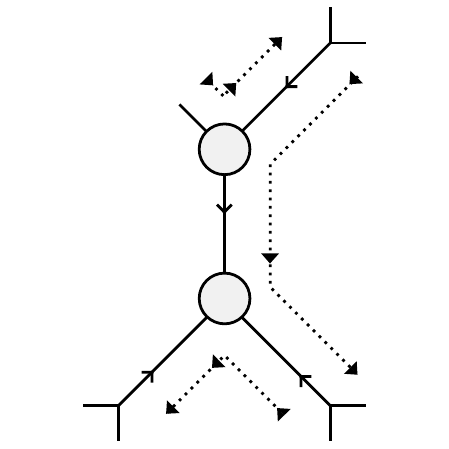 \label{fig:duality4b} }
\caption{Two dual sub-graphs of a target graph with an external edge in common. The subgraphs here each have one less even supermodulus than those in \Fig{fig:duality3}. To find the duality relation we need to fix $\bs z_1$ by satisfying $\hat{P}_\gamma(0|0)= \hat{P}_{\widetilde{\gamma}}(0|0)$.}\label{fig:duality4}
\end{figure}

For example, consider a duality transformation when one of the involved edges is an external edge, as in \Fig{fig:duality4}.
The paths $P_\alpha$ and $P_\beta$ in \Fig{fig:duality4a} and $P_{\widetilde{\alpha}}$, $P_{\widetilde{\beta}}$ in \Fig{fig:duality4b} are identical to those in \Fig{fig:duality3}. Again, solving the equations
\begin{align}
\bs \phi(P_\alpha) & = \bs \phi(P_{\widetilde{\alpha}}) \, ,
&
\bs\phi(P_\beta) & =\bs \phi(P_{\widetilde{\beta}}) \, \label{duality1x}
\end{align}
will ensure that any path traversing the subgraphs is unaffected by the duality transformation.

But now there is a new constraint of a different type: we need the position of $\bs z_1$ --- the position of the puncture corresponding to the external edge $X_1$ --- to be fixed by the duality transformation. Recall that $\bs z_1 = \bs V_1 (0|0)$, where $\bs V_1 = \bs \phi (P_1)$ with $P_1$ some path which starts at $\bs z_1$ and ends at the base point. Any path which starts at $\bs z_1 $ on \Fig{fig:duality4a} can be written with $\hat{P}_\gamma$ as its right-most factor. For \Fig{fig:duality4b}, we can use
the same path, but with the right-most factor changed from $\hat{P}_\gamma$ to $\hat{P}_{\widetilde{\gamma}}$ and any factors of $P_\alpha$ and $P_\beta$ changed to $P_{\widetilde{\alpha}}$ and $P_{\widetilde{\beta}}$, respectively.

Then
\begin{align}
\bs \phi(\hat{P}_\gamma)(0|0) & =\bs \phi( \hat{P}_{\widetilde{\gamma}})(0|0) \label{Pgamma0}
\end{align}
is a sufficient condition to guarantee that $\bs z_1$ is fixed by duality (as long as \eq{duality1x} also holds). The two matrices in \eq{Pgamma0} are given by
\begin{align}
\bs \phi( \hat{P}_\gamma ) & =
\bs \sigma_4^{-1} \,
\bs \rho_b^{-1} \,
\bs \sigma_3^{-1} \,
\bs \rho_a^{-1}
\\
& = \left( \begin{array}{cc|c}
\ve_3 \, \ve_4
& - {\ve_4 \, M^2}/{\ve_3}
& \ve_3 \, \ve_4 \, \varphi_a + \ve_4\, \varphi_b
\\
0
&
1 / \ve_3 \, \ve_4
&
0
\\ \hline
0
&
- \varphi_a - \varphi_b / \ve_3
&
1
\end{array}\right)
\\
\bs\phi(\hat{P}_{\widetilde{\gamma}} )
& = \bs \sigma_b^{-1} \, \bs \rho_1^{-1} \\
& =
\left( \begin{array}{cc|c}
\ve_b & - \ve_b & \ve_b \, \varphi_1
\\
0 & 1 / \ve_b & 0
\\ \hline
0 & - \varphi_1 &  1
\end{array}\right)
\end{align}
and then the images of $0|0$ are computed as
\begin{align}
\bs \phi( \hat{P}_\gamma ) \, \cdot \, (0\,\, 1|0)\tran & = (- \ve_4 \, M^2 /\ve_3 \, \, , \, \, 1 / \ve_3 \, \ve_4 \, | \,  - \varphi_a - \varphi_b / \ve_3 )\tran \\
& \sim - p_4 \, M^2 \, \big| - \ve_3\, \ve_4 \, \varphi_a - \ve_4 \, \varphi_b
\shortintertext{and}
\bs \phi( \hat{P}_{\widetilde{\gamma}}) \, \cdot \, (0\,\, 1|0)\tran
& = ( - \ve_b \, , \, 1 / \ve_b \, | \, - \varphi_1)\tran \\
& \sim - p_b \, \big| \, - \ve_b \, \varphi_1 \, .
\end{align}
It can be easily checked that \eq{dualsupermod} satisfies \eq{Pgamma0}.

This follows automatically from the fact that \eq{dualsupermod} solves $\bs \phi(P_\gamma) = \bs \phi(P_{\widetilde{\gamma}}) $. To see why this is true, note that we have
\begin{align}
\bs \phi(P_\gamma)  & = \bs \phi( \hat{P}_\gamma )\cdot \bs \sigma_1 \, ,
&
\bs \phi(P_{\widetilde{\gamma}})  & = \bs \phi( \hat{P}_{\widetilde{\gamma}} )\cdot \bs \sigma_a \, ,
\end{align}
so
\begin{align}
\bs\phi(\hat{P}_{\widetilde{\gamma}} ) & =
\bs \phi( \hat{P}_\gamma )\cdot \bs \sigma_1 \cdot \bs \sigma_a^{-1} \, .
\end{align}
But $0|0$ is fixed by $\bs \sigma_1 \cdot \bs \sigma_a^{-1}$, and thus \eq{dualsupermod} solves \eq{Pgamma0}.

Note that since $\ve_1$ and $\ve_a$ are not parameters of \Fig{fig:duality4a} and \Fig{fig:duality4b}, respectively, the first equation in \eq{dualsupermod} ($\ve_a = \pm \ve_1 \ve_3 / M$) is to be simply ignored.

For similar reasons, the relevant equations in \eq{dualsupermod} continue to give the relationship between dual sets of pinching supermoduli when two, three or four of the involved edges are external (the edges labelled $t_3$ and $t_c$ must be internal edges). Furthermore, the solutions in section \ref{loopduality} also reduce to give solutions in the cases where one or both of the unconnected edges are taken to be external.

\subsection{Duality involving loops}
\label{loopduality}
\subsubsection{Loop connecting adjacent edges}
\label{adjacentloop}
\begin{figure}
\centering
\subfloat[]{ 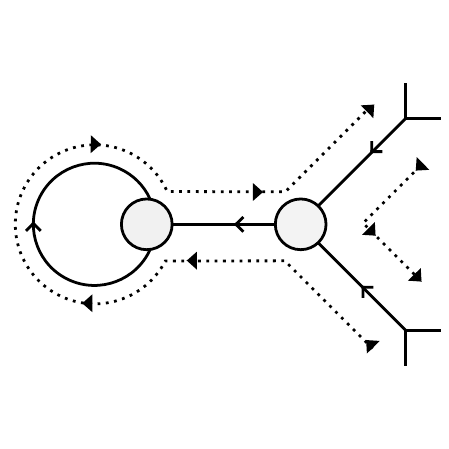 \label{fig:duality5a} }
\subfloat[]{ 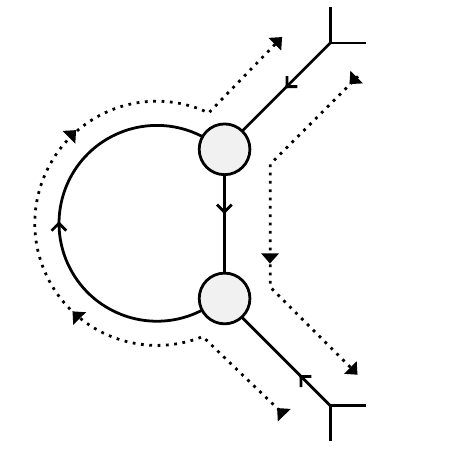 \label{fig:duality5b} }
\caption{Two dual subgraphs in which a loop is involved in the duality.}\label{fig:duality5}
\end{figure}

It is also possible to compute dual pinching supermoduli when a loop is involved in the duality transformation, in the sense that one of the involved edges connects one of the involved vertices back to itself or to the other involved vertex as in \Fig{fig:duality5}.
The paths $P_\beta$ and $P_{\widetilde{\beta}}$ in \Fig{fig:duality5a} and \Fig{fig:duality5b}, respectively, are identical to those in \Fig{fig:duality3}; thus we still need to solve
\begin{align}
\bs \phi (P_\beta) & = \bs \phi(P_{\widetilde{\beta}})  \label{Pbetaeq}
\end{align}
where the matrices for $\bs \phi (P_\beta)$ and $ \bs \phi(P_{\widetilde{\beta}})$ are given in \eq{Pbeta} and \eq{Ptbeta}, respectively. But now we have the paths $P_\delta$ in \Fig{fig:duality5a} and $P_{\widetilde{\delta}}$ in \Fig{fig:duality5b} so we need to solve
\begin{align}
\bs \phi (P_\delta) & = \bs \phi(P_{\widetilde{\delta}})  \label{Pdeltaeq}
\end{align}
These paths are represented by the matrices
\begin{align}
\bs \phi(P_{\delta} )& = \bs \sigma_4^{-1} \, \bs \rho_b^{-1} \, \bs \sigma_3^{-1} \, \bs \rho_a^{-1} \, \bs \sigma_1 \, \bs \rho_a^{-1} \, \bs \sigma_3 \, \bs \rho_b^{-1} \bs \sigma_5 \,
\shortintertext{and}
\bs \phi(P_{\widetilde{\delta}} )& =
\bs \sigma_b^{-1} \,
\bs \rho_1^{-1} \,
\bs \sigma_a \,
\bs \rho_2^{-1} \,
\bs \sigma_e \, .
\end{align}
As always, $\bs \phi( P_\delta)$ is of the form
\begin{align}
\bs \phi(P) & = \left(
\begin{array}{cc|c}
x & y & \xi \\ -1/y & 0 & 0 \\ \hline \xi / y & 0 & 1
\end{array}
\right) \label{pureCWform}
\end{align}
with
\begin{align}
\big(\bs \phi(P_{\delta} )\big)_1{}^1 & = \frac{\ve_3 (1 + p_1 p_3 - \ve_1 (1 +p_3)) \varphi_a \, \varphi_b-(1 + p_3)(1+p_1p_3)}{\ve_1 \, \ve_3^2 \, \ve_5}\,\ve_4 \\
\big(\bs \phi(P_{\delta} )\big)_1{}^2 & = - \, \ve_1 \, \ve_4 \, \ve_5\, p_3 \\
\big(\bs \phi(P_{\delta} )\big)_1{}^3 & = (1-\ve_1)\ve_3\,\ve_4\, \varphi_a + (1 + \ve_1 \, p_3)\ve_4\,\varphi_b  \, .
\end{align}
$\bs \phi ( P _{\widetilde{\delta}})$ is of the same form with
\begin{align}
\big(\bs \phi(P_{{\widetilde{\delta}}} )\big)_1{}^1 & = \frac{1+p_a - \ve_a \, \phi_1 \phi_2}{\ve_a \, \ve_e} \, \ve_b \\
\big(\bs \phi(P_{{\widetilde{\delta}}} )\big)_1{}^2 & = \ve_a\,\ve_b\,\ve_e  \\
\big(\bs \phi(P_{{\widetilde{\delta}}} )\big)_1{}^3 & = \ve_b \, \varphi_1 \, - \, \ve_a \ve_b \, \varphi_2 \, .
\end{align}
\eq{Pbetaeq} and \eq{Pdeltaeq} are solved by:
\begin{align}
\ve_a & = \mp \ve_1 \, \ve_3 \, ,
&
\ve_b & = \pm M\, \ve_4\, ,
&
\ve_c & = \mp \frac{1}{\ve_3} \, ,
&
\ve_e & = \frac{\ve_3\, \ve_5}{M} \, ,
\end{align}
\begin{align}
\varphi_1 & = \pm \frac{\ve_3\, \varphi_a + \varphi_b}{M}
&
\varphi_2 & = - \frac{\varphi_a - \ve_3\, \varphi_b}{M} \, ,
\end{align}
where $M$ is defined in \eq{Mdef}. Notice that we have $\ve_1 = \ve_a \, \ve_c$, corresponding to the multiplier of the loop.
There is another solution
\begin{align}
\ve_a & = \mp \ve_3 \,  ,
&
\ve_b & = \pm \hat{M} \ve_4 \, ,
&
\ve_c & = \mp \frac{1}{\ve_1 \, \ve_3} \, ,
&
\ve_e & = \frac{\ve_1 \, \ve_3 \, \ve_5}{\hat{M}} \, ,
\end{align}
\begin{align}
\varphi_1 & =\mp  \frac{\ve_1\,\ve_3\,\varphi_a - \varphi_b}{\hat{M}} \, ,
&
\varphi_2 & = \frac{\varphi_a + \ve_1 \, \ve_3 \, \varphi_b}{\hat{M}} \, ,
\end{align}
where
\begin{align}
\hat{M}^2  & = 1 + p_1 \, p_3+ \ve_1 \, \ve_3 \, \varphi_a\, \varphi_b\, .
\end{align}
However, this solution has $\ve_a \ve_c = 1/ \ve_1$, so it is not relevant to the pinching limit we want, since both sides of this equation should get small simultaneously as the Schwinger time length of the loop gets large.
\subsubsection{Loop connecting opposite edges}
\label{oppositeloop}
\begin{figure}
\centering
\subfloat[]{ 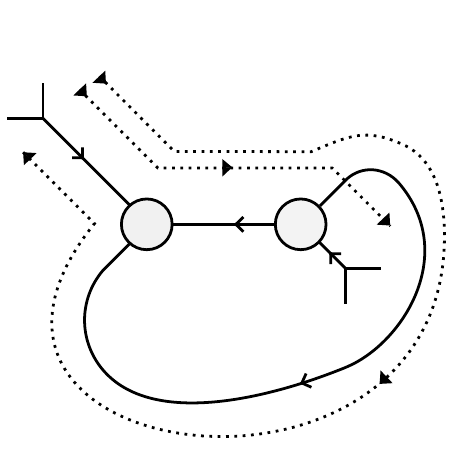 \label{fig:duality6a} }
\subfloat[]{ 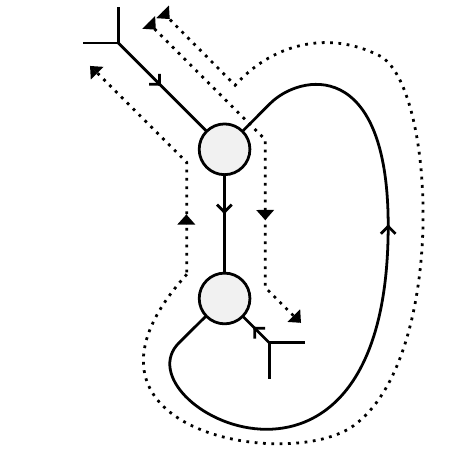 \label{fig:duality6b} }
\caption{Two dual subgraphs with a pair of opposite edges joined to form a loop.}\label{fig:duality6}
\end{figure}

It is also possible to compute the duality relation in the case that two opposite edges are connected in a loop, as depicted in \Fig{fig:duality6}.
To find the duality relation, we need to solve
\begin{align}
\bs \phi (P_\xi) &  = \bs \phi(P_{\widetilde{\xi}}) \, ,
&
\bs \phi ( P _\zeta) & = \bs \phi (P_{\widetilde{\zeta}}) \, . \label{Pzetaxi}
\end{align}
The first two matrices here are given by
\begin{align}
\bs \phi (P_\xi) & =
\bs \sigma_5^{-1} \,
\bs \rho_b \,
\bs \sigma_3^{-1} \,
\bs \rho_a^{-1} \,
\bs \sigma_1
\nonumber \\
& =
\left(
\begin{array}{cc|c}
{\ve_3 \, \ve_5}/{\ve_1} & \ve_1 \, \ve_3 \, \ve_5 & - \ve_3 \, \ve_5 \, \varphi_a  \\
{M^2}/{\ve_1 \, \ve_3\, \ve_5 } & - {\ve_1 \, \ve_3}/{\ve_5} & ({\ve_3 \, \varphi_a + \varphi_b})/{\ve_5} \\
\hline
- ({ \varphi_a - \ve_3\, \varphi_b})/{\ve_1} & \ve_1\, \ve_3\, \varphi_b & 1 + \ve_3\, \varphi_a \, \varphi_b
\end{array}
\right) \\
\bs \phi (P_{\widetilde{\xi}}) & =
\bs \sigma_e^{-1} \,
\bs \rho_2^{-1} \,
\bs \sigma_c \,
\bs \rho_1 \,
\bs \sigma_a \nonumber \\
& =
\left(
\begin{array}{cc|c}
- {\ve_e}/{\ve_a\,\ve_c}
&
- ({1+p_c+\ve_c\,\varphi_1\,\varphi_2})\,\ve_a\,\ve_e/{\ve_c}
&
({ \varphi_1 + \ve_c\, \varphi_2})\,\ve_e/{\ve_c}
\\
{1}/{\ve_a\,\ve_c\,\ve_e}
&
{\ve_a}/{\ve_c \, \ve_e}
&
- {\varphi_1}/{\ve_c \, \ve_e}
\\ \hline
- {\varphi_2}/{\ve_a \, \ve_c}
&
 ({\ve_c \, \varphi_1 - \varphi_2 })\,\ve_a/{\ve_c}
&
1 - {\varphi_1\,\varphi_2}/{\ve_c}
\end{array}
\right) \, .
\end{align}
The second two are given by
\begin{align}
\bs \phi(P_\zeta) & =
\bs \sigma_1^{-1} \,
\bs \rho_a^{-1} \,
\bs \sigma_2 \,
\bs \rho_b^{-1} \,
\bs \sigma_3^{-1} \,
\bs \rho_a^{-1} \,
\bs \sigma_1 \, ,
\\
\bs \phi(P_{\widetilde{\zeta}})
& =
\bs \sigma_a^{-1} \,
\bs \rho_1^{-1} \,
\bs \sigma_c^{-1} \,
\bs \rho_2^{-1} \,
\bs \sigma_b^{-1} \,
\bs \rho_1^{-1} \,
\bs \sigma_a \, ,
\end{align}
which are of the form \eq{pureCWform} (since $P_\zeta$ and $P_{\widetilde{\zeta}}$ contain only clockwise turns), where
\begin{align}
\big( \bs \phi(P_\zeta) \big)_1{}^1 & = \frac{1 + p_2(1+p_3)- \ve_2(1 + \ve_2 \, \ve_3)\varphi_a\,\varphi_b}{\ve_2\, \ve_3} \\
\big( \bs \phi(P_\zeta) \big)_1{}^2 & = p_1 \, \ve_2 \, \ve_3
\\
\big( \bs \phi(P_\zeta) \big)_1{}^3 & = \ve_1(1- \ve_2 \, \ve_3)\varphi_a - \ve_1\,\ve_2 \,\varphi_b  \shortintertext{and}
\big( \bs \phi(P_{\widetilde{\zeta}}) \big)_1{}^1 & = -\frac{1+p_c(1+p_b)+\ve_c(1 - \ve_b\,\ve_c) \varphi_1\,\varphi_2}{\ve_b\, \ve_c}
\\
\big( \bs \phi(P_{\widetilde{\zeta}}) \big)_1{}^2 & = - \ve_a^2 \, \ve_b\, \ve_c
\\
\big( \bs \phi(P_{\widetilde{\zeta}}) \big)_1{}^3 & = \ve_a ( 1 + \ve_b\, \ve_c) \varphi_1 + \ve_a \, \ve_c \, \varphi_2 \, .
\end{align}
Then \eq{Pzetaxi} is solved by
\begin{align}
\ve_a & = \mp \frac{\ve_1 \, \ve_3}{M} \, ,
&
\ve_b & = \mp M^2 \, \ve_2 \, ,
&
\ve_c & = \pm \frac{1}{\ve_3} \, ,
&
\ve_e & = \frac{\ve_3 \, \ve_5}{M} \, ,
\end{align}
\begin{align}
\varphi_1 & = \mp \frac{\ve_3\, \varphi_a + \varphi_b}{M} \, ,
&
\varphi_2 & = - \frac{\varphi_a - \ve_3\, \varphi_b}{M} \, .
\end{align}
where $M$ satisfies \eq{Mdef}.

 \section{The string measure}
 \label{measu}
The Schottky space measure for an $n$-point, $h$-loop bosonic string theory amplitude is given by \cite{DiVecchia:1996uq}
\begin{align}
[\d m]_g^n & = \frac{1}{\d V_{abc}} \prod_{i=1}^{n} \frac{\d z_i}{V_i'(0)} \prod_{j =1}^g \Big[ \frac{\d k_j \, \d u_j \, \d v_j}{ k_j^2(u_j - v_j)^2}(1-k_j)^2 \Big] \nonumber \\
& \hspace{20pt} \times (\det\text{Im}\, \tau)^{- d / 2} {\prod_{\alpha}} ' \Big[\prod_{i=1}^\infty (1 - k_\alpha^n)^{-d}\prod_{j=2}^\infty (1 - k_\alpha^n)^{2}  \Big] \nonumber \\
& \hspace{20pt} \times \text{ factors coming from external states.} \label{schomeas}
\end{align}
$\d V_{abc}$ is the projective-invariant volume element given by
\begin{align}
\d V_{abc} & = \frac{ \d \rho_a \, \d \rho_b \, \d \rho_c}{ (\rho_a - \rho_b)(\rho_b- \rho_c)(\rho_a - \rho_c)} \, , \label{dVabc}
\end{align}
where the three $\rho_i$s have been chosen from among the Koba-Nielsen points $z_j$ and the Schottky fixed points $u_\ell$, $v_\ell$ to be fixed (usually to 0, 1 and $\infty$) to eliminate the projective invariance.
Let us write the first line of \eq{schomeas} as $[\d m_0]_g^n$. Using the parametrization described above, it takes the elegant form
\begin{align}
[\d m_0]_g^n & = \prod_{\text{edges }i}\,\, \frac{\d p_i}{p_i^2}\, \prod_{\text{cl.bord.}\beta} (1 - k_\beta) \label{eq:pimeas}
\end{align}
where the second product is over all closed borders. We define a \emph{border path} in a graph $\Gamma$ to be a path $P$ whose decomposition includes solely either \cw~or \acw~turns. A \emph{closed border} of $\Gamma$ is a closed path $P_\beta$ that is a border path. The multiplier $k_\beta$ of the corresponding M\"obius map $\phi(P_\beta)$ is simply the product of the gluing parameters $p_k$ of the edges $E_k$ traversed by $P_\beta$.

\subsection{Planar surfaces}
\label{planarproof}
We will prove \eq{eq:pimeas} for planar surfaces by the following method: first we will compute it explicitly for some particular choice of parametrization of a planar surface with $g$ loops and $n$ external edges. Then we will show that its form is preserved by `duality transformations' which change the topology of the target skeleton graph $\Gamma$, as in section \ref{DualitySect}.

\begin{figure}
 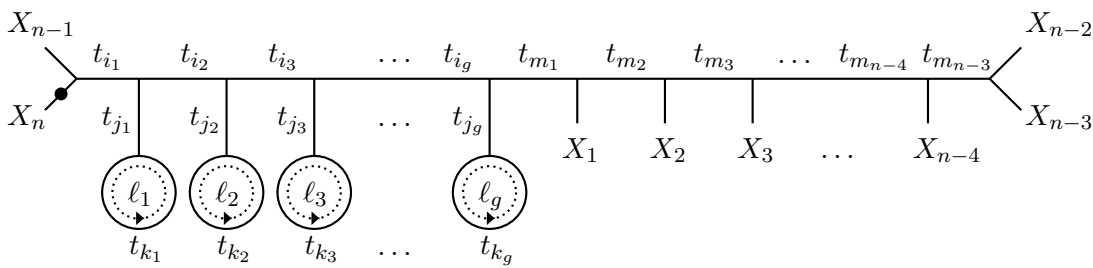
 \caption{An example planar skeleton graph with $g$ loops and $n$ external edges. The large dot indicates our choice of base point.}
 \label{fig:planarex}
 \end{figure}

 Let us choose to parametrize our surface in such a way as to correspond with the skeleton graph in
 \Fig{fig:planarex}.
 Then the surface is parametrized by the $3g-3+n$ variables $\{ p_{i_s},p_{j_s},p_{k_s},p_{m_t}; \, s= 1 , \ldots, g; \, t=1 , \ldots n-3\}$.

 The Schottky group generators $\gamma_s$ are given by
 \begin{align}
 \gamma_s & = U_s \, \sigma_{j_s} \, \rho \, \sigma_{k_s} \, \rho \, \sigma_{j_s} \, U_s^{-1}
 &
 s & = 1 , \ldots , g \, ,
 \shortintertext{where}
 U_s  & = U_{s-1} \, \rho^{-1} \, \sigma_{i_s} \, \rho^{-1} \, ,
 &
 U_0 & = \text{Id} \, .
 \end{align}
 Since $\gamma_s$ is conjugate to $\rho^{-1}\sigma_{k_s}$, its multiplier is
 \begin{align}
 k_s & = p_{k_s} \, .
 \label{exmultcomp}
 \end{align}
 Its fixed points can be computed as
 \begin{align}
 u_s & = \frac{1 + f_{2s} \, p_{i_s} ( 1 + p_{j_s} \,p_{k_s})}{1 + f_{1s}\, p_{i_s}( 1 + p_{j_s} \,p_{k_s}) } \, ;
 &
  v_s & = \frac{1 + f_{2s} \,p_{i_s} ( 1 + p_{j_s})}{1 + f_{1s} \, p_{i_s}( 1 + p_{j_s} ) } \, ,
 \end{align}
 where
 \begin{align}
 f_{rs} & \equiv \begin{cases}
 0 & \text{ if } s < r \\
 1 + p_{i_{s-1}} f_{r\,{(s-1)}} & \text{ if } s \geq r \, ,
 \end{cases} \label{eq:frsdef}
 \end{align}
 for $r = 1,2$.
 The local coordinates at the punctures, $V_t$, are given by
 \begin{align}
 V_1 & = U_g \, \rho^{-1} \, \sigma_{m_1} \, \rho^{-1} \, , \\
 V_t & = V_{t-1} \, \rho^{-1} \, \sigma_{m_t} \, \rho^{-1} \,
 \shortintertext{ for $t  = 2 , \ldots , n-3$,}
 V_{n-2} & = V_{n-3} \, \rho^{-1} \, , \\
 V_{n-1} & = \rho \, , \\
 V_n & = \text{Id} \, .
 \end{align}
 As holomorphic functions, the coordinate charts $V_t$ are given in the form
 \begin{align}
 V_t ( z ) & = \frac{1 + (1-z)\, g_{2t}\, p_{m_t}}{1+ (1-z)\, g_{1t} \, p_{m_t}}
 \end{align}
 for $ t = 1 , \ldots, n-3$, where we've extended the definition of $f_{rs}$ from \eq{eq:frsdef} with
 \begin{align}
 g_{rt} & = \begin{cases}
 f_{r\,(g+1)} & \text{ if } t=1 \\
 1 + p_{m_t} \, g_{r\, (t-1) }  & \text{ if }t> 1.
 \end{cases}
 \end{align}
 For these values of $t$, this gives
 \begin{align}
  z_t = V_t (0) & = \frac{1 + \, g_{2t}\, p_{m_t}}{1+ \, g_{1t} \, p_{m_t}} ;
  &
  V_t '(0) & = \frac{ \prod_{s = 1}^{g} p_{i_s} \prod_{r = 1}^t p_{m_r} }{(1 + g_{1t} p_{m_t})^2} \, , \label{Vtplanarex}
 \end{align}
 where we've used
 \begin{align}
 g_{1t} - g_{2t} & =  \prod_{s = 1}^{g} p_{i_s} \prod_{r = 1}^{t-1} p_{m_r} \, .
 \end{align}
 Note that with this choice of base point, the coordinates of three particular punctures $z_t = V_t(0)$ are `gauge fixed', \ie~independent of the moduli:
 \begin{align}
 V_n(0)& =0 \, ,
 &
  V_{n-1}(0) & = \infty \, ,
 &
  V_{n-2}(0) & = 1 \, .
 \end{align}
 To keep track of the infinity, let us temporarily include a small $\epsilon$ in $V_{n-1}$:
 \begin{align}
 V_{n-1}^{(\epsilon)}(0) & \equiv  1 - \frac{1}{z + \epsilon} \, ,
 \end{align}
 so $V_{n-1}^{(\epsilon)}(z) = 1 - 1 / \epsilon$.
 Since the three Koba-Nielsen variables $z_{n-2}$, $z_{n-1}$ and $z_n$ are independent of the moduli, they must be inserted in the projective-invariant volume element $\d V _{abc}$ \eq{dVabc} as $\rho_a$, $\rho_b$ and $\rho_c$, giving
 \begin{align}
 \d V_{abc}   & = \frac{\d z_{n-2} \, \d z_{n-1} \, \d z_n }{(z_{n-2} - z_{n-1}^{(\epsilon)})(z_{n-1}^{(\epsilon)} - z_n)(z_{n-2}-z_n) } = - \frac{\d z_{n-2} \, \d z_{n-1} \, \d z_n }{(1-\epsilon)/\epsilon^2} \, . \label{dVabcplanarex}
 \end{align}

To write down the measure \eq{schomeas} in terms of our new moduli, we need to compute the Jacobian determinant. Note that $u_s$, $v_s$ and $k_s$ are independent of $p_{i_r}$, $p_{j_r}$ and $p_{k_r}$ for $r> s$; similarly $z_t$ is independent of $p_{m_r}$ for $ r > t$ and $t \leq n-3$. This means that the Jacobian matrix is in upper-triangular block form:
So to compute its determinant we need to evaluate only the entries of the blocks on the diagonal. We can write the Jacobian matrix as $\big(\begin{smallmatrix} A & B \\ 0 & D \end{smallmatrix}\big)$ where $A$ is a $3g \times 3g$ matrix, also in upper-triangular block form:

 \begin{align}
 A & = \left(\begin{array}{cccc}
 A_{11} & A_{12} & \cdots & A_{1g} \\
 0 & A_{22} & \cdots & A_{2g} \\
 \vdots & \vdots & \ddots & \vdots \\
 0 & 0 & \cdots & A_{gg}
 \end{array}
 \right) \, ;
 &
 A_{rs} & = \left(
 \begin{array}{ccc}
 {\partial u_s}/{\partial p_{i_r}} & {\partial v_s}/{\partial p_{i_r}} & 0 \\
 {\partial u_s}/{\partial p_{j_r}} & {\partial v_s}/{\partial p_{j_r}}& 0 \\
 {\partial u_s}/{\partial p_{k_r}}& {\partial v_s}/{\partial p_{k_r}} &  {\partial k_s}/{\partial p_{k_r}}
 \end{array}
 \right) \, ,
 \end{align}
 $D$ is an $(n-3) \times (n-3)$ upper-triangular matrix,
 \begin{align}
 D & = \left(\begin{array}{cccc}
 D_{11} & D_{12} & \cdots & D_{1,(n-3)} \\
 0 & D_{22} & \cdots & D_{2,(n-3)} \\
 \vdots & \vdots & \ddots & \vdots \\
 0 & 0 & \cdots & D_{(n-3),(n-3)}
 \end{array}
 \right) \, ;
 &
 D_{rs} & = \frac{\partial z_{s}}{\partial p_{m_r}} \, ,
 \end{align}
 and the $3g \times (n-3)$ matrix $B$ is not needed to compute the determinant.
  The determinant can be evaluated with the use of the following partial derivatives:
 \begin{align}
 \frac{\partial u_s}{\partial p_{i_s}} & = - \frac{(1 + p_{j_s} p_{k_s})\prod_{r=1}^{s-1} p_{i_r}}{(1 + f_{1s} p_{i_s}(1 + p_{j_s} p_{k_s}))^2} \\
 \frac{\partial u_s}{\partial p_{j_s}} & = - \frac{ p_{k_s}\prod_{r=1}^{s} p_{i_r}}{(1 + f_{1s} p_{i_s}(1 + p_{j_s} p_{k_s}))^2} \\
 \frac{\partial v_s}{\partial p_{i_s}} & = - \frac{(1 + p_{j_s} )\prod_{r=1}^{s-1} p_{i_r}}{(1 + f_{1s} p_{i_s}(1 + p_{j_s}))^2} \\
 \frac{\partial v_s}{\partial p_{j_s}} & = - \frac{ \prod_{r=1}^{s} p_{i_r}}{(1 + f_{1s} p_{i_s}(1 + p_{j_s}))^2} \\
 \frac{\partial k_s}{\partial p_{k_s}} & = 1  \\
 \frac{\partial z_t}{\partial p_{m_t}} & = - \frac{\prod_{ s=1}^{g} p_{i_s}\prod_{ r=1}^{t-1} p_{m_r} }{(1+g_{1t}\, p_{m_t})^2} \label{planarexjaczp}
 \end{align}
 ($\partial u_s / \partial p_{k_s}$ and $\partial v_s / \partial p_{k_s}$ are also unneeded to compute the determinant).
 The minors can be evaluated as
 \begin{align}
 \left| \begin{array}{cc} {\partial u_s}/{\partial p_{i_s}} & {\partial u_s}/{\partial p_{j_s}} \\
 {\partial v_s}/{\partial p_{i_s}} & {\partial v_s}/{\partial p_{j_s}}
 \end{array} \right| & =
 \frac{(1- p_{k_s})\, p_{i_s}\,\prod_{r=1}^{s-1} p_{i_r}^2 }{(1 + f_{1s} p_{i_s}(1 + p_{j_s}))^2(1 + f_{1s} p_{i_s}(1 + p_{j_s}p_{k_s}))^2} \, ,
 \end{align}
 which can be combined with
 \begin{align}
 (u_s - v_s)^2 & = \frac{(1- p_{k_s})^2\, p_{j_s}^2\,\prod_{r=1}^{s} p_{i_r}^2 }{(1 + f_{1s} p_{i_s}(1 + p_{j_s}))^2(1 + f_{1s} p_{i_s}(1 + p_{j_s}p_{k_s}))^2}
 \end{align}
 to give
 \begin{align}
 \frac{(1- k_s)^2}{k_s^2\,(u_s - v_s)^2}\left| \begin{array}{cc} {\partial u_s}/{\partial p_{i_s}} & {\partial u_s}/{\partial p_{j_s}} \\
 {\partial v_s}/{\partial p_{i_s}} & {\partial v_s}/{\partial p_{j_s}}
 \end{array} \right| & = \frac{1 - p_{k_5}}{p_{i_s}\,  p_{j_s}^2\, p_{k_s}^2   } \, . \label{planarexloopcomb}
 \end{align}
 Similarly, $V_t'(0)$ from \eq{Vtplanarex} can be combined with \eq{planarexjaczp} to give
 \begin{align}
 \frac{1}{V_t'(0)} \frac{\partial z_t}{\partial p_{m_t}} & = - \frac{1}{p_{m_t}} \, , \label{planarexexcomb}
 \end{align}
 for $ t  = 1 , \ldots , n-3$. We also have
 \begin{align}
 \frac{1}{V_{n-2}'(0)} & = \prod_{s=1}^{g} p_{i_s} \prod_{t=1}^{n-3} p_{m_t} \, ,
 &
 \frac{1}{V_{n-1}^{(\epsilon)} {}' (0)} & = \epsilon^2 \, ,
&
 \frac{1}{V_{n} ' (0)} & = 1 \, . \label{planarexgfvp}
 \end{align}
 Combining all the factors from \eq{dVabcplanarex}, \eq{planarexloopcomb}, \eq{planarexexcomb} and  \eq{planarexgfvp} and letting $\epsilon \to 0$, we get (up to a sign)
 \begin{align}
 \frac{1}{\d V_{abc}} \prod_{t = 1}^{n} \frac{\d z_t}{V_t'(0)} \prod_{s=1}^g \frac{\d u_s \, \d v_s\,\d k_s}{(u_s - v_s)^2} \frac{ (1 - k_s^2)}{k_s^2} & = \prod_{t = 1}^{n} \frac{\d p_{m_t}}{p_{m_t}^2} \prod_{s=1}^g\frac{\d p_{i_s}}{p_{i_s}^2} \frac{\d p_{j_s}}{p_{j_s}^2} \frac{\d p_{k_s}}{p_{k_s}^2}(1-p_{k_s}) \, .
 \end{align}
 This is exactly of the form of \eq{eq:pimeas}, because the closed borders of the diagram in \Fig{fig:planarex} are precisely the $g$ loops whose multipliers are $p_{k_s}$.

\subsection{Invariance under duality transformations}
\label{dualityt}
We can prove that the pole structure of the Schottky group string measure, \eq{eq:pimeas}, is invariant under the duality transformations of the pinching moduli $p_i$ as described in section \ref{DualitySect}. This implies, in particular, that our calculation of the string measure for the graph in \Fig{fig:planarex} proves the validity of \eq{eq:pimeas} for all planar graphs.

\subsubsection{Invariance under transformations without external edges}
For a duality transformation which doesn't involve external edges, we must simply show that \eq{eq:pimeas} is invariant under the change of variables in \eq{dualsupermod}. For the bosonic case we can write \eq{dualsupermod} in terms of the $p$'s and set the odd supermoduli to 0 to get
\begin{align}
p_a & = \frac{p_1\,p_3}{1+p_3} \,\,
&
p_b & = p_4 \,(1+p_3)\,\,
&
p_c & = \frac{1}{p_3}\,\,
&
p_d & = p_2 \, (1+p_3)\,\,
&
p_e & = \frac{p_3\, p_5}{1+p_3} \, . \label{dualmod}
\end{align}
The factor
\begin{align}
\prod_{\text{cl.bord.}\beta}(1 - k_\beta) \label{borderfactor}
\end{align}
is unchanged.
To see why this true, let us consider an example. Supposed the path $P_\alpha$ in \Fig{fig:duality3a} is part of a closed border, \ie~that there is some path $P_{\alpha ' }$  including only CW turns such that $P_\alpha \cdot P_{\alpha ' }$ is a closed path in some graph $\Gamma$. Its multiplier will be \begin{align}
k & = p_2 \, p_3 \, p_5 \! \! \prod_{E_i \in P_{\alpha '}} \! \! p_i \, .  \label{k235} \end{align}
Then
$P_{\widetilde{\alpha}} \cdot P_{\alpha ' }$ will be a closed path in the dual graph $\hat \Gamma$, where $P_{\widetilde{\alpha}}$ is the path depicted in \Fig{fig:duality3b}. The multiplier of this path will be given by
\begin{align}
\hat k & = p_d \, p_e  \! \! \prod_{E_i \in P_{\alpha '}} \! \! p_i \, .\label{kde} \end{align}
But from the last two equations of \eq{dualmod} we get
\begin{align}
p_d \, p_e & = p_2 \, p_3 \, p_5 \, , \label{pdpe}
\end{align}
which implies that $k$ in \eq{k235} is equal to $\hat k$ in \eq{kde}, and thus the contribution to \eq{borderfactor} is unchanged.

Similarly, from \eq{dualmod} we have
\begin{align}
p_a \, p_b & = p_1 \, p_3 \, p_4 \, ,
&
p_b \, p_c \, p_e & = p_4 \, p_5 \, ,
&
p_a \, p_c \, p_d & = p_1 \, p_2 \, ,
\end{align}
so any closed border in $\Gamma$ will correspond to a closed border in $\hat \Gamma$ with the same multiplier.

The Jacobian determinant of \eq{dualmod} is given by
\begin{align}
\left|\frac{\partial( p_a , \ldots , p_e)}{\partial (p_1 , \ldots , p_5)} \right|
& =
\left|
\begin{array}{ccccc}
p_3/(1+p_3) & 0 & p_1/(1+p_3)^2 & 0 & 0 \\
0 & 0 & p_4 & 1+p_3 & 0 \\
0 & 0 & -1/p_3^2 & 0 & 0 \\
0 & 1 + p_3 & p_2 & 0 & 0 \\
0 & 0 & p_5/(1+p_3)^2 & 0 & p_3/(1+p_3)
\end{array}
\right| \\
& = 1 \, . \label{dualityjac}
\end{align}
Since
\begin{align}
p_a \, p_b \, p_c \, p_d \, p_e & = p_1 \, p_2 \, p_3 \, p_4 \, p_5 \, ,
\end{align}
the form of the measure \eq{eq:pimeas} is unchanged by the duality transformation, which is what we wanted to show.

If the duality transformation includes a loop as in \Fig{fig:duality5} or \Fig{fig:duality6} then it is necessary to use the slightly different transformations as given in sections \ref{adjacentloop} and \ref{oppositeloop}.
\subsubsection{Invariance under transformations involving external edges}
For a duality transformation which does involve external edges, we need to do more work.

The reason for this is that although our prescription for duality transformations \emph{does} ensure that $V_t(0) =  z_t = \hat V_t(0)$ where $V_t$ and $\hat V_t$ are local coordinates around a puncture, it does \emph{not} in general give $V_t'(0) = \hat V_t'(0)$. Therefore since $V_t'(0)^{-1}$ appears on the left-hand-side of \eq{schomeas}, the measure needs to be multiplied by $V_t'(0) / \hat V_t'(0)$ for each external edge involved in the duality transformation.

So to check that \eq{schomeas} is invariant under duality transformations, we need to show that
\begin{align}
\bigg|\prod_{\text{ext.~edges }i} \frac{ V_i'(0) }{\hat V_i'(0)} \bigg| & = \bigg|\prod_{\text{int.~edges }j,\, k} \frac{p_j}{\hat p_k} \bigg|^2\times
\bigg|\frac{\partial \hat p_k}{\partial p_j} \bigg|
\end{align}
holds for all combinations of external edges, where $p_j \in \{ p_a, \ldots , p_e \}$ and $\hat p_k \in \{ p_1 , \ldots , p_5\}$. The cases in which there are 1, 2, 3 or 4 external edges must each be checked separately; furthermore in the case of 2 external edges, there are two sub-cases which differ in whether the edges are adjacent or opposite.

All of these cases can be checked by direct computation; let us give one in detail so that the approach is clear. Suppose there is one external edge involved in the duality transformation; let us use the labels from \Fig{fig:duality4}. First of all, let us note that since we no longer have $p_1$ and $p_a$ as moduli in \Fig{fig:duality4a} and \Fig{fig:duality4b}, respectively, the Jacobian determinant will not be as simple as it was in \eq{dualityjac}. We get
\begin{align}
\left|\frac{\partial( p_b , \ldots , p_e)}{\partial (p_2 , \ldots , p_5)} \right|
& =
\left|
\begin{array}{ccccc}
 0 & p_4 & 1+p_3 & 0 \\
 0 & -1/p_3^2 & 0 & 0 \\
 1 + p_3 & p_2 & 0 & 0 \\
 0 & p_5/(1+p_3)^2 & 0 & p_3/(1+p_3)
\end{array}
\right| = \frac{1 + p_3}{p_3} \, . \label{1xjac}
\end{align}
The product of the moduli is given by
\begin{align}
p_b \, p_c\, p_d\, p_e & = p_2\, (1+p_3)\,p_4 \, p_5 \, . \label{1xprod}
\end{align}
To find the ratio $V_1'(0)/ \hat V_1 '(0)$, we use the fact that we can write
\begin{align}
V_1 & = V_{0} \cdot \phi(\hat P_\gamma) \, ,
&
\hat V_1 & = V_{0} \cdot \phi(\hat P_{\widetilde{\gamma}}) \, .
\end{align}
where $V_{0}$ is some map which is unchanged by the duality transformation, so that
\begin{align}
V_1 & = \hat V_1 \cdot \phi(\hat P_{\widetilde{\gamma}})^{-1} \cdot \phi(\hat P_\gamma) \, .
\end{align}
Using the chain rule, we have
\begin{align}
V_1'(0) & = \hat V_1'\big( \phi(\hat P_{\widetilde{\gamma}}^{-1}\hat P_\gamma)(0)\big) \, \cdot \, \phi(\hat P_{\widetilde{\gamma}}^{-1}\hat P_\gamma)'(0)
\end{align}
but $\phi(\hat P_{\widetilde{\gamma}}^{-1}\hat P_\gamma)(0)=0$ by construction, so
\begin{align}
\frac{V_1'(0)}{ \hat V_1'(0)}  & = \phi(\hat P_{\widetilde{\gamma}}^{-1}\hat P_\gamma)'(0) = \frac{p_3}{1 + p_3} \, . \label{VpRatio1}
\end{align}
We can then combine \eq{1xjac}, \eq{1xprod} and \eq{VpRatio1} to get
\begin{align}
V_1'(0) \frac{\d p_b}{p_b^2} \frac{\d p_c}{p_c^2} \frac{\d p_d}{p_d^2} \frac{\d p_e}{p_e^2}   & = \frac{\hat V_1'(0)}{\phi(\hat P_{\widetilde{\gamma}}^{-1}\hat P_\gamma)'(0)}  \, \left|\frac{\partial( p_b , \ldots , p_e)}{\partial (p_2 , \ldots , p_5)} \right| \, \frac{ \d p_1 \, \d p_2 \, \d p_3 \, \d p_4}{p_2^2 \, (1 + p_3)^3 \, p_4^2 \, p_5^2 } \\
& = \hat V_1'(0)\, \frac{\d p_2}{p_2^2} \frac{\d p_3}{p_3^2} \frac{\d p_4}{p_4^2} \frac{\d p_5}{p_5^2} \, ,
\end{align}
so \eq{eq:pimeas} is preserved by duality transformations involving one external edge.

By similar computations, it is straightforward to show that \eq{eq:pimeas} is invariant under a duality transformation involving two, three, or four external edges.

\subsection{For non-planar diagrams}
The proof in section \ref{planarproof} only establishes \eq{eq:pimeas} for planar diagrams. For non-planar diagrams, we can prove the result by induction on the number of loops $\ell$, by showing that the formula \eq{eq:pimeas} is respected by the process of sewing together a pair of legs on an $(\ell-1)$-loop, $(n+2)$-point diagram to obtain an $\ell$-loop, $n$-point diagram.
\subsubsection{Sewing to form a loop with a closed border}
\label{clborderproof}
The inductive step is performed differently depending on whether or not the sewing creates a closed border path on the diagram (recall that this means that this means a closed path whose turns are either all clockwise or all anti-clockwise).
Let us deal first of all with the case in which a closed border cycle is created.

\begin{figure}
\centering
\subfloat[]{\label{fig:measureproof1b} 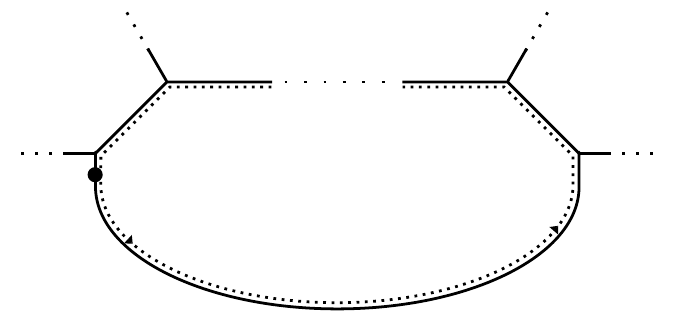}
\subfloat[]{\label{fig:measureproof1a} 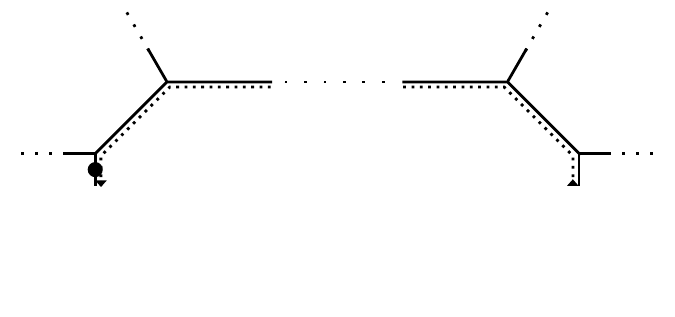}
\caption{Cutting the edge $E_j$ in
\Fig{fig:measureproof1b} leaves a connected, but not closed, border cycle joining two external edges marked $X_1$ and $X_2$ in \Fig{fig:measureproof1a}. To prove \eq{eq:pimeas} inductively we compare the string measures of the two diagrams.
}
\label{fig:measureproof1}
\end{figure}

This case can be further divided into two sub-cases: in the first, the loop being sewn includes at least one edge $E_j$ which also belongs to another, distinct border path (this is not the case, for example, for non-planar diagrams that have a single border cycle). When this is the case, cutting $E_j$ leaves the border open but still with a single component, as in \Fig{fig:measureproof1}.

To perform the inductive step, we compute the ratio of the string measures $[ \d m_0 ]_{\ell}^n$ and $[\d m_0]_{\ell-1}^{n+2}$  for the diagrams in \Fig{fig:measureproof1b} and \Fig{fig:measureproof1a} respectively and show that they differ by the correct factor required by \eq{eq:pimeas}, \ie
\begin{align}
[ \d m_0 ]_{\ell}^n & =  [\d m_0]_{\ell-1}^{n+2}\, \times \,\frac{\d p_j}{p_j^2} \, (1 - k_{\,\text{border}}) \, \label{borderrat}
\end{align}
where $p_j$ is the pinching parameter for the edge which is cut and $k_{\, \text{border}}$ is the multiplier of the closed border cycle in \Fig{fig:measureproof1b}, which is given by
\begin{align}
k_{\,\text{border}} & = p_j \, \times \, \prod_{s=1}^r \, p_{i_s} \, .
\end{align}
The factors in the measures corresponding to the loops that are not cut and to the external edges that are not sewn together will clearly give an identical contribution to both sides of \eq{borderrat}, so we need to compute only the factors that differ. Thus we need to show
\begin{align}
\frac{\d u_\ell\,\d v_\ell\,\d k_\ell}{(u_\ell-v_\ell)^2} \frac{(1-k_\ell)^2}{k_\ell^2} & = \frac{\d z_1 \, \d z_2}{V_1'(0) V_2'(0)} \, \times \, \frac{\d p_j}{p_j^2}\, (1 - k_{\,\text{border}}) \, \label{eq:sewclaim}
\end{align}
where $u_\ell$, $v_\ell$ and $k_\ell$ are the parameters of a Schottky generator corresponding to the newly sewn loop.

The base point (marked by the dot in \Fig{fig:measureproof1b}) lies at the position of the external edge marked $X_1$, so the chart $V_1$ is just the identity:
\begin{align}
V_1 & = \text{Id} \, ,
&
z_1 & = V_1(0) = 0 \, ,
&
V_1'(0) & = 1 \, .
\end{align}
The chart $V_2$ can be found by tracing the path from $X_2$ to the base point indicated in \Fig{fig:measureproof1a} by the dotted line and following the procedure described in section \ref{param} to get
\begin{align}
V_2 & = \rho^{-1} \, \sigma_{i_1} \, \rho^{-1} \, \sigma_{i_2}\, \rho^{-1} \, \ldots \, \rho^{-1} \, \sigma_{i_n} \,
\end{align}
or as a holomorphic function,
\begin{align}
V_2(z) & = \frac{1}{y - x z} \, ,
\end{align}
where
\begin{align}
x & = p_{i_1} \, p_{i_2} \, \ldots \, p_{i_n} \, ,
&
y & = 1+ p_{i_1}(1+p_{i_2}(1 + \ldots (1+p_{i_m})\ldots)) \, .
\end{align}
This gives
\begin{align}
z_2 & = V_2(0) = 1/y \, ,
&
V_2'(0) & = x\, z_2^2 \, .
\end{align}
We need to add a generator $\gamma_\ell$ to the Schottky group to account for the loop being added; we can use the Schottky group element defined by the closed border cycle:
\begin{align}
\gamma_\ell & = V_2 \, \sigma_j \,  \\
\shortintertext{or, as a holomorphic map,}
\gamma(z) & = \frac{z}{y\, z+p_j\, x} \, .
\end{align}
Its multiplier and fixed points are given by
\begin{align}
k_\ell & = p_j\, x = k_{\, \text{border}} \, ,
&
u_\ell & = \frac{1 - k_\ell}{y} \, ,
&
v_\ell & = 0 \, . \label{SchotBdy}
\end{align}
Since $v_\ell$ and $z_1$ are both gauge-fixed to 0, we can delete $\d v_\ell$ and $\d z_1$ from \eq{eq:sewclaim} as they are both cancelled by the projective-invariant volume element $\d V_{abc}$ in \eq{schomeas}.
Then using
\begin{align}
u_\ell & = z_2(1 - x\, p_j) \, & k_\ell & = x \, p_j
\end{align}
we find
\begin{align}
\left|\frac{\partial(u_\ell, k_\ell)}{\partial(z_2,p_j)} \right| & =
\left|
\begin{array}{cc}
1 - p_j x & 0 \\ - z_2 \, x & x
\end{array}\right| = x(1-p_j x) \, .
\end{align}
Inserting this in
\begin{align}\d u_\ell \, \, \d k_\ell & =
\left|\frac{\partial(u_\ell, k_\ell)}{\partial(z_2,p_j)} \right| \, \d z_2 \, \d p_j \, ,
\end{align}
it is easy to verify with simple algebra that \eq{eq:sewclaim} holds.
\subsubsection{Sewing to form a loop without a closed border}
\label{noborderproof}
In section \ref{clborderproof} we gave the details of the inductive step to prove \eq{eq:pimeas} in the case that an $\ell$-loop, $n$-point diagram with $B$ closed border cycles had an internal edge removed to give an $(\ell-1)$-loop, $(n+2)$-point diagram with $(B-1)$ closed border cycles. In this section we deal with the case that both diagrams have the same number of border cycles $B$.

\begin{figure}
\centering
\subfloat[]{\label{fig:measureproof3a} 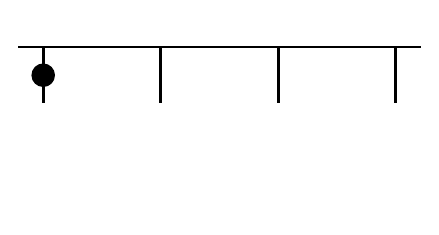}
\subfloat[]{\label{fig:measureproof3b} 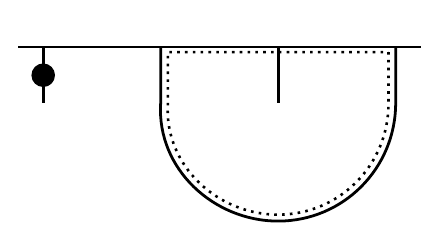}
\caption{
To prove \eq{eq:pimeas} inductively for diagrams belonging to a certain class, we can show that the measures $[\d m_0]_{\ell -1}^{n+2}$ and $[\d m_0]_{\ell }^{n}$ for the diagrams in \Fig{fig:measureproof3a} and \Fig{fig:measureproof3b} have the desired ratio.
}
\label{fig:measureproof3}
\end{figure}

\begin{figure}
\centering
\subfloat[]{\label{fig:measureproof2a} 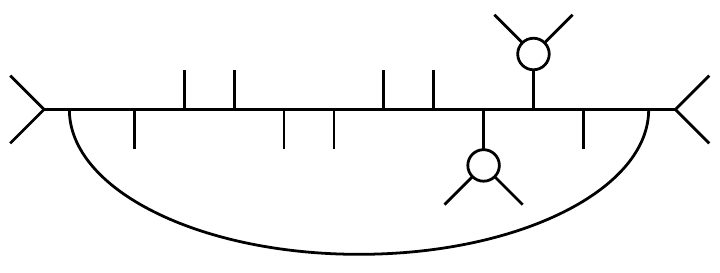}
\subfloat[]{\label{fig:measureproof2b}  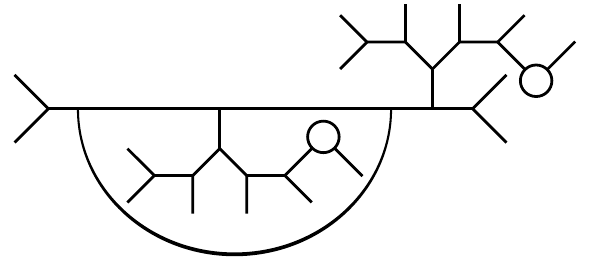}
\caption{A diagram that contains a loop with many edges attached to both its inside and its outside (an example is shown in \Fig{fig:measureproof2a}) is dual to a diagram (\eg~\Fig{fig:measureproof2b}) where the corresponding loop has one edge attached to its inside and two edges attached to its outside.
}
\label{fig:measureproof2}
\end{figure}

The loop being cut must include both clockwise and anti-clockwise turns (or it would have a closed border cycle). Without loss of generality, let us assume that the loop being cut consists of two clockwise turns and one anti-clockwise turn as in \Fig{fig:measureproof3b}.  A diagram with a loop consisting of a complicated sequence of clockwise and anti-clockwise turns (such as \Fig{fig:measureproof2a}) is dual to another diagram (such as \Fig{fig:measureproof2b}) where the corresponding loop is of this sort, and we have proven that \eq{eq:pimeas} is unchanged by duality transformations.
So then we can take the local coordinates of the punctures (before sewing the loop) to be given by:
\begin{align}
V_1 & = \rho^{-1} \, \cdot \, \sigma_{i_1} \, \cdot \, \rho^{-1} \, ,
& \\
V_2 & = V_1 \, \cdot\, \rho^{-1} \, \cdot \, \sigma_{i_2} \, \cdot \, \rho^{-1} \, , \\
V_3 & = V_2 \, \cdot\, \rho^{-1} \, \cdot \, \sigma_{i_3} \, \cdot \, \rho^{-1} \, ,
\end{align}
while the Schottky generator for the loop in \Fig{fig:measureproof2b} (after sewing) is given by
\begin{align}
\gamma & = V_3 \, \cdot \, \sigma_j \, \cdot\, V _1 ^{-1} \, .
\end{align}
We find
\begin{align}
\frac{\d u \, \d v \, \d k}{(u-v)^2}\frac{(1-k)^2}{k^2 } \frac{\d z_2}{V_2'(0)} & = \prod_{m=1}^3 \frac{\d p_{i_m}}{p_{i_m}^2} \times \frac{\d p_j}{p_j^2} \label{nobdyaftersew}
\end{align}
and
\begin{align}
\prod_{m=1}^3 \frac{\d z_{m}}{V_{m}'(0)} & = \prod_{m=1}^3 \frac{\d p_{i_m}}{p_{i_m}^2} \label{nobdybeforesew}
\end{align}
\eq{nobdybeforesew} and \eq{nobdyaftersew} differ by a factor of $\d p_j / p_j^2$, so \eq{eq:pimeas} has been shown by induction.

\subsection{The superstring case}
The corresponding part of the measure for the NS sector of the superstring is given by
\begin{align}
[ \d \bs m_0]_g^n & = \frac{1}{ \d \bs V_{abc}} {  \prod_{t=1}^{n} \frac{ \d \bs z_t}{(D \bs V_t^{\zeta} )(0|0)}  } \prod_{s=1}^g \frac{ \d \bs u_s \, \d \bs v_s\,\d (q_s^2) }{\bs u_s \dotminus \bs v_s} \frac{(1 + q_s)^2 }{q_s^3} \label{NSschomeas}
\shortintertext{with}
\frac{1}{ \d \bs V_{abc}} & = \frac{\sqrt{(\bs a \dotminus \bs b)(\bs b \dotminus \bs c)(\bs c \dotminus \bs a)}}{\d \bs a \, \d \bs b \, \d \bs c} \, \d \bs \Theta_{\bs a\, \bs b, \bs c} \, ,
\end{align}
where $\Theta_{\bs a\, \bs b, \bs c} $ and $\bs a \dotminus \bs b$ have been defined in \eq{oddinvariant} and \eq{superdiff}, respectively. Furthermore, $\bs V_t^{\zeta}$ is the odd part of $\bs V_t$ and $D$ is the superderivative.

When expressed in terms of the pinching parameters of section \ref{supparam}, $[\d \bs m_0]_g^n$ take the following form (cf.~\eq{eq:pimeas}):
\begin{align}
[\d \bs m_0]_g^n & = \prod_{\text{edges }i}\,\,   \prod_{\text{vertices }j}\,\, \prod_{\text{cl.bord.}\beta} \frac{\d p_i}{\ve_i^3}\, \d \varphi_j\,(1 +q_\beta) \label{eq:NSpimeas}
\end{align}
Note that as well as the products over edges and borders, there is now a product over graph vertices, which we have put in 1--1 correspondence with the odd supermoduli $\varphi_j$.

This formula could be proven by applying the same approach as in the previous sections; instead we will show how the computation can be done for one particular diagram.

\subsubsection{The Mercedes-Benz diagram}
\begin{figure}
\centering
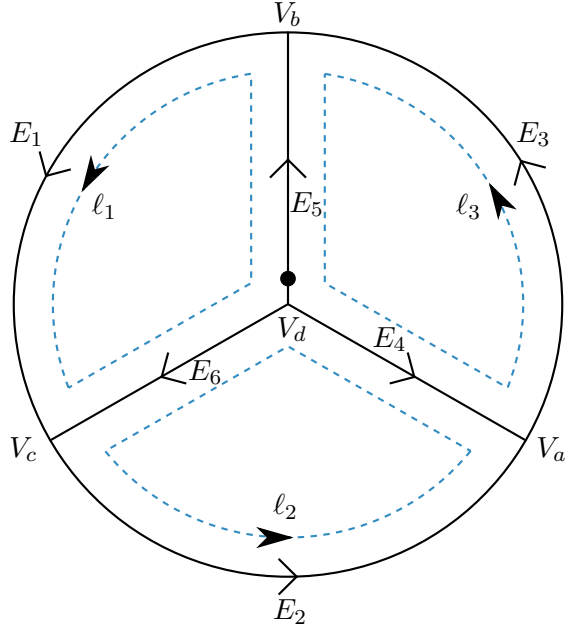
\caption{The symmetric $g=3$, $n=0$ tetrahedral graph.
}
\label{fig:SuperMerc}
\end{figure}

Consider the graph $\Gamma$ depicted in \Fig{fig:SuperMerc}. Let us compute the super Schottky measure $[\d \bs m_0]_{g=3}^{n=0} $ \eq{NSschomeas} in terms of the pinching parameters $(\ve_i | \varphi_j)$ corresponding to this graph.
We get
\begin{align}
\bs \gamma_1 & = \bs \Sigma_{d,-6} \, \bs \Sigma_{c,1} \, \bs \Sigma_{b,5} \, ,  \\
\bs \gamma_2 & = \bs \rho_d^{-1}  \,\, \bs \Sigma_{d,-4} \, \bs \Sigma_{a,2} \, \bs \Sigma_{c,6} \,\, \bs \rho_d \, , \\
\bs \gamma_3 & = \bs \rho_d \, \, \bs \Sigma_{d,-5} \, \bs \Sigma_{b,3} \, \bs \Sigma_{a,4} \, \, \bs \rho_d^{-1} \, ,
\shortintertext{where}
\bs \Sigma_{s,\pm n} & \equiv \bs \rho_{s}^{-1} \, \bs \sigma_n^{\pm 1} \, .
\end{align}
The semimultipliers are given by
\begin{align}
q_1 & = \ve_1 \, \ve_5 \, \ve_6 \, , &
q_2 & = \ve_2 \, \ve_4 \, \ve_6 \, , &
q_3 & = \ve_3 \, \ve_4 \, \ve_5 \, .
\end{align}
The fixed points can be written as eigenvectors satisfying
\begin{align}
\bs \gamma_i \, U_i & = q_i^{-1} \, U_i \, , & \bs \gamma_i \, V_i & = q_i \, V_i \, ,
\end{align}
in the form
\begin{align}
U_i & = \bs \rho_d^{1-i} \, \cdot \, \left(\! \! \! \begin{array}{c} 1 - p_i \\ A_i \\ (1 + x_i)\Psi_i \end{array}\! \! \!\right) \, , &
V_i & = \bs \rho_d^{1-i} \, \cdot \, \left(\! \! \begin{array}{c} 0 \\ 1 \\ 0 \end{array}\! \!\right) \, ,
\end{align}
where
\begin{align}
\Psi_{\alpha \beta \gamma}^{ij} & \equiv \varphi_a + \ve_i\, \varphi_b - \ve_i\, \ve_j \, \varphi_\gamma \, ,
\\
A_{\alpha \beta \gamma}^{ij} & \equiv 1 + p_i(1 + p_j) + \varphi_\alpha \Psi_{\alpha \beta \gamma}^{ij}  - p_i \,\ve_j \varphi_\beta \varphi_\gamma \, .
\end{align}
The 3 super-points that are gauge-fixed are
\begin{align}
\bs v_1 & = 0 | 0  \, ,  & \bs v_2 & = 1 | \varphi_d \, , & \bs v_3 & = \infty | 0 \, ;
\end{align}
their odd super-projective invariant is
\begin{align}
\bs \Theta_{\bs v_1 \bs v_2  \bs v_3 } & = \varphi_d \, .
\end{align}
Thus the standard $(3g-3|2g-2) = (6|4)$ super Schottky moduli used for this worldsheet are
\begin{align}
(q_1,q_2,q_3,u_1,u_2,u_3 | \theta_1, \theta_2 , \theta_3 ,\bs \Theta_{\bs v_1 \bs v_2  \bs v_3 }) .
\end{align}
We can express these in terms of the pinching supermoduli
\begin{align}
(\ve_1,\ldots , \ve_6 | \varphi_a, \ldots , \varphi_d) \, .
\end{align}
To rewrite the super-Schottky measure in terms of the pinching supermoduli we need to compute the Berezinian of the matrix  $\big(\begin{smallmatrix}A & |  B \\ \hline C & | D\end{smallmatrix}\big)$ where
\begin{align}
A & = \left(\begin{array}{ccc}
{\partial q_1}/{\partial \ve_1} & \ldots & {\partial u_3}/{\partial \ve_1} \\
\vdots & \ddots & \vdots  \\
{\partial q_1}/{\partial \ve_6} & \ldots & {\partial u_3}/{\partial \ve_6}
\end{array} \right) \, ;
&
B & = \left(\begin{array}{ccc}
{\partial \theta_1 }/{\partial \ve_1 } & \ldots & {\partial \theta_3}/{\partial \ve_1} \\
\vdots & \ddots & \vdots  \\
{\partial \theta_1 }/{\partial \ve_6 } & \ldots & {\partial \theta_3}/{\partial \ve_6}
\end{array} \right) \, ; \\
C & = \left(\begin{array}{ccc}
{\partial q_1}/{\partial \varphi_a} & \ldots & {\partial u_3}/{\partial \varphi_a} \\
\vdots & \ddots & \vdots  \\
{\partial q_1}/{\partial \varphi_d} & \ldots & {\partial u_3}/{\partial \varphi_d}
\end{array} \right) \, ;
&
D & = \left(\begin{array}{ccc}
{\partial \theta_1 }/{\partial\varphi_a } & \ldots & {\partial \theta_3}/{\partial \varphi_a} \\
\vdots & \ddots & \vdots  \\
{\partial \theta_1 }/{\partial\varphi_d } & \ldots & {\partial \theta_3}/{\partial \varphi_d}
\end{array} \right) \, ;
\end{align}
We can find (with the use of \emph{Mathematica}) that this matrix has
\begin{align}
Ber = \frac{ \det (A - B D^{-1}C)}{\det(D)} & = - 8 \, \frac{(1 + q_4)\, \ve_4^2 \, \ve_5^2\, \ve_6^2 \,  u_1 \, (u_2 - v_2 - \theta_2 \, \phi_2)}{(1 + q_1)(1+q_2)(1+q_3)} \,
\end{align}
where $q_4 \equiv \ve_1 \, \ve_2\, \ve_3$.  Multiplying this by
\begin{align} \prod_{r=1}^2 \frac{1}{\bs u_r \dotminus \bs v_r}\prod_{s=1}^2\left( \frac{1 + q_s}{q_s }\right)^2 \, ,
\end{align}
we get
\begin{align}
[ \d \bs m_0]_3^0 & = \frac{1}{\d \bs V_{\bs v_1 \bs v_2 \bs v_3}   } \prod_{s=1}^g \frac{ \d \bs u_s \, \d \bs v_s\,\d (q_s^2) }{\bs u_s \dotminus \bs v_s} \frac{(1 + q_s)^2 }{q_s^3} \\
& = \d \varphi_d \, \prod_{r=1}^2  \frac{ 1 }{\bs u_r \dotminus \bs v_r} \prod_{s=1}^g \d \bs u_s \, \d (q_s^2) \, \frac{ (1 + q_s)^2 }{q_s^3} \\
& =  \prod_{i=a,b,c,d} \!\!\!\! \d \varphi_i \prod_{j=1}^6 \frac{\d \ve_j}{\ve_j^3} \, \prod_{\ell=1}^4 ( 1- q_\ell) \, .
\end{align}
This is exactly of the form given in \eq{NSschomeas}. In section 4.2 of \cite{Playle:2016hie} a similar computation was given for the $g=2$, $n=2$ diagram obtained by cutting one of the edges of \Fig{fig:SuperMerc}.

\section{Directions for further research}

\subsection{Feynman graphs with gluons}
The pinching parametrization allows us to arrive at the worldline Green's function as the $\alpha ' \to 0$ limit of the worldsheet Green's function, and thus we can see exactly how the various Feynman graphs for scalar $\Phi^3$ theory arise from the infrared dynamics of string theories with scalar fields in their spectrum. In \cite{Magnea:2015fsa}, it was shown how to isolate not only different Feynman graph topologies within the string measure, but even further to distinguish propagators belonging to different fields (such as gluons, ghosts and scalars). Whereas the results of \cite{Magnea:2013lna,Magnea:2015fsa} were worked out at length only for the two loop vacuum diagram, the use of pinching parameters would allow a similar analysis to be extended to Feynman graphs of arbitary topology. However, our proof in section \ref{graph} that the pinching parameters give a diagram-by-diagram matching between string theory and QFT is only valid for the scalar sector. We do know that at two loops the pinching parameters allow us to correctly obtain Feynman graphs involving gluons, because the pinching parametrization is a generalization of the parametrization used in \cite{Magnea:2013lna,Magnea:2015fsa}.  However, it is possible that for graphs with more complicated topologies the correspondence ceases to hold for gluons.

It would be interesting, then, to investigate whether Feynman diagrams with gluons can be reproduced with the use of the pinching parametrization, possibly with modification.

\subsection{Ramond pinching parameters}

The construction described in section \ref{supparam} provides a pinching parametrization for superstring worldsheets with states from the NS sector --- which correspond to spacetime bosons --- propagating along the plumbing fixtures. To describe a complete superstring theory we would have to allow states from both the Ramond (R) and NS sectors to propagate. The proper inclusion of the R sector --- necessary for the description of spacetime fermions --- would require a different type of plumbing fixture to be used (see section 6.2.1 of \cite{Witten:2012ga}). The inclusion of R plumbing fixtures would require the introduction of a number of novel features. First of all, R punctures are best described by non-superconformal coordinates $z|\zeta$ in which the superderivative takes the form $D_\zeta^* = \partial_\zeta + \zeta \, f(z)\, \partial_z$, where $f(z)$ is some polynomial vanishing at the puncture.
Next, an SRS disc with three punctures can have zero or two R punctures, and unlike the NNN disc, the NRR disc has no supermoduli (odd or even).
 Lastly, a plumbing fixture between two R punctures must have one Grassmann-even parameter and one Grassmann-odd parameter, whereas an NS plumbing fixture just has an even one ($\ve_i$ in our case). Whereas an NS plumbing fixture is an \osp~mapping between the charts at its two ends, this is not the case for the R plumbing fixture, and so the transition functions obtained will not belong to \osp.

One possible pinching parametrization that implements these constraints is the following. Let us put three charts on the NRR disc,  $z_i | \zeta_i$, $i = 0,1,2$. Let $z_0 | \zeta_0 $ be a superconformal chart centred on the NS puncture at $z_0|\zeta_0 = 0|0$. Then we can define the other two charts by
\begin{align}
z_0 |\zeta _0  = \rho(z_1) \Big | \frac{\zeta_1}{\sqrt{z_1}} \, = \rho^{-1} (z_2) \Big|  \frac{\sqrt{z_2}}{z_2-1} \, \zeta_2 \, , \label{rrhodef}
\end{align}
where $\rho$ and $\rho^{-1}$ are defined in \eq{rhoanalytic} and \eq{rhoi}, respectively. It follows from the chain rule that the superderivative takes on the following form:
\begin{align}
\partial_{\zeta_0} + \zeta_0 \partial_{z_0} & = z_1^{1/2} (\partial_{\zeta_1} + \zeta_1 \, z_1 \, \partial_{z_1}) = (z_2^{1/2} - z_2^{-1/2})(\partial_{\zeta_2} + \zeta_2 \, z_2 \, \partial_{z_2}) \, ,
\end{align}
so the charts $z_1|\zeta_1$ and $z_2 |\zeta_2$ define standard R punctures at their origins (cf.~Eq.~(4.4) of \cite{Witten:2012ga}). Sewing two such R punctures can be effected with the map
\begin{align}
\sigma_{p | \kappa}^{\pm} (z|\zeta) \equiv - p \, \frac{1 + \kappa \,\zeta}{z} \Big | \pm \ii \, (\zeta + \kappa) \, , \label{rsigmadef}
\end{align}
which preserves the superderivative. Here $p$ is a Grassmann-even parameter of the sewing and $\kappa$ is a Grassmann-odd parameter. Note that the inverse of $\sigma_{p | \kappa}^{\pm} $ is given by
\begin{align}
\Big(\sigma_{p | \kappa}^{\pm} \Big)^{-1} & = \sigma_{p | \mp \,\ii \, \kappa}^{\mp} \, .
\end{align}
Of course, the purely bosonic parts of \eq{rrhodef} and \eq{rsigmadef} reproduce \eq{rhodef} and \eq{sigmadef}.
Unfortunately, the pinching parametrization defined by these equations does not enjoy the nice properties of the parametrization defined in section \ref{supparam}. For example, if two target graphs are related by a duality transformation as in section \ref{DualitySect}, the two corresponding pinching parametrizations may no longer be obtained from each other simply by changing variables.

It could be very valuable to construct a pinching parametrization for the R sector (whether the one defined by \eq{rrhodef} and \eq{rsigmadef} or otherwise), and to write down, for example, the superstring measure in terms of these supermoduli.

\section*{Acknowledgements}
The authors would like to thank R.~Russo and L.~Magnea for collaboration on related projects.  This work was supported by the Compagnia di San Paolo contract
``MAST: Modern Applications of String Theory'' \verb=TO-Call3-2012-0088=.
\appendix
\section{Proofs of formulae}
\label{schotapp}
\subsection{(Semi)multiplier}
\label{multiplierproof}
In this section we prove \eq{eq:multexpr} and \eq{loopsemi}, \ie~that to leading order the (semi)multiplier of a (super) Schottky group element is simply the (signed) product of the (NS) pinching parameters associated to the edges in the graph that are included in the corresponding loop.

We will write out the proof for the NS case \eq{loopsemi}; the bosonic case \eq{eq:multexpr} follows as a corollary because any Schottky group can be embedded in a split super Schottky group (\ie~one with odd constants $\theta_j$ set to 0), and then the multipliers are the squares $k_\alpha = q_\alpha^2$ of the corresponding semimultipliers.

Any super-Schottky group element $\bs \gamma_\alpha$ written down according to the procedure above is conjugate to one of the form
\begin{align}
\bs \gamma _N & = \bs \sigma_{\ve_{i_1}}^{\, r_1} \, \bs \rho_{\theta_{j_1}}^{\, s_1} \, \bs \sigma_{\ve_{i_2}}^{\, r_2} \, \bs \rho_{\theta_{j_2}}^{\, s_2} \cdots \sigma_{\ve_{i_N}}^{\, r_N} \, \bs \rho_{\theta_{j_N}}^{\, s_N} \, , \label{gamForm}
\end{align}
where the $r_n$'s and $s_n$'s are all $\pm1$ and the $\ve_{i_n}$'s and $\theta_{i_n}$'s are not necessarily distinct (in case the loop traverses an edge or vertex more than once). Since the semimultiplier is a conjugacy class invariant, we can use \eq{gamForm} to compute it without loss of generality.

Let us define
\begin{align}
\Pi_{\bs \gamma_N} & = \prod_{n=1}^{N} r_n \, s_n \, \ve_{i_n} \, . \label{Pidef}
\end{align}
Then \eq{loopsemi} can be restated as
\begin{align}
q_{\bs \gamma_N} & = \Pi_{\bs \gamma_N} ( 1 + {\cal O}(\ve_{i_n})) \, .  \label{loopsemirs}
\end{align}
The overall sign $\prod_{n} r_n \, s_n$ in \eq{loopsemirs} matches the sign we described geometrically in \eq{loopsemi} because each edge $E_{i_n}$ traversed against the marked direction corresponds to a factor of $\bs \sigma_{\ve_{i_n}}^{-1}$ in \eq{gamForm} so $r_n = -1$, and each clockwise move around a vertex $V_{j_n}$ corresponds to a factor of $\bs \sigma_{\theta_{j_n}}^{-1}$ in \eq{gamForm} so $s_n = -1$.

Note that to invert $\bs \gamma_N$, as well as reversing the order of the factors in \eq{gamForm}, we would multiply each of the exponents by $(-1)$, so \eq{Pidef} implies $\Pi_{\bs \gamma_N^{-1}} = \Pi_{\bs \gamma_N}$ because $(-1)^{2N}=1$.

To compute the semimultiplier, we first show that $\bs \gamma$ is always of the form
\begin{align}
\bs \gamma_N &  = \begin{cases}
\Pi_{\bs \gamma_N}^{-1} \left(\begin{array}{cc|c}
\Pi_{\bs \gamma_N}^2 + P_{2N-1}& P_{2N-2} & P_{2N-1}\, \Omega \\
-1 + P_{2N-2} & 1 +P_{2N-2}  & P_{2N-3}\,\Omega - \theta_N \\ \hline
P_{2N-1}\,\Omega & P_{2N-3}\,\Omega & \Pi_{\bs \gamma_N} + P_{2N-2}\,\Omega^2
\end{array}\right)
& \text{ if } s_N = 1 \\
\phantom{.}
\\
\Pi_{\bs \gamma_N}^{-1} \left(\begin{array}{cc|c}
\Pi_{\bs \gamma_N}^2 + P_{2N-1} & -\Pi_{\bs \gamma_N}^2 + P_{2N-1}& P_{2N-1} \,\Omega \, + \, \Pi_{\bs \gamma_N}^2\, \theta_N \\
P_{2N-2} & 1 + P_{2N-2} & P_{2N-2}\, \Omega \\ \hline
P_{2N-1}\, \Omega & P_{2N-1}\, \Omega & \Pi_{\bs \gamma_N} \, + \, P_{2N-1} \, \Omega^2
\end{array}\right)
& \text{ if } s_N =- 1 \, ,
\end{cases} \label{gamGenPoly}
\end{align}
where the notation $P_{n}$ has been used to stand for any arbitrary polynomial in the $\ve_i$'s of degree at most $n$ with constant term zero, whose coefficients are valued in the Grassmann algebra generated by the $\theta_i$'s.
None of the various polynomials $P_n$ necessarily coincide with each other, and $P_n \equiv 0 $ if $n \leq 0$. $\Omega$ denotes the odd-parity subspace of the Grassmann algebra generated by the $\theta_i$'s.

We can prove \eq{gamGenPoly} by induction on $N$. We can take all of the $r_n$'s to be $+1$ without loss of generality since $\bs \sigma_{\ve_{i_n}}^{\pm 1} = \bs \sigma_{\pm \ve_{i_n}}$ so the exponents may be absorbed into the signs of the $\ve_{i_n} \!\!$'s. In the $N=1$ case, Eqs.~(\ref{gamForm}) and (\ref{brhodef}) -- (\ref{bsigdef}) give us
\begin{align}
\bs \gamma_1 & =  \begin{cases}
\ve_{i_1}^{-1}  \left(\begin{array}{cc|c}
\ve_{i_1}^2 & 0& 0 \\
-1 & 1 &  - \theta_{j_1} \\ \hline
- \ve_{i_1} \, \theta_{j_1} & 0 & \ve_1
\end{array}\right)
& \text{ if } s_1 = 1 \\
\phantom{.}
\\
(- \ve_{i_1})^{-1}  \left(\begin{array}{cc|c}
\ve_{i_1}^2 & -\ve_{i_1}^2 & \ve_{i_1}^2 \, \theta_1\\
0 & 1 &  0 \\ \hline
0 & \ve_{i_1} \, \theta_{j_1}  & - \ve_{i_1}
\end{array}\right)
& \text{ if } s_1 =- 1 \, ,
\end{cases} \label{gamN1}
\end{align}
which is of the form \eq{gamGenPoly} with $\Pi_{\bs \gamma_1} = s_1 \, \ve_1$.

For the inductive step, we need to multiply each of the matrices in \eq{gamGenPoly} on the right by both of the matrices in \eq{gamN1} (after replacing $\ve_{i_1}, \theta_{j_1} \mapsto \ve_{i_{N+1}}, \theta_{j_{N+1}}$). It can be verified that the resulting matrices are still of the form in \eq{gamGenPoly}, with the replacement $N \mapsto N+1$. To check this, note that $P_m \, \ve_{i_n}$ may be written as $P_{n+1}$, that any polynomial $P_m$ can also be written as $P_n$ if $m \leq n$, and that we can write $P_n+P_n = P_n$.

Now we know the general form of an \osp~element given by \eq{gamForm}, we can compute the semimultiplier.
 The \emph{supertrace} of an \osp~matrix $\bs \gamma = (\bs \gamma_{ij})$ is defined to be
\begin{align}
\text{sTr}(\bs \gamma) & = \bs \gamma_{11} + \bs \gamma_{22} - \bs \gamma_{33} \, . \label{sTr}
\end{align}
Any hyperbolic \osp~map $\bs \gamma$ satisfies
\begin{align}
\text{sTr}(\bs \gamma) & = q_{\bs \gamma} + q_{\bs \gamma}^{-1} - 1 \, , \label{strsemim}
\end{align}
where $q_{\bs \gamma}$ is the semimultiplier. To compute the semimultiplier of a hyperbolic \osp~map, we can compute the supertrace and solve \eq{strsemim} for $q_{\bs \gamma}$.

Both matrices in \eq{gamGenPoly} have supertraces of the form
\begin{align}
\text{sTr}(\bs \gamma_N) & = \Pi_{\bs \gamma_N} + \frac{1+ P_{2N-1}}{\Pi_{\bs \gamma_N}} - 1 \, , \label{sTrForm}
\end{align}
(after simplifying $P_{2N-1} + P_{2N-2} - P_{2N-2} \Omega^2 \mapsto P_{2N-1} $ in the first case and  $P_{2N-1} + P_{2N-2} - P_{2N-1} \Omega^2 \mapsto P_{2N-1} $ in the second case).

Equating the right-hand-sides of \eq{strsemim} and \eq{sTrForm}, we get a quadratic equation for $q_{\bs \gamma_N}$, with solution
\begin{align}
q_{\bs \gamma_N} & = \frac{\Pi_{\bs \gamma_N}^2 + P_{2N-1} + 1}{2 \Pi_{\bs \gamma_N}} \left( 1 - \sqrt{1 - \Big(\frac{2 \Pi_{\bs \gamma_N}}{\Pi_{\bs \gamma_N}^2 + P_{2N-1} + 1}  \Big)^2} \right)
\end{align}
(the other root is $q_{\bs \gamma_N}^{-1}$). Then expanding the square root as a power series in $\Pi_{\bs \gamma_N}$,
\begin{align}
q_{\bs \gamma_N} & = \frac{\Pi_{\bs \gamma_N}^2 + P_{2N-1} + 1}{2 \Pi_{\bs \gamma_N}} \left( 1 - 1  + \frac{1}{2} \Big(\frac{2 \Pi_{\bs \gamma_N}}{\Pi_{\bs \gamma_N}^2 + P_{2N-1} + 1}  \Big)^2 + { \cal O}( \Pi_{\bs \gamma_N}^3 )  \right) \\
& =  \frac{ \Pi_{\bs \gamma_N}}{\Pi_{\bs \gamma_N}^2 + P_{2N-1} + 1}  \, + \, { \cal O}( \Pi_{\bs \gamma_N}^3 )   \, ,
\end{align}
and lastly we can expand the denominator in the pinching parameters yielding
\begin{align}
q_{\bs \gamma_N} & = \Pi_{\bs \gamma_N}( 1 + { \cal O} ( \ve_i ) ) \, ,
\end{align}
which is what we wanted to show.

Of course, the proof of the bosonic case \eq{eq:multexpr} is analogous, and can be obtained by setting all Grassmann-odd parameters to 0 and considering only the \psl~subgroup of \osp.

\subsection{Period matrix}
\label{pmproof}
We want to prove the period matrix formula
\begin{align}
(\tau_{ij}) & = \frac{1}{2 \pi \ii} \sum_{k} \langle \ell_i , \ell_j \rangle^k \, \log ( \pm \, p_k) \, +  \,  + {\cal O} (p_n) \, . \label{graphpmtheorem}
\end{align}
Our starting point is the Schottky group formula for the period matrix, \eq{pmseries}.
To compute the cross-ratio in \eq{pmseries}, we can use
\begin{align}
 \gamma_\alpha( u_j) & =   u_{\alpha  j  \bar \alpha}
 &
 \gamma_\alpha( v_j) & =  v_{\alpha  j  \bar \alpha}
\end{align}
where $u_{\alpha  j  \bar \alpha}$ (respectively $v_{\alpha  j  \bar \alpha}$) is the attractive (repulsive) fixed point of $\gamma_{\alpha  j  \bar \alpha} \equiv \gamma_\alpha \gamma_j \gamma_{\alpha}^{-1}$.
If we write $\eta_{\alpha \beta}$ to denote the following cross-ratio of the fixed points of two Schottky group elements $\gamma_\alpha$, $\gamma_\beta$:
\begin{align}
\eta_{\alpha \beta} & \equiv \frac{u_\alpha - v_\beta}{u_\alpha - u_\beta} \frac{v_\alpha - u_\beta}{v_\alpha - v_\beta} \, , \label{etacr}
\end{align}
then \eq{pmseries} can be rewritten as \begin{align}
\tau_{ij} & = \frac{1}{2 \pi \ii } \Big( \delta_{ij} \, \log k_i \, - \, {}^{(i)} {\sum_{\gamma_\alpha}}'{}^{(j)} \log \eta_{i, \alpha j \bar \alpha} \Big) \label{pmseriesnocr} \, .
\end{align}

Now  we will show how the cross-ratio $\eta_{i, \alpha j \bar \alpha}$ on the right-hand-side of \eq{pmseriesnocr} can be expressed in terms of multipliers, which can ultimately be expressed in terms of pinching parameters depending on the geometry of the graph $\Gamma$, thanks to the results of section \ref{multiplierproof}.

\begin{figure}
\centering
\subfloat[]{ 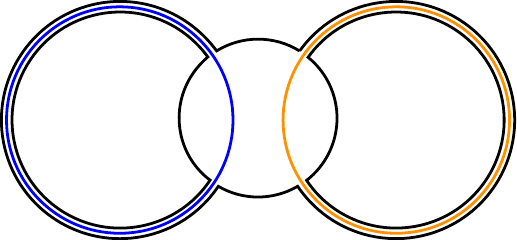 \label{fig:crossratplus} }
\subfloat[]{ 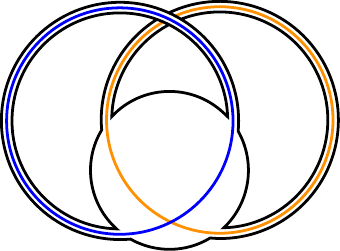 \label{fig:crossratminus} }
\caption{
 Ignoring all the other loops and external edges, if two loops $\ell_\alpha$, $\ell_\beta$ don't cross each other as in \Fig{fig:crossratplus} then $\eta_{\alpha \beta} > 0$; if they do cross as in  \Fig{fig:crossratminus} then $\eta_{\alpha \beta} < 0$.
}\label{fig:crossrat}
\end{figure}

Let us note that the multiplier $k_{\alpha \beta}$ of the product of two Schottky group elements $\gamma_\alpha$, $\gamma_\beta$ is given by:
\begin{align}
k_{\alpha \beta} & = \frac{(1 - \eta_{\alpha \beta})^2}{\eta_{\alpha \beta}^2} \, k_\alpha \, k_\beta + {\cal O}(k_\alpha)^2 + {\cal O}( k_\beta)^2  \, . \label{kofprod}
\end{align}
We can use \eq{kofprod} to express the cross-ratio \eq{etacr} in terms of multipliers:
\begin{align}
\eta_{\alpha \beta} & = \frac{\pm \sqrt{k_{\alpha}k_{\beta}/k_{\alpha \beta}}}{1 \pm\sqrt{k_{\alpha}k_{\beta}/k_{\alpha \beta}}}  + {\cal O}(k_\alpha) + {\cal O}( k_\beta) \, . \label{etaab}
\end{align}
The sign choice in \eq{etaab} depends on the ordering of the fixed points $\{u_\alpha, v_\alpha, u_\beta, v_\beta\}$ on the projective line, which is determined by the topology of the two loops $\ell_\alpha$ and $\ell_\beta$. If the two loops don't cross, as in \Fig{fig:crossratplus}, then $\eta_{\alpha \beta } > 0$; if they do cross as in \Fig{fig:crossratminus}) then we have $\eta_{\alpha \beta } < 0$.

With this, the period matrix can be expressed as a function of multipliers, and thus pinching parameters, by making the replacement
\begin{align}
\log \eta_{i, \alpha j \bar \alpha}  & = \log \frac{\pm \sqrt{k_{i}k_{j}/k_{i \alpha j \bar \alpha}}}{1 \pm \sqrt{k_{i}k_{j}/k_{i \alpha j \bar \alpha}}}  + {\cal O}(p_i)^2\, \label{etaiaja}
\end{align}
 in \eq{pmseriesnocr}, where $k_{i\alpha j \bar \alpha} \equiv k(\gamma_i \gamma_\alpha \gamma_j \gamma_\alpha^{-1} )$ and so on, and we have used the fact that $k_{\alpha j \bar \alpha} = k_j$.

 Only finitely many Schottky group elements give a contribution to \eq{pmseries} that doesn't vanish to first order in the $p_i$'s. Let's compute one of the off-diagonal elements---say $\tau_{12}$ without loss of generality. Let's compute the contribution to the series coming from the identity element (which is always included in the summation). We have
 \begin{align}
 \log \eta_{12} & = \log \frac{\pm \sqrt{k_1k_2/k_{12}}}{1 \pm \sqrt{k_1k_2/k_{12}}}  + {\cal O}(k_1) + {\cal O}(k_2)  \, . \label{cre12}
 \end{align}
 Let's suppose, first of all, that the loops $\ell_1$ and $\ell_2$ do not have any edges in common. Let us take the base point to be on $\ell_1$ (there is no loss of generality since the period matrix elements are \psl~invariant). Then $\gamma_1$ depends only on the pinching parameters in the loop $\ell_1$ while $\gamma_2$ is of the form $\gamma_2 = V^{-1} \, \wt \gamma_2 \, V$, where $\wt \gamma_2$ depends only on the pinching parameters in the loop $\ell_2$ and $V = \phi(P)$ is a matrix moving the base point along a path $P$ to a point on the loop $\ell_2$. Then we have
 \begin{align}
 \gamma_1\gamma_2 & = \gamma_1\,  V^{-1}  \, \wt \gamma_2 \, V
\,
\end{align}
so, with the results of section \ref{multiplierproof}, we have
\begin{align}
k_{12} = k(\gamma_1\, \gamma_2) & =  k_1 \,k_2 \prod_{E_i \in P} p_i^2 \, \big( 1 \, + \, {\cal O}(p_j) \, \big) \, .
\end{align}
Plugging this in \eq{cre12} we get
\begin{align}
\log \eta_{12} = {\cal O}(p_i)
\end{align}
in the case of no intersecting edges. One can check that the contribution from non-identity Schottky group elements vanishes.

Next, suppose that there is a connected chain of edges $E_{i_1}, E_{i_2}, \ldots, E_{i_n}$ lying on both $\ell_1$ and $\ell_2$; and that both loops cross these edges in \emph{opposite} directions. Assume the base point is at the start of the chain as in \Fig{fig:PMproof1}. Then the Schottky group element $\gamma_1 \gamma_2$ corresponds to a closed path $P_{12}$ which includes all of the edges in $\gamma_1$ and $\gamma_2$ except from $E_{i_1}\, \ldots \, E_{i_n}$. Thus we
get
\begin{align}
k_{12} = k(\gamma_1 \gamma_2) & = \frac{k_1\, k_2}{p_{i_1}^2 \cdots p_{i_n}^2} \big(1 + {\cal O}(p_j) \big) \, .
\end{align}
Putting this in \eq{cre12}, we find
\begin{align}
\log \eta_{12} & =\log \big( \pm \, p_{i_1} \,p_{i_2} \,\cdots\, p_{i_n} \big) +{ \cal O} (p_j) \, .
\end{align}
\begin{figure}
\centering
\subfloat[]{ 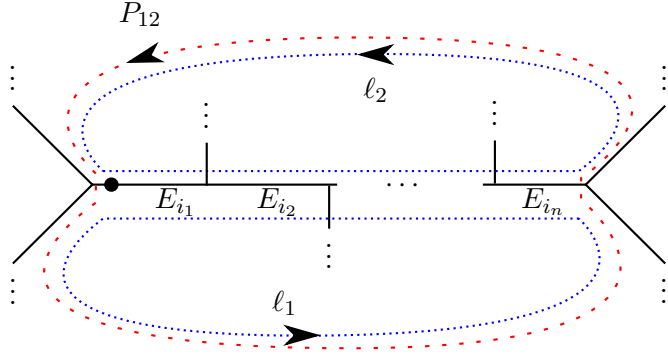 \label{fig:PMproof1} }
\caption{
Two loops $\ell_1$ and $\ell_2$ from the homology basis (shown as blue dotted curves) intersect along a chain of edges $E_{i_1}, \ldots , E_{i_n}$. Here, their orientations are opposite where they intersect. The red dashed curve is the reduced path $P_{12} = P_1 \cdot P_2$ (where $P_j$ is the path corresponding to the loop $\ell_j$). It crosses all the edges that $\ell_1$ and $\ell_2$ do, with the exception of $E_{i_1}, \ldots , E_{i_n}$.
}\label{fig:PMproof1}
\end{figure}
 Now suppose, on the other hand, that $\ell_1$ and $\ell_2$ cross $E_{i_1}\, \ldots \, E_{i_n}$ in the \emph{same} direction. Then $\ell_1$ and $\ell_2^{-1}$ cross in opposite directions, so the above result has to be modified by replacing $\eta_{12}$ with $\eta_{1 \bar 2}$ to get
\begin{align}
\log \eta_{1 \bar 2} & = \log \big( \pm \, p_{i_1} \,p_{i_2} \,\cdots\, p_{i_n} \big) +{ \cal O} (p_j) \, . \label{oppdirexpr}
\end{align}
We want to compute $\eta_{12}$ not $\eta_{1\bar 2}$, but since $\gamma_2^{-1}$ and $\gamma_2$ can be obtained from each other by fixing the multiplier $k_2$ and swapping the fixed points (\ie~$u_{\bar 2} = v_2$, $v_{\bar 2} = u_2$), we have
\begin{align}
\eta_{1 \bar 2} & = \frac{1}{\eta_{12}} \, ,
\end{align}
and thus in this case
\begin{align}
\log \eta_{12} & = \, - \, \log \big(\pm\, p_{i_1} \,p_{i_2} \,\cdots\, p_{i_n} \big) +{ \cal O} (p_j) \, .
\end{align}

Now suppose that $\ell_1$ and $\ell_2$ intersect more than once, \ie~that they both traverse $E_{i_1}, \ldots , E_{i_{n_1}}$, then separate, then come together again to intersect on a second connected chain $E_{j_1} , \ldots , E_{j_{n_2}}$, as in \Fig{fig:PMproof2}. Suppose that they cross $E_{j_1} , \ldots , E_{j_{n_2}}$ in opposite directions. Then there is a contribution from the Schottky group identity which is the same as in \eq{oppdirexpr}, but now contributions from other Schottky group elements must be added. If we follow $\ell_2$ from $E_{i_{n_1}}$ to $E_{j_1}$ and then $\ell_1$ from $E_{j_1}$ back to $E_{i_{n_1}}$, then we get a loop $\ell_\alpha$ corresponding to a closed path $P_{\alpha}$ starting and ending at the base point, and a corresponding Schottky group element $\gamma_\alpha$. The loop $\ell_{1 \alpha 2 \bar \alpha}$ crosses each edge as many times as $\ell_1$ and $\ell_2$ do, with the exception of the edges $E_{j_1}, \ldots , E_{j_{n_2}}$. Using our expression for Schottky group multipliers in terms of pinching parameters \eq{eq:multexpr}, we get
\begin{align}
k_{1 \alpha 2 \bar \alpha } = k(\gamma_1 \gamma_\alpha \gamma_2 \gamma_\alpha^{-1}) \, & = \, \frac{k_1 \, k_2}{p_{j_1}^2 \cdots p_{j_{n_2}}^2} \big( 1 + {\cal O}(p_i) \big)
\intertext{which we can insert into \eq{etaiaja} finding}
\log \eta_{1,\alpha 2 \bar \alpha} & = \log \big( \pm \, p_{j_1} \cdots p_{j_{n_2}} \big) \, \, + \, {\cal O}(p_i) \, , \label{etaafterloop1}
\end{align}
which (recalling \eq{pmseriesnocr}) is the contribution to the period matrix entry $\tau_{12}$ coming from the Schottky group element $\gamma_\alpha$.
\begin{figure}
\centering
\subfloat[]{ 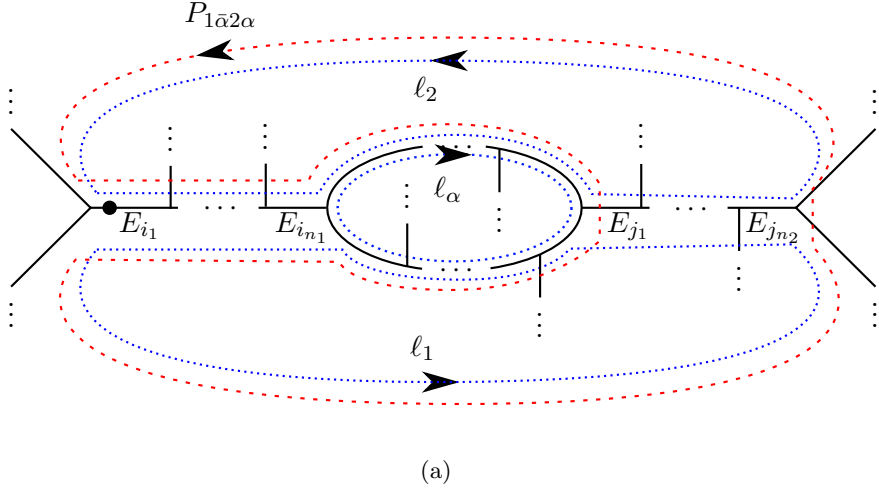 \label{fig:PMproof2} }
\caption{
Two loops $\ell_1$ and $\ell_2$ from the homology basis (shown as blue dotted curves) could intersect as in \Fig{fig:PMproof1} before splitting apart and then joining together to intersect again in a chain $E_{j_1}, \ldots , E_{j_{n_2}}$ of edges. The loop $\ell_\alpha$ (also shown with blue dots) is formed by tracing $\ell_2$ from $E_{i_{n_1}}$ to $E_{j_1}$ and then tracing $\ell_1$ back to $E_{i_{n_1}}$. The red dashed curve is the reduced path $P_{1  \alpha 2\bar \alpha} = P_1 \cdot P_\alpha \cdot P_2\cdot P_\alpha^{-1}$. It crosses all the edges that $\ell_1$ and $\ell_2$ do, with the exception of $E_{j_1}, \ldots , E_{j_{n_2}}$.
}\label{fig:PMproof2}
\end{figure}

If $\ell_1$ and $\ell_2$ traverse $E_{j_1} , \ldots , E_{j_{n_2}}$ in the same direction then, as above, we get
\begin{align}
\log \eta_{1,\alpha 2 \bar \alpha} & = \, - \,  \log \big(\pm\, p_{j_1} \cdots p_{j_{n_2}} \big) \, \, + \, {\cal O}(p_i) \, . \label{etaafterloop2}
\end{align}

The computation of \eq{etaafterloop1} and \eq{etaafterloop2} is valid for arbitrary numbers of intersections between $\ell_1$ and $\ell_2$, with each one getting a contribution from a different Schottky group element $\gamma_\alpha$.
The only Schottky group elements $\gamma_\alpha$ which give contributions to \eq{pmseries} that don't vanish at leading order are those that arise in this manner. In fact, in the remaining cases, the path $\ell_{1 \alpha 2 \bar \alpha} = \ell_1 \ell_\alpha \ell_2 \ell_\alpha^{-1}$ does not have any cancellations and \eq{eq:multexpr} gives simply
\begin{align}
k_{1 \alpha 2 \bar \alpha} & = k_1\,k_2\,k_\alpha^2 \, \big(1 + {\cal O}(p_i) \big) \, ,
\end{align}
which can be substituted in \eq{etaiaja} to give
\begin{align}
\log \eta_{1,\alpha 2 \bar \alpha} & = \log \frac{1}{1 + k_\alpha} + {\cal O}(p_i ) = {\cal O}(p_i) \, .
\end{align}

\begin{figure}
\centering
\subfloat[]{ 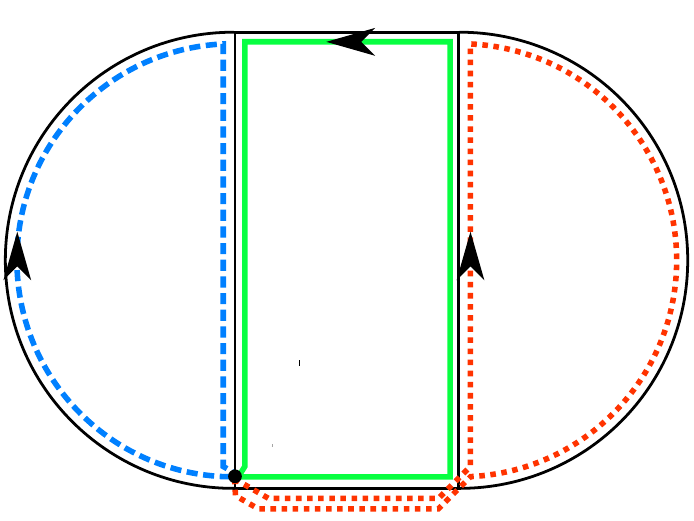 \label{fig:PMproof3a} } \\
\subfloat[]{ 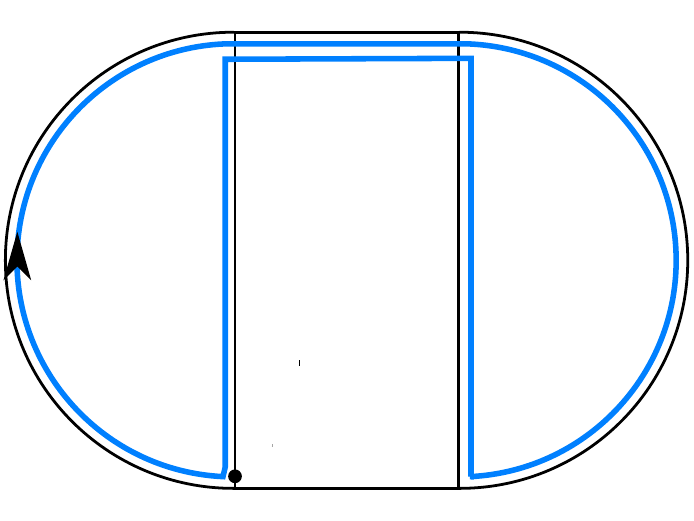 \label{fig:PMproof3d} }
\caption{
A 3-loop example illustrating that if $\ell_1$ and $\ell_2$ have no edges in common then the reduced loop $\ell_{1 \alpha 2 \bar \alpha}$ always includes all the edges in the loops $\ell_1$ and $\ell_2$ even if there are some intermediate cancellations. \Fig{fig:PMproof3a} shows the three paths $P_1$, $P_2$ and $P_\alpha$.
}\label{fig:PMproof3}
\end{figure}

Now suppose that $\ell_1$ and $\ell_2$ have no edges in common. Then for any loop $\ell_\alpha$, we find
\begin{align}
\frac{k_{1 \alpha 2 \bar \alpha} }{k_1 k_2} = {\cal O}(p_i)^2 \, , \label{nointer}
\end{align}
which can be substituted in \eq{etaiaja} to give
\begin{align}
\log \eta_{1,\alpha 2 \bar \alpha} &= {\cal O}(p_i) \, .
\end{align}
The general proof of \eq{nointer} is quite cumbersome. It is easy to check in the sub-case that none of the loops ($\ell_1$, $\ell_2$ and $\ell_\alpha$) have edges in common with each other. It can also be checked in the sub-case that $\ell_1$ and $\ell_2$ may have some edges in common with $\ell_\alpha$: it can be verified that any edges from $\ell_1$ cancelled in the product $\ell_1 \ell_\alpha$ are recovered from $\ell_{\bar \alpha}$, and so on, such that $\ell_{1 \alpha 2 \bar \alpha}$ contains all of the edges in $\ell_1$ and $\ell_2$ plus more (see \Fig{fig:PMproof3}).

 This proves that \eq{graphpmtheorem} holds for the off-diagonal elements of the period matrix.

The leading contribution to the diagonal elements comes entirely from the term $\frac{1}{2 \pi \ii } \, \delta_{ij} \log k_i$ in \eq{pmseries}, since
\begin{align}
\log k_i & = \sum_{E_j \in \ell_i} \log p_j \, + \, {\cal O}(p_j) \, . \label{diagcont}
\end{align}
The second term in \eq{pmseries} vanishes at leading order for the diagonal elements because $\log \eta_{1, \alpha 1 \bar \alpha}$ is always ${\cal O}(p_i)$ for all Schottky group elements $\gamma_\alpha$. For $\gamma_\alpha$ such that the corresponding loop $\ell_\alpha$ has no edges in common with $\ell_1$, this claim can be checked easily.

\begin{figure}
\centering
\subfloat[]{ 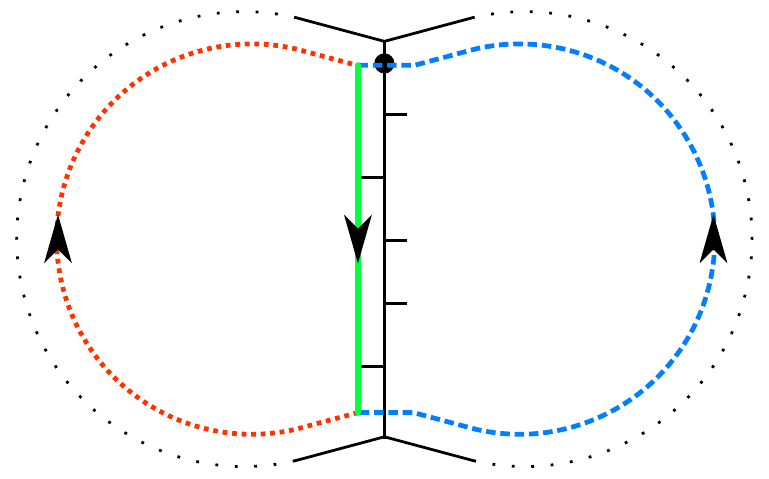 \label{fig:PMproof4a} } \\
\subfloat[]{ 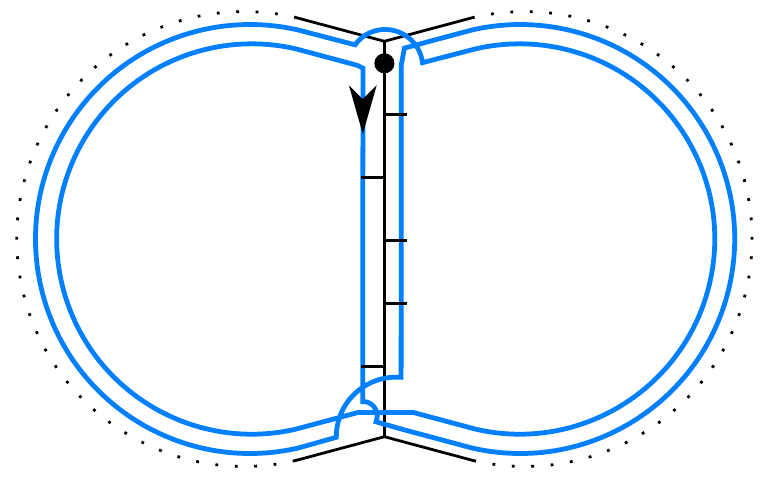 \label{fig:PMproof4b} }
\caption{
The path $\ell_{1 \alpha 1 \bar \alpha}$ always includes all of the edges in $\ell_1$ at least twice. This figure illustrates why this is true in the case that $\ell_1$ and $\ell_\alpha$ have an edge in common.
}\label{fig:PMproof4}
\end{figure}

The proof is a bit more involved in the case where $\ell_\alpha$ and $\ell_1$ have at least one edge (or connected chain of edges) in common. Let us suppose the base point is at one of the ends of $\ell_1 \cap \ell_\alpha$ such that the paths corresponding to the loops are given by $P_1 = P_1 ' \cdot P_I$ and $P_\alpha = P_\alpha ' \cdot P_I$, where $P_I$ is a path traversing the connected component of the intersection (see \Fig{fig:PMproof4a}).
Then
\begin{align}
P_1 \, P_\alpha\, P_1\, P_{\bar \alpha} & = P_1 ' \, P_I \, P_\alpha' \, P_I \, P_1' \, P_{\bar \alpha}' \, .
\end{align}
Since $P_I$ and $P_1'$ are each traversed twice, and together they comprise all the edges in the loop $\ell_1$ (as shown in \Fig{fig:PMproof4b}), we see that
\begin{align}
k_{1 \alpha 1 \bar \alpha} & = k_1 ^2 \,\Big( \prod_{E_i \in P_{\alpha}'} p_i ^2 \Big)\big(1 + {\cal O}(p_i ) \big) \, . \label{k1a1bara1}
\end{align}
We've assumed here that $P_\alpha$ and $P_1$ cross $P_I$ in the same direction, but if not then we could simply swap $P_\alpha \leftrightarrow P_{\bar \alpha}$ without changing the result since $\ell_{1 \alpha 1 \bar \alpha } = \ell_{1 \bar \alpha 1 \alpha}$.
\eq{etaiaja} gives us
\begin{align}
\log \eta_{1,\alpha 1 \bar \alpha} & = \log \frac{1}{1 \pm \sqrt{k_{ 1 \alpha 1 \bar \alpha}}/k_1}  + {\cal O}(p_i)  \, , \label{loge1a1a}
\end{align}
into which \eq{k1a1bara1} can be inserted giving
\begin{align}
\log \eta_{1,\alpha 1 \bar \alpha} & =  \,  - \! \prod_{E_i \in P_{\alpha}'} p_i \, + \, {\cal O}(p_j ) \, = \, {\cal O}(p_j) \, .
\end{align}
So at leading order, the only contribution to the diagonal elements of the period matrix $\tau_{ii}$ comes from \eq{diagcont}.

This complete the proof that the graph period matrix arises as the $\alpha ' \to 0$ limit of the period matrix when using the pinching parametrization along with \eq{pteq}.

\subsubsection{An example requiring non-identity Schottky group elements}
Let us consider an example where the computation of the period matrix requires summing over more Schottky group elements than just the identity, even at leading order. Consider the graph shown in \Fig{fig:4loopPM1}.
\begin{figure}
\centering
\subfloat[]{ 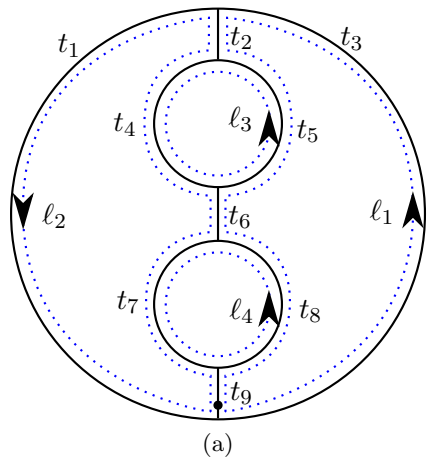 \label{fig:4loopPM1} }
\caption{
A 4-loop graph with a homology basis marked. This example shows the necessity of including contributions from non-identity Schottky group elements in the period matrix formula \eq{pmseries} even at leading order.
}\label{fig:4loopPM1}
\end{figure} The graph period matrix $(\theta)_{ij}$ with the indicated loop basis is
\begin{align}
\theta_{ij} & = \left(\begin{array}{cccc}
t_2 + t_3+t_5 + t_6 + t_8 +t_9 & \! \!\! \!\! \!\! \! \! \!\! \!-t_2 - t_6 - t_9 & \! \!\! \! - t_5 &\! \!\! \! - t_8 \\
- t_2 - t_6 - t_9 & \! \!\! \!\! \!\! \! \! \!\! \! t_1 + t_2 + t_4 + t_6 + t_7 + t_9 & \! \!\! \!- t_4 & \! \!\! \!- t_7 \\
- t_5 & \! \!\! \!\! \!\! \! \! \!\! \!- t_4 & \! \!\! \!t_4 + t_5 &\! \!\! \! 0 \\
- t_8 & \! \!\! \!\! \!\! \! \! \!\! \! - t_7 &\! \!\! \! 0 & \! \!\! \!t_7 + t_8
\end{array}\right).\label{eq:4loopPM}
\end{align}
We want to see how the Schottky group formula reproduces \eq{eq:4loopPM}.
Let us choose the following set of Schottky generators corresponding to the loop basis and base point indicated:
\begin{align}
\gamma_1 \, & = \, \sigma_9 \, \cdot \, \rho^{-1} \, \cdot \, \sigma_8 \, \cdot \, \rho^{-1} \, \cdot \, \sigma_6 \, \cdot \, \rho^{-1} \, \cdot \, \sigma_5 \, \cdot \, \rho^{-1} \, \cdot \, \sigma_2 \, \cdot \, \rho^{-1} \, \cdot \, \sigma_3 \, \cdot \, \rho^{-1} \, ,  \\
\gamma_2 \, & = \, \rho^{-1} \, \cdot \, \sigma_1 \, \cdot \, \rho^{-1} \, \cdot \, \sigma_2 \, \cdot \, \rho^{-1} \, \cdot \, \sigma_4 \, \cdot \, \rho^{-1} \, \cdot \,\sigma_6 \, \cdot \, \rho^{-1} \, \cdot \,\sigma_7 \, \cdot \, \rho^{-1} \, \cdot \,\sigma_9 \, ,  \\
\gamma_3  \,  & =  \,  (\rho    \,    \sigma_6       \rho^{-1}       \sigma_7     \rho^{-1}       \sigma_9 )^{-1}\,  \cdot   \,      \rho^{-1}       \sigma_4 \,       \rho^{-1}   \sigma_5     \, \cdot \,  (\rho  \,      \sigma_6       \rho^{-1}       \sigma_7       \rho^{-1}       \sigma_9 ) \\
\gamma_4 \, & = \, (\rho  \, \cdot \, \sigma_9)^{-1} \, \cdot \, \rho^{-1}  \, \cdot \, \sigma_7  \, \cdot \, \rho^{-1}  \, \cdot \, \sigma_8  \, \cdot \, (\rho  \, \cdot \, \sigma_9) \, .
\end{align}
The multipliers of these generators are
\begin{align}
k_1 & = p_2 \, p_3\, p_5\, p_6 \, p_8 \, p_9 \, ,
&
k_2 & = p_1 \, p_2 \, p_4 \, p_6 \, p_7 \, p_9 \, ,
&
k_3 & = p_4 \, p_5 \, ,
&
k_4 & = p_7 \, p_8 \, .
\end{align}
Clearly, the diagonal elements of \eq{eq:4loopPM} are given by
\begin{align}
\theta_{ii} & =  - \alpha' \log k_i \,
\end{align}
after making the replacement \eq{pteq}.
The off-diagonal entries come from the second term in \eq{pmseries}. We can compute all the non-vanishing contributions as
\begin{align}
(-\log \widehat \eta_{ij}) & = \left( \begin{array}{cccc}
0 & - \log p_9 & 0 & - \log p_8 \\
- \log p_9 & 0 & - \log p_4 & - \log p_7 \\
0 & - \log p_4 & 0 & 0 \\
- \log p_8 & - \log p_7 & 0 & 0
\end{array}\right) \, + \, {\cal O}(p_n) \label{4loopPMod1}
\\
(-\log \widehat \eta_{i, \bar 4 j 4}) & = \left( \begin{array}{cccc}
0 & 0 & 0 & 0 \\
- \log p_6 & 0 & 0 & 0 \\
- \log p_5 & 0 & 0 & 0 \\
0 & 0 & 0 & 0
\end{array}\right)\, + \, {\cal O}(p_n)\label{4loopPMod2}
\\
(-\log \widehat \eta_{i,  4 j\bar 4}) & = \left( \begin{array}{cccc}
0 & - \log p_6  & - \log p_5 & 0 \\
0& 0 & 0 & 0 \\
0 & 0 & 0 & 0 \\
0 & 0 & 0 & 0
\end{array}\right)\, + \, {\cal O}(p_n)\label{4loopPMod3}
\\
(-\log \widehat \eta_{i,  34 j\bar 4\bar 3}) & = \left( \begin{array}{cccc}
0 & - \log p_2  & 0 & 0 \\
0& 0 & 0 & 0 \\
0 & 0 & 0 & 0 \\
0 & 0 & 0 & 0
\end{array}\right)\, + \, {\cal O}(p_n)\label{4loopPMod4}
\\
(-\log \widehat \eta_{i,  \bar 4\bar 3 j 34}) & = \left( \begin{array}{cccc}
0 & 0 & 0 & 0 \\
- \log p_2 & 0 & 0 & 0 \\
0 & 0 & 0 & 0 \\
0 & 0 & 0 & 0
\end{array}\right)\, + \, {\cal O}(p_n)  \, ,\label{4loopPMod5}
\end{align}
where we've used the short-hand notation $\widehat \eta_{i , \alpha j \bar \alpha}$ to mean $\widehat \eta_{i , \alpha j \bar \alpha} \equiv 1$ if $i=j$ and $\gamma_\alpha = \text{id}$ or if the left-most factor of  $\gamma_\alpha$ is $\gamma_i^{\pm n}$ or the right-most factor is $\gamma_j^{\pm n}$; otherwise $\widehat\eta_{i , \alpha j \bar \alpha} \equiv \eta_{i , \alpha j \bar \alpha}$ as defined in \eq{etacr}. Clearly, when all these terms are inserted in the sum \eq{pmseriesnocr} along with the diagonal elements, we can use \eq{pteq} to arrive at
\begin{align}
\theta_{ij} & = \lim_{\alpha' \to 0} \text{Im}(2 \pi \alpha' \tau_{ij}) \,
\end{align}
for the graph period matrix in \eq{eq:4loopPM}.

To see an application of the construction in \Fig{fig:PMproof2} which was used to arrive at \eq{etaafterloop1} and \eq{etaafterloop2} in the proof above, let us consider one particular entry in the period matrix---say, $\tau_{12}$. It is clear from Eqs.~(\ref{4loopPMod1})--(\ref{4loopPMod5}) that the important contributions to $\tau_{12}$ come from
\begin{align}
\log{\eta_{12}} & \sim \log p_9  \, ,&
\log{\eta_{1,4 2 \bar 4}} & \sim \log p_6 \, ,  &
\log{\eta_{1,342\bar 4 \bar 3}} & \sim \log p_2  \, .  \label{t12contribs}
\end{align}
To see why these are the important Schottky group elements, recall from the proof above that each cross-ratio $\log \eta_{1,\alpha 2 \bar \alpha}$ can be computed with \eq{etaiaja} by finding the multiplier of $k_{1 \alpha 2 \bar \alpha}$ and noting how it differs from $k_1 k_2$ at leading order.

\begin{figure}
\centering
\subfloat[]{ 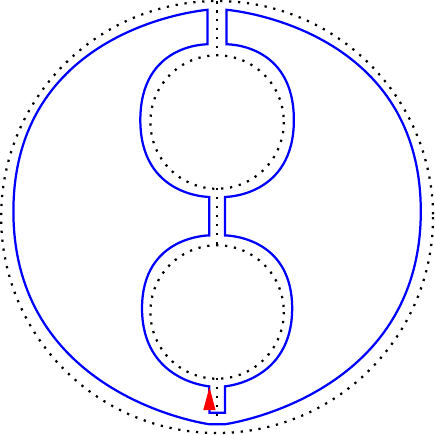 \label{fig:4loopPM2a} }
\subfloat[]{ 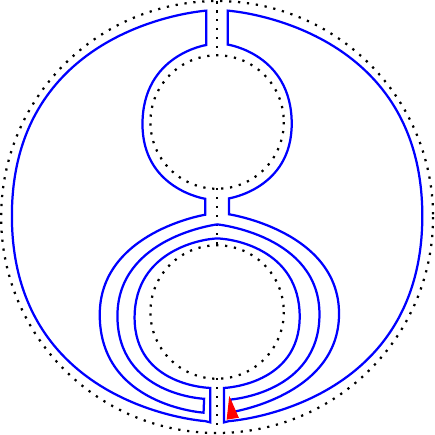 \label{fig:4loopPM2b} }
\subfloat[]{ 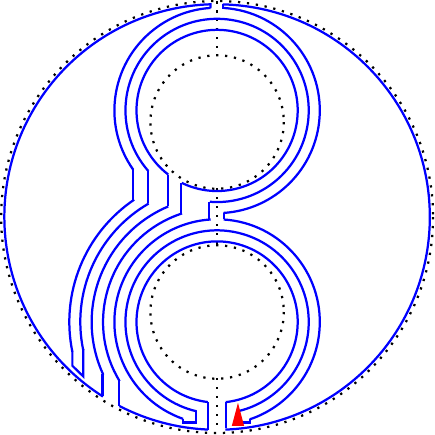 \label{fig:4loopPM2c} }
\caption{Starting from the red arrows, the blue curves indicate paths of the form $P_{1 \alpha 2 \bar \alpha} \equiv P_1 \cdot P_\alpha \cdot P_2 \cdot P_\alpha^{-1}$ for $\gamma_\alpha = \text{id}$ (\Fig{fig:4loopPM2a}), $\gamma_\alpha = \gamma_4^{-1}$ (\Fig{fig:4loopPM2a}) and $\gamma_\alpha = \gamma_4^{-1} \gamma_3^{-1}$ (\Fig{fig:4loopPM2a}), for the marked graph in \Fig{fig:4loopPM1}. }\label{fig:4loopPM2}
\end{figure}

\begin{figure}
\centering
\subfloat[]{ 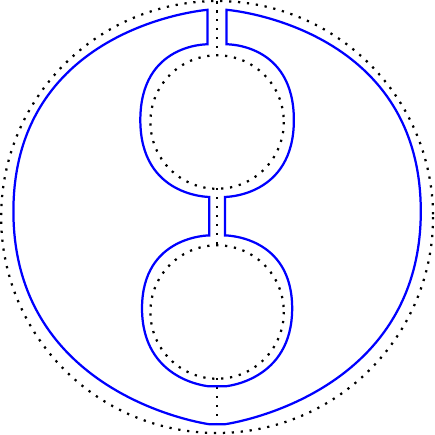 \label{fig:4loopPM3a} }
\subfloat[]{ 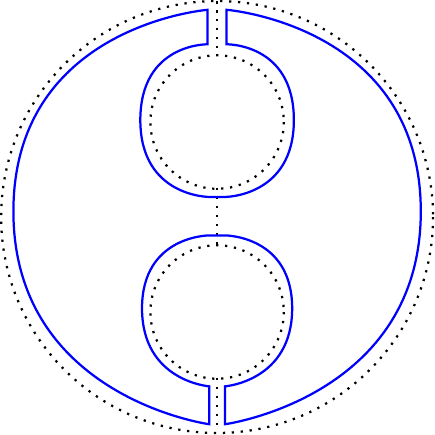 \label{fig:4loopPM3b} }
\subfloat[]{ 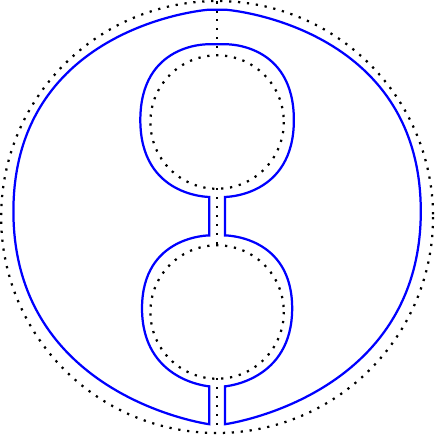 \label{fig:4loopPM3c} }
\caption{
By deleting pairs of consecutive edges that are inverses of each other in the paths in \Fig{fig:4loopPM2}, we arrive at the loops shown here.
}\label{fig:4loopPM3}
\end{figure}

In \Fig{fig:4loopPM2} we have drawn the paths $P_{1 \alpha 2 \bar \alpha} \equiv P_1 \cdot P_\alpha \cdot P_2 \cdot P_\alpha^{-1}$ for the three relevant Schottky group elements, namely $\gamma_\alpha = \text{id}$, $\gamma_\alpha = \gamma_4^{-1}$ and $\gamma_\alpha = \gamma_4^{-1} \gamma_3^{-1}$.
These paths can be reduced by removing pairs of consecutive edges that are inverses of each other; the resulting reduced paths (shown in \Fig{fig:4loopPM3}) can be used to read off the multipliers $k_{1 \alpha 2 \bar \alpha}$. These reduced paths cross the same edges that $\ell_1$ and $\ell_2$ do, with the exception of one edge in each case ($E_9$ in \Fig{fig:4loopPM3a}, $E_6$ in \Fig{fig:4loopPM3b} and $E_2$ in \Fig{fig:4loopPM3c}).
In accordance with the calculation in the above proof (particularly  \eq{etaafterloop1}), this explains why at leading order $\tau_{12}$ gets the three non-zero contributions listed in \eq{t12contribs}, with one corresponding to each of the three respective edges where $\ell_1$ and $\ell_2$ intersect.

We have seen above that the necessity of including non-identity Schottky group elements in \eq{pmseries} comes from pairs of loops $\ell_i$, $\ell_j$ whose intersections have $>1$ connected components. Although $\ell_1$ and $\ell_3$ intersect in just one edge ($E_5$), the non-vanishing contribution to $\tau_{13}$ comes from $\gamma_\alpha = \gamma_4^{-1}$ in \eq{4loopPMod3}, not from $\gamma_\alpha = \text{id}$. This is because we have used the same base point (indicated on \Fig{fig:4loopPM1}) for each period matrix entry $\tau_{ij}$, instead of using \psl~invariance to choose a new one common to $\ell_i$ and $\ell_j$ in each case.

\subsection{Green's function}
\label{greenfapp}

In this section we prove that the worldline Green's function arises as the $\alpha ' \to 0$ limit of the worldsheet Green's function, as described in section \ref{graph}.
In section \ref{primeformleading} and \ref{primeformrest} we show \eq{sClaim}, then in section \ref{greenrem} we prove \eq{vClaim}.

\subsubsection{Leading factor}
\label{primeformleading}
In this section we show that when expressed in terms of pinching parameters, the Schottky-Klein prime form converges to the length of a path in the graph, as in \eq{sClaim}.
We will begin by considering the first factor $(x_1-x_2)/(\sqrt{\d x_1\,}\sqrt{ \d x_2})$ in \eq{primeform}. It turns out that this is the only factor which gives a non-vanishing contribution at leading order in the pinching parameters. The computation will be re-used in section \ref{primeformrest} to show that the rest of the factors vanish.

Choose a coordinate system at $X_1$ (\ie~a path from $X_1$ to the base point) and pick some path $P$ from $X_2$ to $X_1$. This determines a coordinate system at $X_2$: let $W= \phi(P)$ (where $\phi$ is the homomorphism from paths to M\"obius maps defined in section \ref{param}) and take
\begin{align}
V_2 & = V_1 \cdot W \, , \label{V2simp}
\end{align}
then the factor we want to compute is given by
\begin{align}
 \frac{V_1(0) - V_2(0)}{\sqrt{ V_1'(0) V_2'(0)}} & \equiv
 \frac{V_1(0) - (V_1\!\cdot  W) (0)}{\sqrt{ V_1'(0) (V_1\!\cdot  W)'(0)}} \equiv f(V_1,W) \, . \label{fdef}
\end{align}
If we write
\begin{align}
W & = W_N \equiv \rho^{\pm 1} \sigma_{{i_1}} \rho^{\pm 1} \sigma_{{i_2}} \rho^{\pm 1} \ldots \rho^{\pm 1} \sigma_{{i_N}} \rho^{\pm 1} \, ,
\end{align}
where $N$ is the number of edges in the path $P$, then we can denote the sum of the logs of the pinching parameters in this path as
\begin{align}
S_P & = \sum_{E_i \in P} \log p_i \, .  \label{SPdef}
\end{align}
Then we claim that
\begin{align}
 \log | f(V_1, W)| & = - \frac{1}{2} S_P  + {\cal O}(p_{i_m}) \, . \label{ssimpsum}
\end{align}

We can show this by induction on $N$ (the number of edges in the path $P$ between $X_1$ and $X_2$). Let's start with the $N=0$ base case in which the two external edges $X_1$ and $X_2$ must be on the same cubic vertex and thus
\begin{align}
W= W_{0} & = \rho^{\pm 1} \, .
\end{align}
Let us write $V_1  =  V = \big( \begin{smallmatrix} a & b \\ c & d \end{smallmatrix} \big)$ with $ad-bc=1$, so $z_1  = b / d$ and $V_1' (0) = 1/d^2$.

 First we can check the case with $W_{0}= \rho$, which gives  $z_2= a/c$ and $V_2'(0) = 1/c^2$. Then
 \begin{align}
\log \frac{|z_1 - z_2|}{\sqrt{\log V_1'(0) V_2'(0)}} & = \log \frac{| b/d - a/c |}{\sqrt{{1}/{c^2d^2}}} = \log|ad-bc| =  0 \, .
 \end{align}
 Alternatively, if $W _{0} = \rho^{-1}$ so  $z_2 = \frac{a+b}{c+d}$ and $V_2'(0) = 1/(c+d)^2$, then
 \begin{align}
\log \frac{|z_1 - z_2|}{\sqrt{\log V_1'(0) V_2'(0)}}  & = \log \frac{ |a/c - (a+b)/(c+d)|}{\sqrt{(1/(c+d)^2)(1/c^2)}} = 0 \, ,
 \end{align}
 Since the sum in \eq{ssimpsum} has no terms for $N=0$ it should vanish, so the base case has been checked.

It remains to do the inductive step: for $m \geq 1$ let us denote by $C(m)$ the statement
\begin{align}
\log \Big| \frac{ (V\!\cdot W_m)(0) - V(0)}{\sqrt{(V\!\cdot W_m)'(0) V'(0)}} \Big| & = - \frac{1}{2 } \sum_{n=1}^m \log p_{i_n} \, \, + \, \, {\cal O} (p_{i_\ell}) \, ,
\end{align}
then we want to show that $C(m) \Rightarrow C(m+1)$. This is equivalent to
\begin{align}
\log \Big| \frac{ (V\!\cdot W_{m+1})(0) - V(0)}{ (V\!\cdot W_{m})(0) - V(0)} \sqrt{\frac{(V\!\cdot W_{m})'(0)}{(V\!\cdot W_{m+1})'(0) }} \Big| & = - \frac{1}{2} \log p_{i_{m+1}} + { \cal O}(p_{i_n}) \, \label{intermed1} .
\end{align}
We can show this by expanding $(V\!\cdot W_{m+1})(0)$ and $(V\!\cdot W_{m+1})'(0)$ as power series in $p_{i_{m+1}}$.
By factorizing $(V\!\cdot W_{m+1})(z) $ as
\begin{align}
(V\!\cdot W_{m+1})(z) = (V\!\cdot W_{m})(  \sigma_{i_{m+1}} \cdot \rho^{\pm 1} (z)) \label{VWfac}
\end{align}
we can power expand:
\begin{align}
(V\!\cdot W_{m+1})(z)  & = (V\!\cdot W_{m} )(0) + (V\!\cdot W_{m} )'(0) \,  ( \sigma_{i_{m+1}} \cdot \rho^{\pm 1})(z) + \ldots \, .
\label{VWz}
\end{align}
Similarly, Using the chain rule on \eq{VWfac} gives
\begin{align}
 (V\!\cdot W_{m+1})'(z) & = (V\!\cdot W_{m})'(  \sigma_{i_{m+1}} \cdot \rho^{\pm 1} (z)) \, \times \, (  \sigma_{i_{m+1}} \cdot \rho^{\pm 1})'(z) \, , \label{VWderiv}
\end{align}
and the first factor $(V\!\cdot W_{m})'$ can be power-expanded as above,
\begin{align}
(V\!\cdot W_{m+1})'(z) & = \Big( (V \!\cdot W_{m})'(0) \, + \,  (  \sigma_{i_{m+1}} \cdot \rho^{\pm 1} (z))\, (V \!\cdot W_{m})''(0) + \ldots \Big)
\label{VWpz}
 \\
& \hspace{180pt} \times (  \sigma_{i_{m+1}} \cdot \rho^{\pm 1})'(z) \, .
\nonumber
\end{align}
To find the dependence of $(V\!\cdot W_{m+1})(0)$ and $(V\!\cdot W_{m+1})'(0)$ on $p_{i_{m+1}}$, we can use
\begin{align}
(\sigma_{i_{m+1}} \rho )(z)  & =p_{i_{m+1}} \,  \frac{ z}{1-z} \label{srz}
\shortintertext{and}
(\sigma_{i_{m+1}} \rho ^{-1} )(z) & = - p_{i_{m+1}}(1-z) \, , \label{srpz}
\end{align}
to see
\begin{align}
(\sigma_{i_{m+1}} \rho )(0) & = 0 \, ,
&
(\sigma_{i_{m+1}} \rho^{-1}  )(0) & = - p_{i_{m+1}} \, ;
\end{align}
we will use the observation that these are both of the general form
\begin{align}
(\sigma_{i_{m+1}} \rho^{\pm 1} )(0) & = {\cal O}(p_{i_{m+1}}) \, . \label{sratzero}
\end{align}
In both \eq{srz} and \eq{srpz} the derivative at $z=0$ is the same:
\begin{align}
(\sigma_{i_{m+1}} \rho^{\pm 1} )' (0) & = p_{i_{m+1}} \, .  \label{srpatzero}
\end{align}
Now, putting $z=0$ in \eq{VWz} and inserting \eq{sratzero} we see that
\begin{align}
(V\!\cdot W_{m+1})(0)  & = (V\!\cdot W_{m} )(0) + {\cal O}(p_{i_{m+1}}) \, . \label{VW0}
\end{align}
Similarly, putting $z=0$ in \eq{VWpz} and inserting \eq{srpatzero} and \eq{sratzero} we see that
\begin{align}
(V\!\cdot W_{m+1})'(0)  & = p_{i_{m+1}} \, \Big( (V\!\cdot W_{m} )'(0) \,  +  \, {\cal O}(p_{i_{m+1}}) \, \Big) \, . \label{VWp0}
\end{align}
Now, with the use of \eq{VW0}, the first factor inside the log on the left-hand-side of \eq{intermed1} becomes
\begin{align}
\log \Big| \frac{ (V\!\cdot W_{m+1})(0) - V(0)}{ (V\!\cdot W_{m})(0) - V(0)} \Big| & = \log \Big|1 \, + \, {\cal O}(p_{i_{m+1}}) \,  \frac{1}{ (V\!\cdot W_{m})(0) - V(0)} \Big| \nonumber \\
& = {\cal O}(p_{i_{m+1}}) \, . \label{term1im1}
\end{align}
Similarly, with the use of \eq{VW0} and \eq{VWp0} the other term on the left-hand side of \eq{intermed1} becomes
\begin{align}
\log \Big| \sqrt{\frac{(V\!\cdot W_{m})'(0)}{(V\!\cdot W_{m+1})'(0) }} \Big|& =
\log \Big| \sqrt{\frac{1}{p_{i_{m+1}}}\big(1  + {\cal O}(p_{i_{m+1}}) \big) }\Big| \nonumber \\
& = - \frac{1}{2} \log p_{i_{m+1}} \, + \, {\cal O}(p_{i_{m+1}}) \, . \label{term2im1}
\end{align}
Adding \eq{term1im1} and \eq{term2im1}, we can check that \eq{intermed1} holds for all $m \geq 0$ and thus by induction, \eq{ssimpsum} holds for all $N$.

This proves that
\begin{align}
\log | E(x_1, x_2) | & = - \frac{1}{2} S_P \, + \, {\cal O}(p_{i}) \, .
\end{align}

\subsubsection{The rest of the prime form factors}
\label{primeformrest}
The rest of the factors in the prime form are of the form
\begin{align}
F_{\alpha}(x_1,x_2) & \equiv \frac{x_1 - \gamma_\alpha(x_2)}{x_1 - \gamma_\alpha(x_1)} \frac{x_2 - \gamma_\alpha(x_1)}{x_2 - \gamma_\alpha(x_2)} \, .
\end{align}
These factors don't contribute at order ${\cal O}(p_i)$,
\begin{align}
\log F_{\alpha}(x_1,x_2) & = {\cal O}(p_i) \, .  \label{Falphavan}
\end{align}
To see why this is true, first note that since $F_\alpha(x_1,x_2) \equiv F_\alpha(V_1(0),V_2(0))$ is a cross-ratio, it is a projective invariant so we can make a change of coordinates to replace $V_1$, $V_2$ and $\gamma_\alpha$ with
\begin{align}
\hat V_1 & = \text{id} \, ,
&
\hat V_2 & = V_1^{-1} V_2 \, ,
&
\hat \gamma_\alpha = V_1^{-1} \gamma_\alpha V_1  \, . \label{primehatdef}
\end{align}
Then we can write $F_\alpha(x_1,x_2)$ in the form
\begin{align}
F_\alpha(x_1,x_2) & = \frac{\text{id}(0)- (\hat \gamma_\alpha \hat V_2)(0)}{\text{id}(0)- \hat \gamma_\alpha(0)} \cdot \frac{\hat V_2 (0)- \hat \gamma_\alpha(0)}{\hat V_2(0)- (\hat \gamma_\alpha \hat V_2)(0)} \, .
\end{align}
In terms of the function
$f(V,W)$ defined in \eq{fdef},
$F_\alpha(x_1,x_2)$ may be expressed as
\begin{align}
F_\alpha(x_1,x_2)& = \frac{f(\text{id},\hat \gamma_\alpha  \hat V_2)}{f(\text{id}, \hat \gamma_\alpha )}\frac{f(\hat V_2, \hat V_2^{-1} \hat \gamma_\alpha)}{f(\hat V_2, \hat V_2^{-1} \hat \gamma_\alpha \hat  V_2)} \, .
\end{align}
Using \eq{ssimpsum} and the results of section \ref{primeformleading}, we can express $F_\alpha(x_1,x_2)$ in terms of the `lengths' of some paths in the graph:
\begin{align}
\log |F_\alpha(x_1, x_2)|
& = - \frac{1}{2}\Big| \hat S_{\alpha 2} - \hat S_{\bar 2 \alpha 2} +  \hat S_{\bar 2 \alpha} - \hat S_{\alpha } \Big|+ {\cal O}(p_i) \, , \label{pathedgesum}
\end{align}
where $\hat{S}_{\bar 2 \alpha} $ means $S_{\hat P_{2}^{-1} \hat P_\alpha}$ and so on, with $\hat P_2$ and $\hat P_\alpha$ being the paths corresponding to the maps $\hat V_2$ and $\hat \gamma_2$ defined in \eq{primehatdef}.

The sum over edges on the right-hand side of \eq{pathedgesum} vanishes. In fact, the path edge sums $S_\alpha$ obey the sum rule
\begin{align}
S_{\beta \delta} & = S_{\beta} + S_{\delta } - 2 S_{\widehat{\beta \delta}} \, ,
\end{align}
where $S_{\widehat{\beta \delta}}$ means the contribution from any right-most factors of the path $P_\beta$ that cancel left-most factors of the path $P_\delta$. Then
\begin{align}
\widehat{S}_{\alpha 2} & = \widehat S_\alpha + \widehat S_2 - 2 \widehat S_{\widehat{\alpha 2}} \\
\widehat{S}_{\bar 2 \alpha 2} & =\widehat  S_{\bar 2} +\widehat  S_\alpha +\widehat  S_2 - 2\widehat  S_{\widehat{\bar 2 \alpha}} - 2\widehat  S_{\widehat{\alpha 2}} \\
\widehat{S} _{\bar 2 \alpha}  & =\widehat  S_{\bar 2} +\widehat  S_\alpha - 2 \widehat S_{\widehat{\bar 2 \alpha} }
\end{align}
and so
\begin{align}
\widehat{S}_{\alpha 2} - \widehat S_{\bar 2 \alpha 2} + \widehat S_{\bar 2 \alpha} - \widehat{S}_\alpha & = 0 \, .
\end{align}
This shows that \eq{primeformrest} doesn't contribute at leading order.

\subsubsection{The second term in the Green's function}
\label{greenrem}
The second term in the Green's function \eq{wsgf} ensures that ${\cal G}(x_1,x_2)$ is single-valued on $\Sigma \times \Sigma$ \cite{d'hokerlecture}. For our purposes, this means that the Green's function is independent of the choice of path $P$ from $X_2$ to $X_1$ used to define the coordinate chart $V_2$. Therefore certainly when we find $\lim_{\alpha ' \to 0}\alpha'{\cal G}(x_1,x_2)$, we must also find an expression that is independent of the path $P$.

We have proven in the previous section that in the $\alpha ' \to 0 $ limit, the first term in ${\cal G}(x_1,x_2)$ asymptotes to $s/2$, where $s$ is the sum of the Schwinger parameters $t_i$ in the path $P$ (\eq{sdef}). This means that the second term in the Green's function should asymptote to some expression
\begin{align}
f(\Gamma, P, t_i) & \equiv \lim_{\alpha ' \to 0} \alpha '  \Big( \int_{x_1}^{x_2} \vec{\omega} \Big) \cdot ( 2 \pi \text{Im} \tau)^{-1} \cdot  \Big( \int_{x_1}^{x_2} \vec{\omega} \Big)
\end{align}
such that the combination
\begin{align}
-\frac{1}{2}s + \frac{1}{2} \, f(\Gamma, P, t_i)
\end{align}
is independent of which path $P$ is chosen between external edges $X_1$ and $X_2$ on $\Gamma$, and vanishes when the two external edges coincide.

The function which satisfies this symmetry requirement is
\begin{align}
f(\Gamma, P, t_i) & = \vec v \,{} ^{\text{t}}  \cdot \, \theta^{-1} \, \cdot \, \vec v \, ,
\end{align}
where the graph period matrix and the vector $\vec v$ associated to the path $P$ are defined in \eq{thetadef} and \eq{vdef}, respectively.

This completes the proof of the claim that the worldline Green's function arises as the limit of the worldsheet Green's function,
\begin{align}
G(X_1, X_2) & = - \, \lim_{\alpha ' \to 0}  \alpha' {\cal G}(x_1, x_2) \, ,
\end{align}
and in particular this is an indirect proof that $\int_{x_1}^{x_2} \vec \omega$ has the asymptotics given in \eq{vClaim}, since we already checked in section \ref{pmproof} that the asymptotics of $\tau_{ij}$ are given by \eq{thetaClaim}.

\bibliographystyle{jhep}
\bibliography{rs}
\end{document}